\documentclass[10pt,journal,compsoc]{IEEEtran}
\usepackage{acronym}
\usepackage[british]{babel}
\usepackage{graphicx}
\usepackage{subfig}
\usepackage{amsmath}
\usepackage{amsthm}
\usepackage{amssymb}
\usepackage{algpseudocode}
\usepackage{listings}
\usepackage{array}
\usepackage{balance}
\usepackage{enumitem}
\usepackage{mdframed}
\usepackage{url}
\usepackage[ruled]{algorithm2e} 
\usepackage{multirow}
\usepackage{rotating}
\usepackage[table]{xcolor}
\usepackage{mathtools}
\usepackage{fancyhdr}
\usepackage{tikz}
\usepackage{hyperref}
\hypersetup{
    colorlinks=true,
    linkcolor=blue,
    filecolor=blue,      
    urlcolor=blue,
    citecolor=blue
}

\usepackage{multirow}
\usepackage{xcolor}
\usepackage{multicol}
\usepackage{textcomp}
\usepackage{color, colortbl}
\usepackage{colortbl}

\definecolor{LightCyan}{rgb}{0.8,0.93,1}
\definecolor{Lightgray}{rgb}{0.90,0.98,1.00}

\def\checkmark{\tikz\fill[scale=0.4](0,.35) -- (.25,0) -- (1,.7) -- (.25,.15) -- cycle;} 

\usepackage{stackengine}

\newif\ifshowcomments
\showcommentstrue

\ifshowcomments
\newcommand{\mynote}[2]{\fbox{\bfseries\sffamily\scriptsize{#1}}
	{\small$\blacktriangleright$\textsf{\emph{#2}}$\blacktriangleleft$}}
\else
\newcommand{\mynote}[2]{}
\fi

\newcolumntype{P}[1]{>{\centering\arraybackslash}p{#1}}

\ifCLASSINFOpdf
\else
\fi
\fancypagestyle{AllPages}{
}
\fancypagestyle{FirstPage}{
}
\begin{document}
%
\title{Blockchain Consensus Algorithms: A Survey}
%
%
%

\author{Md Sadek Ferdous,~\IEEEmembership{Member,~IEEE,}
        Mohammad~Jabed~Morshed~Chowdhury, Mohammad~A.~Hoque,~\IEEEmembership{Member,~IEEE}, and~ Alan Colman
\thanks{M. S. Ferdous is with Shahjalal University of Science and Technology, Sylhet 3114, Bangladesh and Imperial College London, London SW7 2AZ, U.K. E-mail: sadek-cse@sust.edu.}
\thanks{\par \noindent Mohammad Jabed Morshed Chowdhury is with La Trobe University, Melbourne, Victoria-3086, Australia. E-mail: m.chowdhury@latrobe.edu.au.}
\thanks{\par \noindent Mohammad A. Hoque is with University of Helsinki, 3835 Helsinki, Helsinki Finland. E-mail: mohammad.a.hoque@helsinki.fi.}%
\thanks{\par \noindent Alan Coleman is with Swinburne University of Technology, Hawthorn, Australia. E-mail: acolman@swin.edu.au.}}%
\maketitle
\thispagestyle{FirstPage} 

\begin{abstract}
In recent years, blockchain technology has received unparalleled attention from academia, industry, and governments all around the world. It is considered a technological breakthrough anticipated to disrupt several application domains touching all spheres of our lives. The sky-rocket anticipation of its potential has caused a wide-scale exploration of its usage in different application domains. This has resulted in a plethora of blockchain systems for various purposes. However, many of these blockchain systems suffer from serious shortcomings related to their performance and security, which need to be addressed before any wide-scale adoption can be achieved. A crucial component of any blockchain system is its underlying consensus algorithm, which in many ways, determines its performance and security. Therefore, to address the limitations of different blockchain systems, several existing as well novel consensus algorithms have been introduced. A systematic analysis of these algorithms will help to understand how and why any particular blockchain performs the way it functions. However, the existing studies of consensus algorithms are not comprehensive. Those studies have incomplete discussions on the properties of the algorithms and fail to analyse several major blockchain consensus algorithms in terms of their scopes. This article fills this gap by analysing a wide range of consensus algorithms using a comprehensive taxonomy of properties and by examining the implications of different issues still prevalent in consensus algorithms in detail. The result of the analysis is presented in tabular formats, which provides a visual illustration of these algorithms in a meaningful way. We have also analysed more than hundred top crypto-currencies belonging to different categories of consensus algorithms to understand their properties and to implicate different trends in these crypto-currencies. Finally, we have presented a decision tree of algorithms to be used as a tool to test the suitability of consensus algorithms under different criteria. 
\end{abstract}

\begin{IEEEkeywords}
Blockchain, Distributed Consensus, Proof of Work, PoW, Proof of Stake, PoS, Delegated Proof of Stake, DPoS.
\end{IEEEkeywords}

%
\IEEEpeerreviewmaketitle

\section{Introduction}
\label{sec:introduction}

In the last few years, blockchain has received wide-spread attention among the industry, the Government, and academia alike. This interest has been piqued by the success of Bitcoin \cite{nakamoto2008bitcoin} that was introduced in 2008.  While crypto-currencies have emerged as the principal and the most popular application of blockchain technology, many enthusiasts from different disciplines have identified and proposed a plethora of applications of blockchain in a multitude of application domains \cite{pilkington201611, crosby2016blockchain}. The possibility of exploiting blockchain in so many areas has created huge anticipation surrounding blockchain systems. Indeed, it is regarded as one of the fundamental technologies to revolutionise the landscapes of the identified application domains.

A blockchain system is, fundamentally, a distributed system that relies on a consensus algorithm that ensures agreement on the states of certain data among distributed nodes. A consensus algorithm is the core component that directly dictates how such a system behaves and the performance it can achieve. Distributed consensus has been a widely studied research topic in distributed systems, however, with the advent of blockchain, it has received renewed attention. A wide variety of crypto-currencies targeting different application domains has introduced an array of unique requirements that can only be satisfied by their corresponding consensus mechanisms. This fact has fuelled the need not only to examine the applicability of existing consensus algorithms in newer settings, but also to innovate novel consensus algorithms. Consequently, several consensus algorithms have emerged, each of which possesses interesting properties and unique capabilities.

As the characteristics of various types of blockchain systems are fundamentally dependent on the consensus algorithms they use, a  systematic analysis of existing consensus algorithms is required.  It is necessary to examine, compare, and contrast these algorithms.  There have a been a number of attempts aiming to fulfil this goal can be found in \cite{Cachin2017, BanoSOK2017, WangConsensus2019, baligaCon2017, sankar2017survey, seibold2016consensus, mukhopadhyay2016brief}. 
In particular, the works carried out by Cachin et al. \cite{Cachin2017} and Bano et al. \cite{BanoSOK2017} are noteworthy as they represent the pioneer works in this scope. Cachin et al., in their work, have explored different aspects of distributed systems and consensus and focused on consensus algorithm deployed in blockchain systems that are not to open to the public. On the other hand, the focus of the work by Bano et al. is more general in the sense they have explored consensus algorithms used both in public as well as private systems. Another exceptional work is by Wang et al. \cite{WangConsensus2019} in which the authors have presented a comprehensive survey of different aspects of consensus, mining, and blockchains in a detailed fashion.

However, all these works have some major shortcomings. For example, the factors upon which the consensus algorithms have been analysed are not comprehensive. Importantly, a wide range of consensus algorithms and their internal mechanisms utilised in many existing crypto-currencies have not been considered at all. In addition, all of these studies have failed to capture the practical interrelation between blockchain systems (mostly crypto-currencies) and their corresponding consensus algorithms. All in all, there is a pressing need for a study that analyses a wide range of existing consensus algorithms and the blockchain systems in a practical-oriented way and synthesises this analyses into a conceptual framework in a concise yet comprehensive manner. The principal motivation of this article is to fill in this gap.







\vspace{2mm}
\noindent \textbf{Contributions.} The main contributions of the article are presented below:
\begin{itemize}
    \item A novel taxonomy of consensus properties, capturing different aspects of a consensus algorithm, has been created. In this taxonomy, consensus algorithms have been categorised in two major categories: incentivised and non-incentivised algorithms, which have been again sub-divided as per different considerations. Consensus algorithms belonging to each sub-category analysed together using the taxonomy of consensus properties.
    \item The analysis of each sub-category has been summarised in tabular formats so as to visually represent it in a comprehensible way.
    \item For each category (and the sub-category, if any), the corresponding blockchain systems (predominantly crypto-currencies) have been analysed as well. The analysis result has been presented in a concise fashion, which can be used to understand the inter-relation between these systems and their underlying consensus algorithms.
    \item The major issues in each category of consensus algorithm have been examined in detail, and their implications have been further analysed.
    \item Over hundred crypto-currencies, belonging to different consensus algorithms, have been examined to understand their different properties. These properties then have been utilised to analyse and identify different trends among these crypto-currencies.
    \item Finally, a decision tree of consensus algorithms have been presented. This tree can be utilised to test the suitability of a consensus algorithm under certain criteria.
\end{itemize}
In short, with these contributions, this article represents one of the most comprehensive studies of blockchain consensus algorithms as of now. 

\vspace{2mm}
\noindent \textbf{Structure.} In Section \ref{sec:backConsensus} we present a brief background on distributed consensus, highlighting its different components, types and properties. Section \ref{sec:back} outlines a brief presentation on blockchain covering its different aspects such as types, properties, layers. A taxonomy of consensus algorithms and their underlying properties is presented in Section \ref{sec:conTaxProp}. Section \ref{sec:incentivised} and Section \ref{sec:hybrid} analyse different incentivised consensus algorithm whereas Section \ref{sec:nonIncentivConsensus} examines the different non-incentivised consensus algorithms. Finally, we conclude in Section \ref{sec:disccusion} with a detailed discussion on different issues involving the analysed consensus algorithms and the corresponding crypto-currencies. 

\section{Background: Distributed Consensus}
\label{sec:backConsensus}
Consensus mechanisms in distributed systems have been a well studied research problem for nearly three decades. Such mechanisms enable consensus to be achieved regarding a shared state/data among a set of distributed nodes. The need for a shared state originated the notion of replicated database systems in order to ensure resilience against node failures within a network. Such database systems ensure that data is not lost when one or more nodes fail to function in an excepted fashion. 

The notion of the replicated database can be generalised with the concept of State Machine Replication (SMR) \cite{SMR1990}. The core idea behind SMR is that a computing machine can be expressed as a deterministic \textit{state machine}.  The machine accepts an input message, performs its predefined computation, and might produce an output/response. These actions essentially change its state. SMR conceptualises that such a state machine, with an initial state, can be replicated among different nodes. If it can be ensured that all the participating nodes receive the same set of input messages in the exact same order (the phenomenon known as \textit{atomic broadcast}), then each node would be able to evolve the states of its state machine individually in exactly the same fashion. This can guarantee consistency and availability regarding the state of the machine (as well as data it holds)  among all (applicable) nodes even in the presence of node failures. Once this occurs, it can be said that a distributed consensus has emerged among the participating nodes. It is imperative that a protocol is defined to ensure timely dissemination and atomic broadcast of input messages among the nodes and, in many ways, dictates how a distributed consensus is achieved and maintained. Hence, such a protocol is aptly called a consensus protocol.

Designing and deploying a consensus protocol is a challenging task as it needs to consider several crucial issues such as resiliency against node failures, node behaviour, network partitioning, network latency, corrupt or out-of-order inputs, and so on \cite{baligaCon2017}.  Schneider pointed out that there are two crucial requirements to reach and maintain consensus among distributed nodes. The first requirement is a deterministic state machine. The second requirement is a \textit{consensus protocol} to disseminate inputs in a timely fashion and to ensure atomic broadcast among the participating nodes. At the same time, the consensus protocols must ensure the properties of the atomic broadcast \cite{CachinIntro2011,Hadzilacos1993,Cachin2017, BanoSOK2017}. The properties of atomic broadcast in distributed consensus is illustrated in Table~\ref{tab:atomic}.
%
%

\begin{table}[t]
  \begin{tabular}{m{18mm}|m{60mm}}
    \hline
    \rowcolor[gray]{.6}
     \centering\textbf{Properties } &  \centering    \textbf{Note} \tabularnewline [2ex]
      \hline
      \hline
     Validity & This guarantees that if a message is broadcast by a valid node, it will be correctly included within the consensus protocol. \\\hline
    \rowcolor[gray]{.90}  Agreement  & This is to guarantee that if a message is delivered to a valid node, it will ultimately be delivered to all valid nodes. \\\hline
     Integrity & This is to ensure that a message is broadcast only once by a valid node.\\\hline
   \rowcolor[gray]{.90}   Total Order & This is to ensure that all nodes agree to the order of all delivered messages. \\\hline
\end{tabular}
  \caption{Atomic broadcast Properties of Distributed Consensus Protocols.}
  \label{tab:atomic}
\end{table}

\begin{table}[t]
  \begin{tabular}{p{18mm}|p{60mm}}
    \hline
        \rowcolor[gray]{.6}
   \centering\textbf{Properties } & \centering\textbf{Note}  \tabularnewline [2ex] \hline
   \hline
     Safety/ Consistency &  A consensus protocol is considered safe (or consistent) only when all nodes produce the same valid output, according to the protocol rules, for the same atomic broadcast. \\\hline
     \rowcolor[gray]{.9} Liveness/ availability & If all non-faulty participating nodes produce an output (indicating the termination of the protocol), the protocol is considered live.  \\\hline
    Fault Tolerance & It exhibits the network's capability to perform as intended in the midst of node failures.\\\hline
\end{tabular}
  \caption{Properties of Distributed Consensus Protocols.}
  \label{tab:conprotocols}
\end{table}

One way to achieve the design goals of such a protocol is to make certain assumptions under which the protocol is proved to function properly. These assumptions influence the critical characteristics of a consensus protocol. Next, we explore two sets of  widely-used assumptions for any distributed consensus protocol. 

The first set of assumptions are about the underlying networking type. Dwork et al. categorised three types of networks exhibiting different properties: synchronous, asynchronous, and partially/eventually synchronous \cite{DworkCon1990}. The latency involved in delivering a message to all nodes in a synchronous network is bound by some time denoted as $\Delta$. On the other hand, the latency in an asynchronous network cannot be reliably bound by any$\Delta$. Finally, in a partially/eventually synchronous network, it is assumed that the network will eventually act as a synchronous network, even though it might be asynchronous over some arbitrary period of time.
%
%
%
%

The second set of assumptions is about the different properties of a consensus protocol. According to \cite{baligaCon2017}, a consensus protocol should have the following three properties; namely consistency, availability, and fault tolerance. These properties are elaborated in Table~\ref{tab:conprotocols} A well-known theorem, by Fischer, Lynch and Paterson \cite{FLPImpos1985}, called \textit{FLP Impossibility} has shown that a deterministic consensus protocol cannot satisfy all three properties described above in an asynchronous network. It is more common to tend to favour safety and liveness over fault tolerance in the domain of distributed system applications. A related theorem is the CAP theorem \cite{CAPTheorem2000}, which states that a shared replicated datastore (or, more generally, a replicated state machine) cannot achieve both consistency and availability when a network partitions in such a way that an arbitrary number of messages might be dropped. 
%
%
%
%

In addition to the above assumptions, there are two major fault-tolerance models within distributed systems: crash failure (or tolerance) and Byzantine failure \cite{BanoSOK2017, baligaCon2017, Cachin2017}. The crash failure model deals with nodes that simply fail to respond due to some hardware or software failures. It may happen any time without any prior warning, and the corresponding node remains unresponsive until further actions are taken. Byzantine failure, on the other hand, deals with nodes that misbehave due to some software bugs or because of the nodes being compromised by an adversary. This type of failure was first identified and formalised by Leslie Lamport in his seminal paper with a metaphorical Byzantine General's problems \cite{LamportByzantine1982}. A Byzantine node can behave maliciously by arbitrarily sending deceptive messages to others, which might affect the security of distributed systems. Hence, such nodes are mostly relevant in application with security implications.
%
%

To handle these two failure models, two corresponding major types of consensus mechanisms have emerged: Crash-tolerant consensus and Byzantine consensus \cite{Cachin2017}. Next, we briefly discuss each of them, along with their associated properties.

\begin{enumerate}
    \item \textbf{Crash-tolerant consensus:} Algorithms belonging to this class aim to guarantee the atomic broadcast (total order) of messages within the participating nodes in the presence of certain number of node failures. These algorithms utilise the notion of views or epochs, which imply a certain duration of time or events. A leader is selected for each epoch who takes decisions regarding the atomic broadcast, and all other nodes comply with its decision. In case a leader fails due to a crash failure, the protocols elect a new leader to function. The best known algorithms belonging to this class can continue to function if the following condition holds: $t < n/2$ where $t$ is the number of faulty nodes and $n$ is the total number of participating nodes \cite{Cachin2017}. Examples of some well-known crash-tolerant consensus protocol are: Paxos \cite{LamportPart1998, LamportPaxos2001}, Viewstamped Replication \cite{OkiVSR1988}, ZooKeeper \cite{HuntZookeeper2010}, and Raft \cite{OngaroRaft2014}.
    \item \textbf{Byzantine consensus:} This class of algorithms aim to reach consensus amid of certain nodes exhibiting Byzantine behaviour. Such Byzantine nodes are assumed to be under the control of an adversary and behave unpredictably with malicious intent. Similar to any crash-tolerant consensus protocol, these protocols also utilise the concept of views/epochs where a leader is elected in each view to order messages for atomic broadcast, and other honest nodes are assumed to follow the instructions from the leader. One of the most well-known algorithms under this class is called Practical Byzantine Fault Tolerant (PBFT), which can achieve consensus in the presence of a certain number of Byzantine nodes under an eventual synchronous network assumption \cite{PBFT2002}. The tolerance level of PBFT is $f < n/3$, where $f$ the number of Byzantine nodes and $n$ denotes the number of total nodes participating in the network \cite{Cachin2017}. As we will explore later, PBFT algorithms have been widely utilised in different blockchain systems.
\end{enumerate}




\section{Background: Blockchain}
\label{sec:back}
In this section, we present a brief introduction to the blockchain technology and it related terminologies. At the centre of the blockchain technology is the blockchain itself stored by the nodes of a P2P network. A blockchain is a type of distributed ledger consisting of consecutive blocks chained together following a strict set of rules. Here, each block is created at a predefined interval, or after an event occurs, in a decentralised fashion by means of a consensus algorithm. Within each block, there are transactions by which a value is transferred in case of crypto-currencies or a data is stored for other blockchain systems. The consensus algorithm guarantees several data integrity related properties (discussed below) in blockchain. 

Even though the terms blockchain and DLT (Distributed Ledger Technology) are used inter-changeably in the literature, there is a subtle difference between them which is worth highlighting. A blockchain is just an example of a particular type of ledger, there are other types of ledger. When a ledger (including a blockchain) is distributed across a network, it can be regarded as a Distributed Ledger. 

Since the blockchain technology has been introduced with Bitcoin, it will be useful to understand how Bitcoin works. In Section \ref{sec:back:subsec:bitcoin}, we discuss a brief primer of Bitcoin and its associated terminologies. Then, we describe different properties and types of blockchains in Section \ref{sec:back:subsec:properties} and Section \ref{sec:back:subsec:type} respectively. Finally, we present the concept of blockchain layers in Section \ref{sec:back:subsec:layers}.

\subsection{Bitcoin}
\label{sec:back:subsec:bitcoin}

The Bitcoin network consists of nodes within a P2P (Peer-to-Peer) network. Each node needs to download the Bitcoin software to connect to the network. There are different types of nodes in the network, with miner nodes and general nodes being the major ones. A general node is mostly used by users to transfer bitcoin in the network, whereas a miner node is a special node used for mining bitcoins (see below).

Each user within a node needs to utilise wallet software to create identities. An identity in the Bitcoin network consists of a private/public key pair, and a bitcoin address is derived from the corresponding public key. A sender needs to know such an address of the receiver to transfer any bitcoin. Bitcoin is transferred between two entities using the notion of a transaction where the sender utilises a wallet software for this purpose. This transaction is propagated to the network, which is collected by all miner nodes. Each miner node combines these transactions into a block and then engages in solving a cryptographic puzzle, with other miners, in which it tries to generate a random number which satisfies the required condition (the random number must be less than a target value called the \textit{difficult target}). When a miner successfully solves the puzzle, that miner is said to have generated a valid block which is then propagated in the network. The Bitcoin protocol generates a certain amount of new Bitcoins for each new valid block and rewards  the miner for its effort in creating the block. Other miners validate this newly mined block and then add it to the blockchain. Each new block refers to the last block in the chain, which in turn refers to its previous block, and so on. The very first block in the chain, known as the \textit{genesis block}, however, has no such reference. 

The decentralised nature of this mining process might result in multiple valid blocks generated by different miners and propagated at the same time in the network. All of them are added to the blockchain and they refer to the same last block in the chain. Consequently,  multiple branches emerge from the same blockchain. This is a natural phenomenon in blockchain and is aptly known as \textit{fork}. The fundamental goal of the corresponding consensus protocol is to resolve this fork so that only one branch remains and other branches are discarded. The consensus algorithm utilised in Bitcoin follows a simple rule: it lets the branches grow. As soon as one branch grows longer than the others (more specifically, the total cumulative computational effort of one branch exceeds the others), all miners select the longest branch (or the branch with the highest computational effort), discarding all other branches. Such a branch is known as the main branch and other branches are known as orphan branches. Only the miners in the main branch are entitled to receive their Bitcoin rewards. When a fork is resolved across the network, a distributed consensus emerges in the network.

The frequency of Bitcoin block generation depends on the difficulty parameter, which is adjusted after $2016$ blocks. The protocol adjusts the difficulty parameter in such a way that a block is generated in every 10 minutes on average. However, the Bitcoin reward is halved after every $210,000$ blocks, or approximately after every $4$ years. At the initial stage, the reward for generating a valid block had been $50$ Bitcoins, which was  halved to $25$ Bitcoins in $2012$ and $12.5$ Bitcoins in $2016$. The next halving will occur in $2020$ where the reward will be reduced to $6.25$ bitcoins per block. This geometric reduction in every four years underlines a maximum total supply of $21$ million of Bitcoins. It is expected that this supply will be exhausted in the year of $2140$ when the rewarded bitcoin will be infinitesimally small for each block.

The process of Bitcoin protocol is presented in Figure \ref{fig:bitcoinblock}.

\begin{figure}
\caption{Block Generation Process of Bitcoin} %

%
%
\begin{mdframed}[backgroundcolor=black!7,rightline=false,leftline=false]
\begin{enumerate}[leftmargin=2mm,label=\Alph*]
\item Each miner collects transactions that are broadcast in the network and uses her hashpower to try to generate a block via repeated invocation of a hash function on data. The data  consists of the transactions that she saw fit to include, the hash of the previous block, her public key address, and a nonce.
\item When a miner succeeds in generating a block, meaning that the hash of her block data is smaller than the current difficulty target, she broadcasts her block to the network.
\item The other miners continue to extend the blockchain from this new  block, only if they find that this block is valid, i.e., it refers the hash of the previous block of the longest chain and meets the current difficulty target.
\item The block reward (newly minted coins) and the fees from the transactions go to the miner's public address. This means that only the miner can spend those by signing with her corresponding private key. 
\item The difficulty level readjusts according to the mining power of the participates, by updating the hash target value every 2016 blocks ($\approx$2 weeks) so that blocks get generated once every 10 minutes on average.
\item The block reward starts at 50 coins and halves in every 210, 000 blocks, i.e., about every 4 years.
\end{enumerate}
\end{mdframed}
\label{fig:bitcoinblock}
\end{figure}

\subsection{Properties of blockchain}
\label{sec:back:subsec:properties}
A blockchain exhibits several properties that make it a suitable candidate for several application domains \cite{ChowdhuryDLTComp2019}. The properties are discussed below.
\begin{itemize}
    \item \textbf{Distributed consensus on the chain state}: One of the crucial properties of any blockchain is its capability to achieve a distributed consensus on the state of the chain without being reliant on any trusted third party. This opens up the door of opportunities to build and utilise a system where states and interactions are verifiable by the miners in public blockchain systems or by the authorised entities in private blockchain systems.
    \item \textbf{Immutability and irreversibility of chain state}: Achieving a distributed consensus with the participation of a large number of nodes ensures that the chain state becomes practically immutable and irreversible after a certain period of time. This also applies to smart-contracts and hence enabling the deployment and execution of immutable computer programs. 
    \item \textbf{Data (transaction) persistence}: Data in a blockchain is stored in a distributed fashion,  ensuring data persistence as long as there are participating nodes in the P2P network.
    \item \textbf{Data provenance}: The data storage process in any blockchain is facilitated by means of a mechanism called the transaction. Every transaction needs to be digitally signed using public key cryptography,  which ensures the authenticity of the source of data. Combining this with the immutability and irreversibility of a blockchain provides a strong non-repudiation instrument for any data in the blockchain.
    \item \textbf{Distributed data control}: A blockchain ensures that data stored in the chain or retrieved from the chain can be carried out in a distributed manner that exhibits no single point of failure. 
    \item \textbf{Accountability and transparency}: Since the state of the chain, along with every single interactions among participating entities, can be verified by any authorised entity, a blockchain promotes accountability and transparency.
\end{itemize}

\subsection{Blockchain type}
\label{sec:back:subsec:type}
Depending on the application domains, different blockchain deployment strategies can be pursued. Based on these strategies, there are predominantly two types of blockchains, namely Public and Private blockchain, as discussed below:
\begin{itemize}
    \item \textbf{Public blockchain}: A public blockchain, also known as the \textit{Unpermissioned or permissionless Blockchain}, allows anyone to participate in the blockchain to create and validate blocks as well as to modify the chain state by storing and updating data through transactions among participating entities. This means that the blockchain state and its transactions, along with the data stored is transparent and accessible to everyone. This raises privacy concerns for particular scenarios where the privacy of such data needs to be preserved.
    \item \textbf{Private blockchain}: A private blockchain, also known as the \textit{Permissioned Blockchain}, has a restrictive notion in comparison to its public counterpart in the sense that only authorised and trusted entities can participate in the activities within the blockchain. By allowing only authorised entities to participate in activities within the blockchain, a private blockchain can ensure the privacy of chain data, which might be desirable in some use-cases.
\end{itemize}


\subsection{Blockchain Layers}
\label{sec:back:subsec:layers}
\begin{figure}    
\includegraphics[width=1\linewidth]{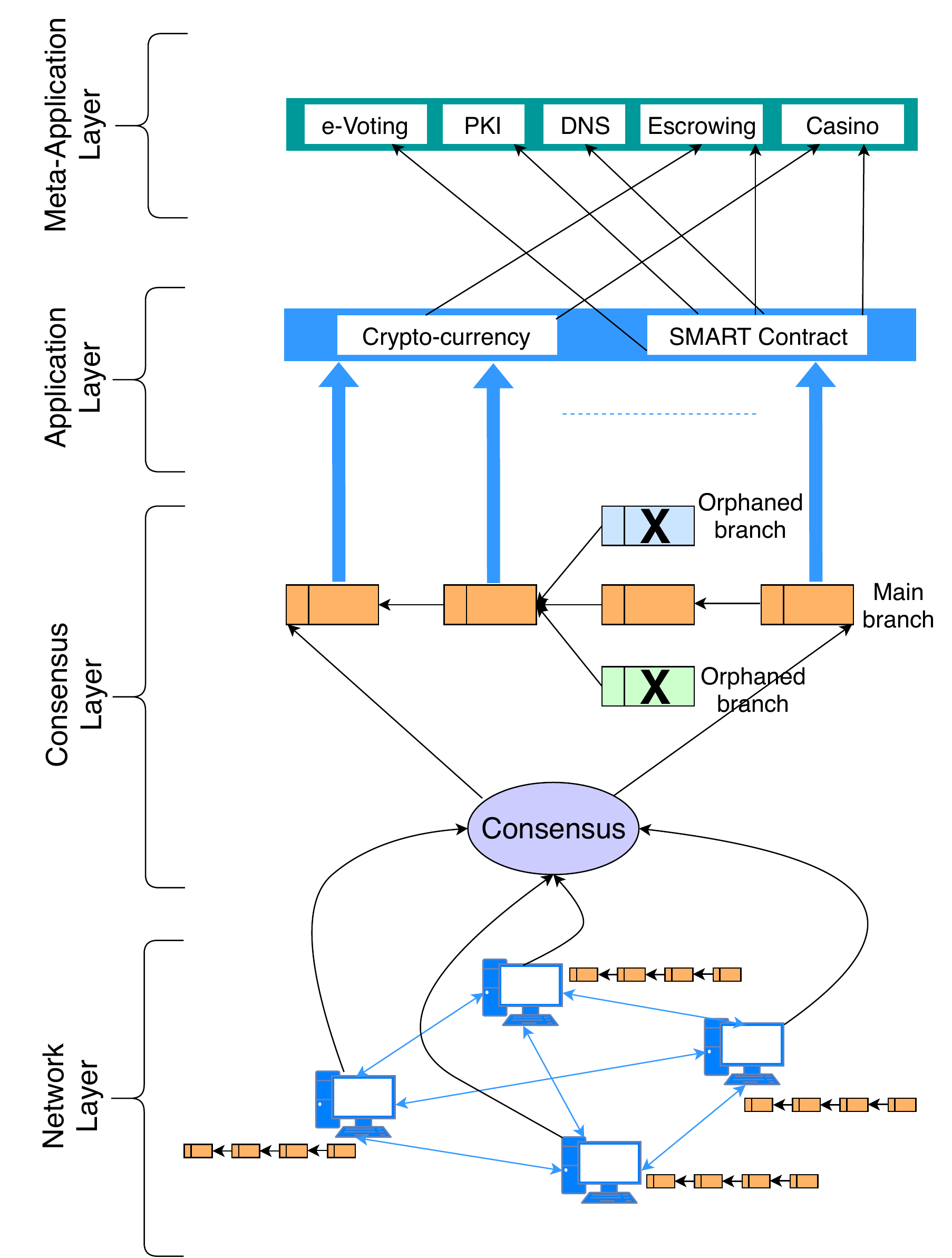}      
  \caption{Blockchain Layers}
  \label{fig:blockchainLayers}
   \end{figure}

There are several components in a blockchain system whose functionalities range from collecting transactions, propagating blocks, mining, achieving consensus and maintaining the ledger for its underlying crypto-currencies, and so on. These components can be grouped together according to their functionalities using different layers similar to the well-known TCP/IP layer. In fact, there have been a few suggestions to design a blockchain system using a layered approach \cite{JoiLayer2016, XiaoLayer2016}. The motivation is that a layered design will be much more modular and easier to maintain. For example, in case a bug is found in a component of a layer in a blockchain system, it will only affect the functionalities of that corresponding layer while other layers remain unaffected. For example, David et al. \cite{XiaoLayer2016} suggest four layers: consensus, mining, propagation, and semantic. However, we believe that the proposed layers do not reflect the proper grouping of functionalities. For example, consensus and mining should be part of the same layer as mining can be considered an inherent part of achieving consensus. In addition to this, some blockchain systems might not have any mining algorithms associated with it. In this paper,  we, therefore, will define four layers (Figure \ref{fig:blockchainLayers}): network, consensus, application, and meta-application. The functionalities of these layers are briefly presented below.

\vspace{1mm}
\noindent
\textbf{Meta-Application Layer:} The functionalities of the meta-application layer in a blockchain system (see Figure \ref{fig:blockchainLayers}) is to provide an overlay on top of the application layer to exploit the semantic interpretation of a blockchain system for other purposes in other application domains. For example, Bitcoin has been experimented to adopt in  multiple application domains, such  as  DNS like  decentralised  naming  system  (Namecoin  \cite{Namecoin2018}),  decentralised  immutable  time-stamped  hashed  record  (Proof  of  Existence  \cite{PoE2018}), and  decentralised  PKI (Public Key Infrastructure), such as Certcoin  \cite{Certcoin2018}.  

\vspace{1mm}
\noindent
\textbf{Application Layer:} The application layer (in Figure \ref{fig:blockchainLayers}) defines the semantic interpretation of a blockchain system. An example of a semantic interpretation would be to define a crypto-currency and then set up protocols for how such a currency can be exchanged between different entities. Another example is to establish protocols to maintain a state machine embodying programming capabilities within the blockchain, which can be exploited to create and deploy immutable code (the so-called \textit{smart contract}). The application also defines the rewarding mechanism, if any, in the blockchain system.

\vspace{1mm}
\noindent
\textbf{Consensus Layer:} The consensus layer, as presented in Figure \ref{fig:blockchainLayers}, is responsible for providing the distributed consensus mechanism in the blockchain that essentially governs the order of the blocks. A critical component of this layer is the proof protocol (e.g., proof of work and proof of stake) that is used to verify every single block, which ultimately is used to achieve the required consensus in the system. 

\vspace{1mm}
\noindent
\textbf{Network Layer:} The components in the network layer are responsible for handling network functionalities which include joining in the underlying P2P network, remaining in the network by following the underlying networking protocol, disseminating the current state of the blockchain to newly joined nodes, propagating and receiving transactions and blocks and so on.

\section{Consensus taxonomy \& properties}
\label{sec:conTaxProp}
 
With the introduction and advancement of different blockchain systems, there has been a renewed interest in distributed consensus with the consequent innovation of different types of consensus algorithms. These consensus algorithms have different characteristics and functionalities. In this section, we first distinguish between two major types of consensus and then present a taxonomy of their properties. Later, in Section \ref{sec:incentivised} and \ref{sec:hybrid}, we explore numerous crypto currencies and discuss incentivised consensus algorithms. Similarly, we focus on non-incentivised consensus and the blockchain applications in Section  \ref{sec:nonIncentivConsensus}.

 Consensus mechanisms used by the various blockchain systems can be classified based on the reward mechanism that participating nodes might receive. Therefore, we first classify the consensus mechanisms in blockchain systems into two categories: incentivised and non-incentivised algorithms. 

\vspace{1mm}\noindent
\textbf{Incentivised Consensus.} Some consensus algorithms reward participating nodes for creating and adding a new block in the blockchain. Such algorithms belong to this category. These algorithms are exclusively used in public blockchain systems and the reward provided acts as an incentive for participating nodes to behave accordingly and to follow the corresponding consensus protocol rigorously.

\vspace{1mm}\noindent
\textbf{Non-incentivised Consensus.} Private blockchain systems deploy a type of consensus algorithms that do not rely on any incentive mechanism for the participating nodes to create and add a new block in the blockchain. Such algorithms belong to this category. With the absence of any reward mechanism, these nodes are considered trusted as only authorised (allowed) nodes can participate in the block creation process of the consensus algorithm.


\subsection{Consensus properties}
\label{subsec:properties}
Each consensus algorithm has different characteristics and serves different purposes. To compare these disparate groups of consensus algorithms, we need to define evaluation criteria. In this section, we present this evaluation criteria in the form of taxonomies of consensus properties. These properties have been collected from existing researches, such as \cite{BanoSOK2017, Cachin2017}, and compiled as a taxonomy in this work.

\begin{figure}
    \centering
    \includegraphics[width=1\linewidth]{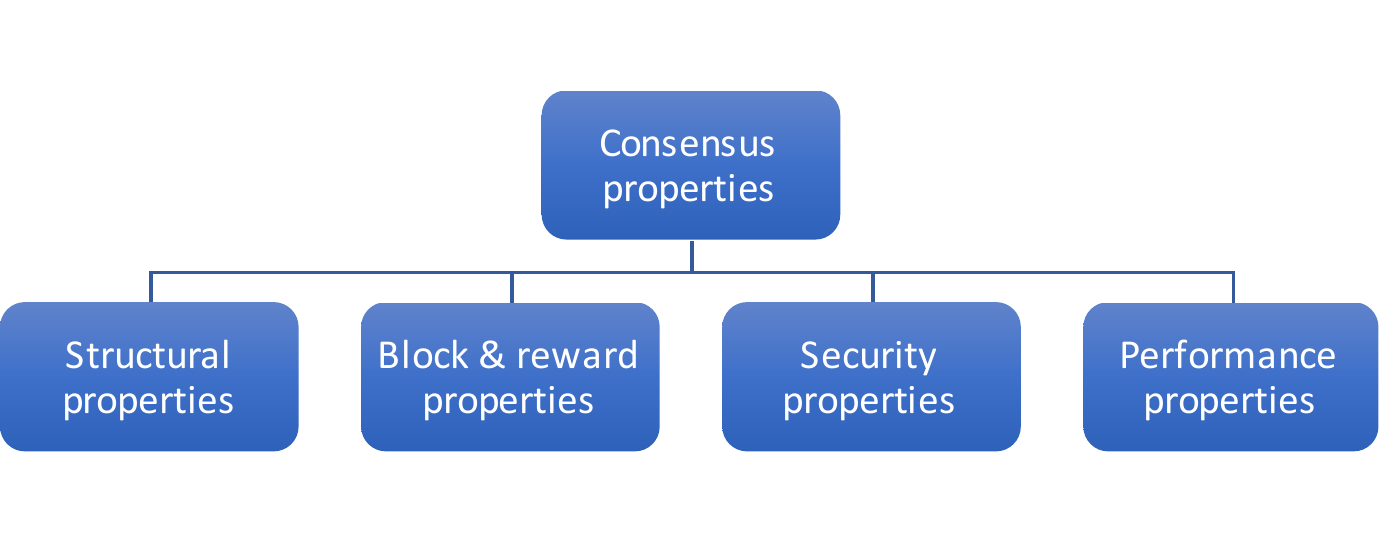}
    \caption{Taxonomy of consensus properties.}
    \label{fig:conProTax}
\end{figure}

The taxonomy is presented in Figure \ref{fig:conProTax}. According to this taxonomy, a consensus mechanism has four major groups of properties: \emph{Structural}, \emph{Block \& reward}  , \emph{Security} and \emph{Performance} properties. Each of these properties is briefly discussed below.

\subsubsection{Structural properties}
\begin{figure}
\centering
\includegraphics[width=1\linewidth]{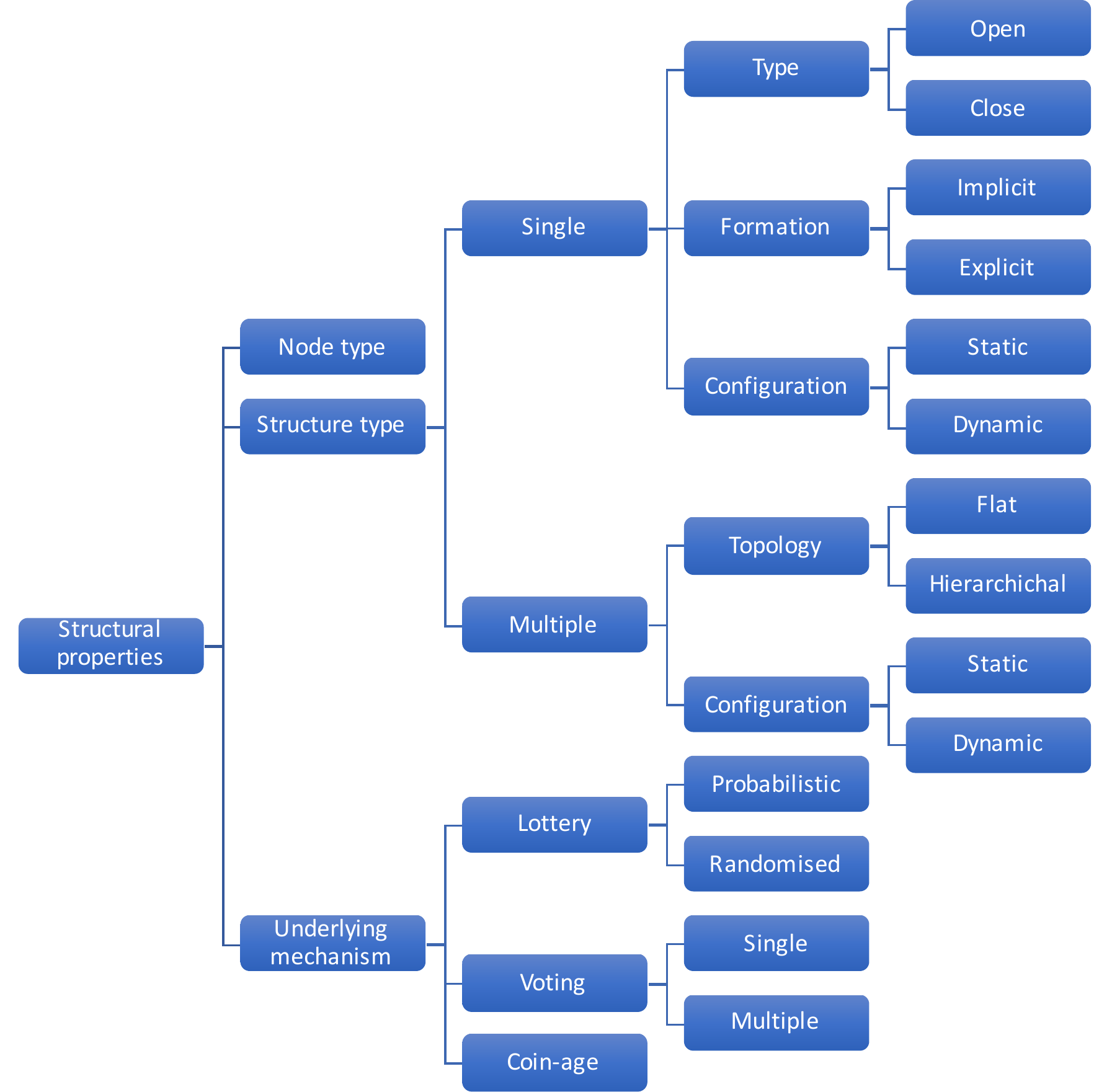}      
  \caption{Taxonomy of structural properties.}
  \label{fig:conStrucTax}
   \end{figure}

Structural properties define how different nodes within a blockchain network are structured to participate in a consensus algorithm. These properties can be sub-divided into different categories as illustrated in Figure \ref{fig:conStrucTax}. We briefly describe each of these categories below.

\begin{itemize}
    \item \textbf{Node types:} It refers to \emph{different types of nodes} that a consensus algorithm is required to engage with to achieve its consensus. The types will depend on the consensus algorithm which will be presented in the subsequent section. 
	\item \textbf{Structure type:} It refers to \emph{the ways different nodes are structured} within the consensus algorithm using the concept of a committee. The committee itself can be of two types: single and multiple committees. Each of these committees is described below.
	\item \textbf{Underlying mechanism:} It refers to the \emph{specific mechanism that a consensus algorithm deploys to select a particular node}. The mechanism can utilise lottery, the age of a particular coin or a voting mechanism. A lottery can utilise either a cryptography based probabilistic mechanism or other randomised mechanisms. In a voting mechanism, voting can be carried out either in a single or multiple rounds. The coin-age, on the other hand, utilises a special property, which depends on how long a particular coin has been owned by its owner. 
\end{itemize}
Next, we explore different types of voting committees for existing consensus algorithms. 
 
\textbf{Single committee.} A single committee refers to a special group of nodes among the participating nodes which actively participate in the consensus process by producing blocks and extending the blockchain. Each single committee can have different properties. Next, we briefly explore these properties. 

\begin{itemize}
	\item \textbf{Committee type:} A committee can be open or close. A committee is open if it is \textit{open} to any participating nodes or closed if it is restricted to a specific group of nodes.
    \item \textbf{Committee formation:} A committee can be formed either implicitly or explicitly. An implicit formation does not require the participating nodes to follow any additional protocol rules to be in the committee, whereas an explicit formation requires a node to follow additional protocol steps to be a part of the committee.
    \item \textbf{Committee configuration:} A committee can be configured in a static or a dynamic fashion. 
    	\begin{itemize}
    	\item \textbf{Static:} In a static configuration, the members of the committee are pre-selected and fixed. No new members can join and participate in the consensus process.
        \item \textbf{Dynamic:} In a dynamic configuration, the committee members are defined for a time-frame (known as epoch), after which new members are added, and old members are removed based on certain sets of criteria. In such a committee, nodes are selected using a voting mechanism where voting is carried either in a single or multiple rounds.  Some consensus algorithms, however, do not specify any specific time-frame, and hence, members can join or leave any time at will. Nodes in such configuration are selected using a lottery mechanism which utilises either a cryptography based probabilistic mechanism or other randomised mechanisms. 
    	\end{itemize}
\end{itemize}


\textbf{Multiple committee.} It has been observed that the time it takes to achieve consensus in a single committee tends to increase as the number of the member starts to increase \cite{BanoSOK2017}, thereby reducing performance. To alleviate this problem, the concept of multiple committee has been introduced, where each committee consists of different validators \cite{BanoSOK2017}. A multiple committee can have different properties. Next, we explore two properties.


 \begin{itemize}
 	\item \textbf{Topology:} It refers to the way different committees are organised. For example, the topology can be \textit{flat} to indicate that different committees are at the same level or can be \textit{hierarchical} where the committees can be considered in multiple layered levels. 
    \item \textbf{Committee configuration:} In addition, like a single committee, the multiple committees can be configured in a static or dynamic way.
 \end{itemize}

\subsubsection{Block \& reward properties}
Properties under this category can be utilised as quantitative metrics to differentiate different crypto-currencies. The properties are (Figure \ref{fig:conBlockTax}): genesis date, block reward, total supply, formula, and block creation time. These properties  do not necessarily characterise different consensus algorithm directly, however, most of them (except the genesis date) have a direct and indirect impact on how consensus is achieved in a particular crypto-currency based blockchain system. For example, block reward incentivises miners to act accordingly by solving a cryptographic puzzle, which is then ultimately used to achieve consensus. The properties are described below:


\begin{figure}
    \centering
    \includegraphics[width=1\linewidth]{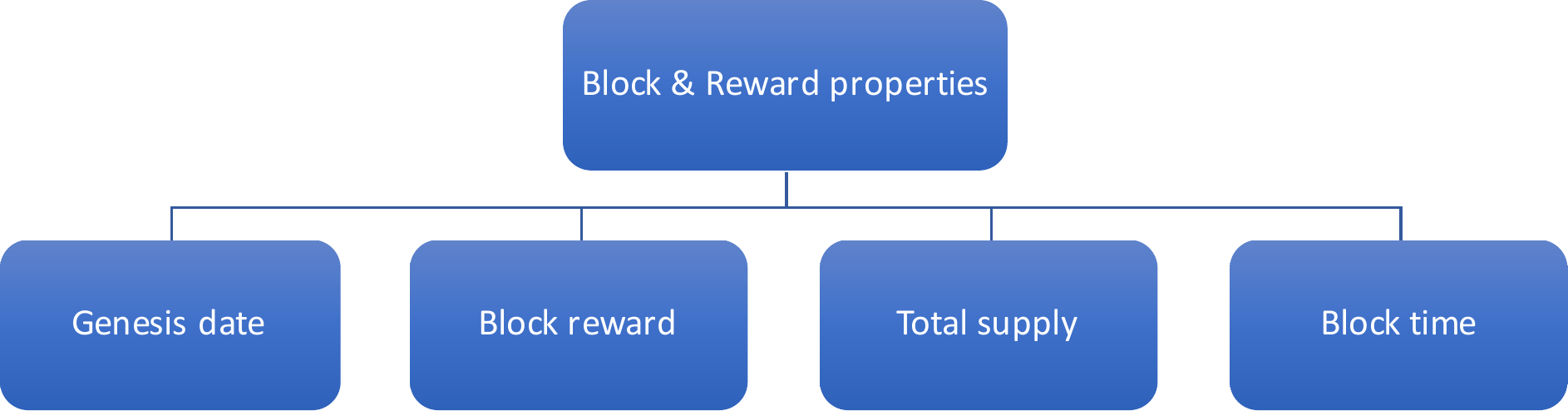}
    \caption{Taxonomy of block \& reward properties.}
    \label{fig:conBlockTax}
\end{figure}

\begin{itemize}
	\item \textbf{Genesis date} represents the timestamp when the very first block was created for a particular crypto-currency.
    \item \textbf{Block reward} represents the reward a node receives for creating a new block.
    \item \textbf{Total supply} represents the total supply of a crypto-currency. 
    \item \textbf{Block time} represents the average block creation time of a crypto-currency.
\end{itemize}

\subsubsection{Security properties}
A consensus algorithm must satisfy a number of security properties as shown in (Figure \ref{fig:conSecTax}) and are described below: 
\begin{figure}
    \centering
    \includegraphics[width=.8\linewidth]{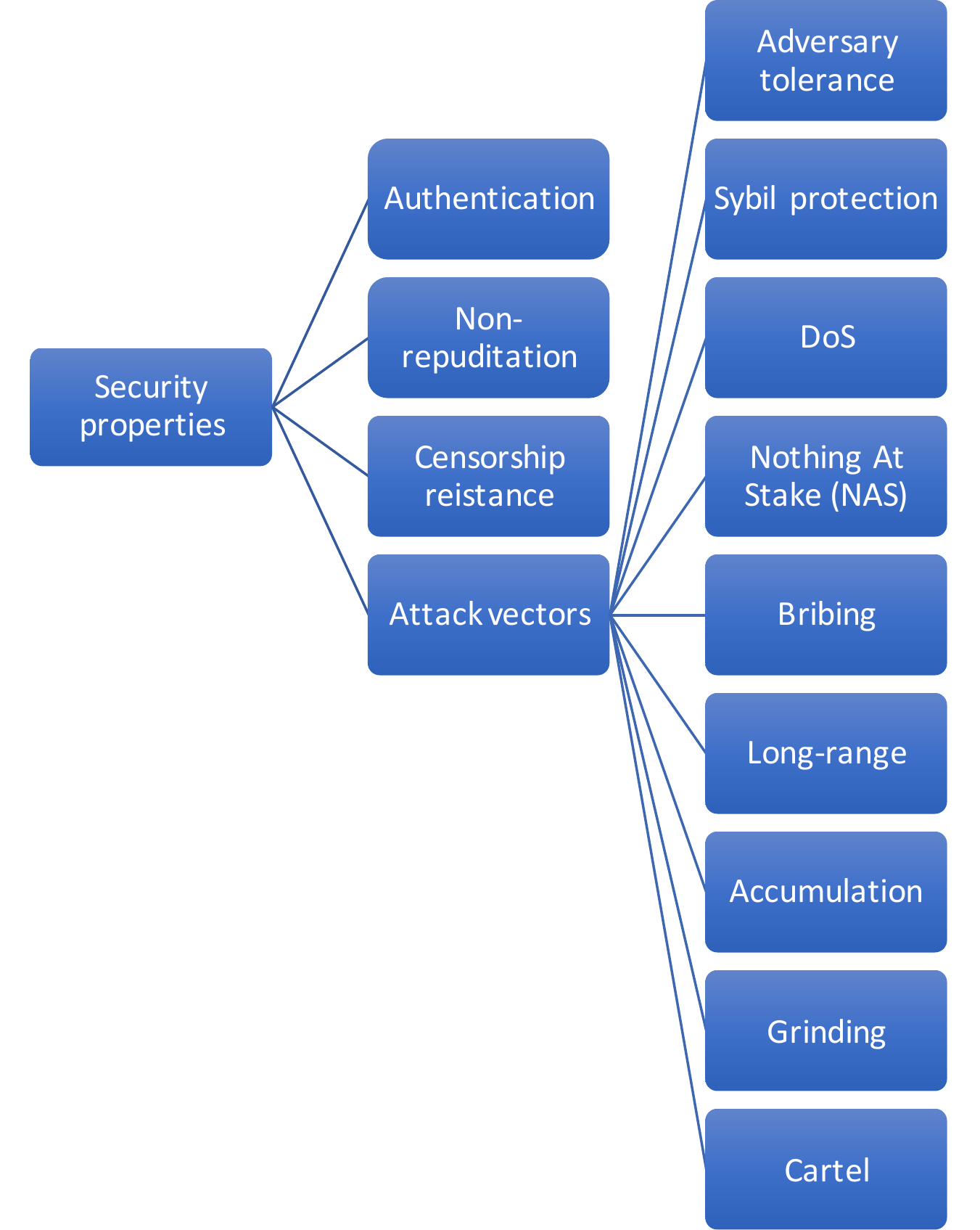}
    \caption{Taxonomy of security properties.}
    \label{fig:conSecTax}
\end{figure}



\begin{itemize}	
    \item \textbf{Authentication:} This implies if nodes participating in a consensus protocol need to be properly verified/authenticated.
    \item \textbf{Non-repudiation:} This signifies if a consensus protocol satisfies non-repudiation.
    \item \textbf{Censorship resistance:} This implies if the corresponding algorithm can withstand against any censorship resistance.
    \item \textbf{Attack vectors}: This property implies the attack vectors applicable to a consensus mechanism. Here, we present a set of attack vectors that are applicable to any consensus algorithm. The other attack vectors presented in Figure \ref{fig:conSecTax} are applicable to a specific class of consensus algorithm. Therefore, we will discuss them in the upcoming sections, when we explore such algorithms.
    \begin{itemize}
      \item \textbf{Adversary tolerance:} This signifies the maximum byzantine nodes supported/tolerated by the respective protocol.
        \item \textbf{Sybil protection:} In a Sybil attack \cite{sybil2002}, an attacker can duplicate his identity as required in order to achieve illicit advantages. Within a blockchain system, a sybil attack implicates the scenario when an adversary can create/control as many nodes as required within the underlying P2P network to exert influence on the distributed consensus algorithm and to taint its outcome in her favour.
        
        \item \textbf{DoS (Denial of Service) resistance:} This implies if the consensus protocol has any built-in mechanism against DoS attacks.
    \end{itemize}
\end{itemize}

\subsubsection{Performance properties}
\begin{figure}
    \centering
    \includegraphics[width=1\linewidth]{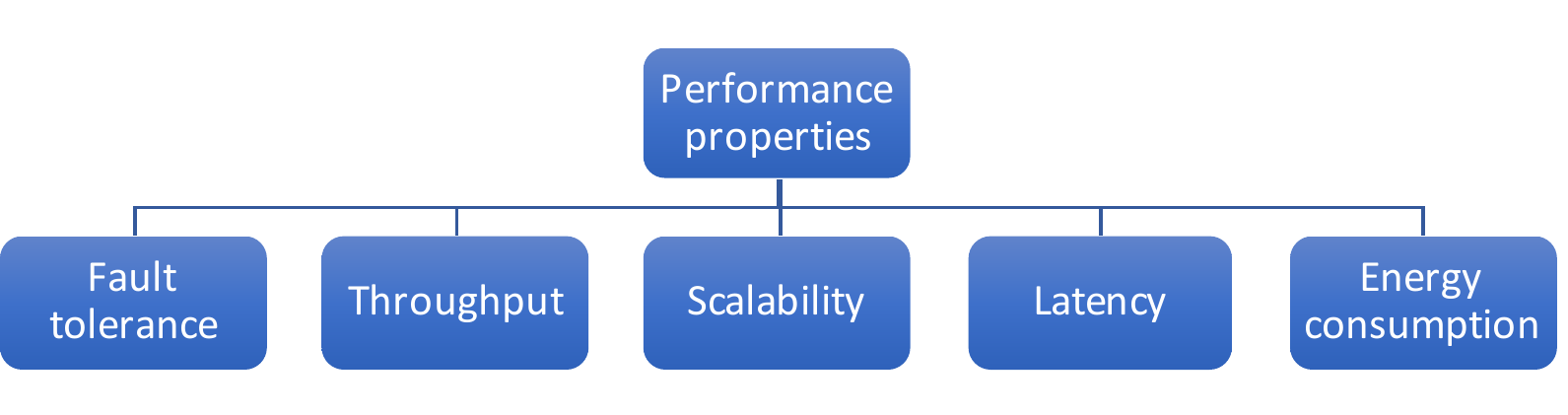}
    \caption{Taxonomy of performance properties.}
    \label{fig:conPerTax}
\end{figure}

The properties belonging to this group can be utilised to measure the quantitative performance of a consensus protocol. A brief description of each property is presented below with its illustration in Figure \ref{fig:conPerTax}

\begin{itemize}
	\item \textbf{Fault tolerance:} signifies the maximum faulty nodes the respective consensus protocol can tolerate.
    \item \textbf{Throughput:} implies the number of transactions the protocol can process in one second. 
	\item \textbf{Scalability:} refers to the ability to grow in size and functionalities with- out degrading the performance of the original system \cite{Ferdous:IoT:2019}.
	\item \textbf{Latency (Finality):} refers to "\textit{the time it takes from when a transaction is proposed until consensus has been reached on it}" \cite{BanoSOK2017}. It is also known as finality.
	\item \textbf{Energy consumption:} indicates if the algorithm (or the utilising system) consumes a significant amount of energy.
\end{itemize}

\section{Incentivised Consensus: PoW \& PoS}
\label{sec:incentivised}
In this section, we explore different incentivised consensus algorithms. Such algorithms can be grouped in three major categories: Proof of Work (PoW), Proof of Stake (PoS), and Hybrid Consensus. Among them, this section discusses PoW and PoS algorithms in Section \ref{sec:incentivised:subsec:pow} and Section \ref{sec:incentivised:subsec:pos} respectively. For readability, hybrid algorithms are presented in Section \ref{sec:hybrid}.
\subsection{Proof of Work (PoW)}
\label{sec:incentivised:subsec:pow}
A Proof of Work (PoW) mechanism involves two different parties (nodes): prover (requestor) and verifier (provider). The prover performs a resource-intensive computational task intending to achieve a goal and presents it to a verifier or a set of verifiers for validation that requires significantly less resource. The core idea is that the asymmetry, in terms of resource required, between the proof generation and validation acts intrinsically as a deterrent measure against any system abuse. 


Within this aim, the idea of PoW was first presented by Dwork and Naor in their seminal article in 1993 \cite{dwork1992pricing}. They put forward the idea of use PoW to combat email spamming. According to their proposal, an email sender would be required to solve a resource-intensive mathematical puzzle and attach the solution within the email as a proof that the task has been performed. The email receiver would accept an email only if the solution can be successfully verified.

Within the blockchain setting, a similar concept has been adopted. Each PoW mechanism is bound to a threshold, known as the difficulty parameter in many blockchain systems. The prover would carry out the computational task in several rounds until a PoW is generated that matches the required threshold, and every single round is known as a single proof attempt. 

PoW has been the most widely-used mechanism to achieve a distributed consensus among the participants regarding the block order and the chain state. In particular, a PoW mechanism in a blockchain serves two critical purposes:

\begin{itemize}
    \item A deterrent mechanism against the \textit{Sybil Attack}. In PoW, every mining node would require a significant monetary investment to engage in a resource-intensive PoW mechanism during the block creation process. To launch a Sybil attack, the monetary investment of an attacker will be proportional to the number of Sybil identities, which might outweigh any advantage gained from launching a Sybil attack.
    \item The PoW mechanism is used as an input to a function which ultimately is used to achieve the required distributed consensus when a fork happens in a blockchain \cite{bates2017lightweight}.
\end{itemize}

We differentiate between three major classes of PoW consensus mechanisms: \textit{Compute-bound} PoW,  \textit{Memory-bound} PoW and \textit{Chained} PoW. Each of these is explored in the following sections.

\subsubsection{Compute-bound PoW}
\label{subsubsec:computePoW}
A \textit{Compute-bound PoW}, also known as \textit{CPU-bound PoW}, employs a CPU-intensive function that carries out the required computational task by leveraging the capabilities of the processing units (e.g., CPU/GPU), without relying on the main memory of the system. These particular characteristics facilitate the scenario in which the computation can be massively optimised for faster calculation using Application-specific Integrated Circuit (ASIC) rigs. This has drawn criticisms among the crypto-currency enthusiasts as general people cannot participate in the mining process with their general purpose computers, and the mining process is mostly centralised among a group of mining nodes. 

\vspace{2mm}
Hashcash by Back et  al.~\cite{back2002hashcash} is the earliest example to leverage a PoW mechanism in practical systems. Similar to the proposal of Dwork and Naor in \cite{dwork1992pricing}, Hashcash is also designed to combat spams. In this scheme, the email sender would require to generate a SHA-1 hash with a certain property using as the input a number of information including recipient's email address and date. The property dictates that the generated hash must have at least 20 bits of leading zeroes. Generating an SHA-1 hash with this property would require the senders to engage in several proof attempts in a pseudo-random fashion. Once the hash is generated, it is added within the email header. The verification on the recipient's side is rather trivial, which requires comparing a newly generated hash using the required information with the supplied hash. If they match, it proves that the email sender has engaged in the required amount of computational work. The effectiveness of this approach of fighting spams depends on the hypothesis that spammers rely on the revenue model requiring a mere amount of cost to send a single email. When they would need to engage in such a computationally intensive task for sending every single email, the aggregated associated cost might heavily affect their profit margin and thus deter them from spamming.

\vspace{2mm}
Nakamoto consensus is the compute-bound PoW consensus algorithm leveraged in Bitcoin. It is based on the approach of Hashcash, modified to be applied within the blockchain setting. As discussed in Section \ref{sec:back:subsec:bitcoin}, all mining nodes (miners) compete with each other to generate a valid block by finding a solution smaller than the difficulty target. Similar to the idea of HashCash, the miners need to engage in several proof attempts, until the solution is found. In each of these proof attempts, each miner generates a hash using either the SHA-256 or SHA-256d (a double hashing mechanism using SHA-256) algorithm and checks if the generated hash is smaller than the difficulty target. The effect of this distributed engagement is that forks happen, and then the Nakamoto consensus algorithm is utilised to resolve the fork and to achieve a network-wide distributed consensus. The reader is referred back to Section \ref{sec:back:subsec:bitcoin} (and Figure \ref{fig:bitcoinblock}) for a brief description of Nakamoto consensus. 

Currently, there are many crypto-currencies that utilise the Nakamoto consensus algorithm. Table \ref{tab:powconsensus} shows the top 10 of such currencies according to their market capitalisation as rated by CoinGecko \footnote{https://www.coingecko.com/} (a website which tracks different activities related to crypto-currencies) as of July 24, 2019. The table also presents their Block and reward properties as presented in Figure \ref{fig:conBlockTax}. It is to be noted that information regarding the properties in Table \ref{tab:powconsensus} for these (and other subsequent) currencies has been collected by consulting their corresponding whitepapers, websites and introductory announcements on Reddit website \footnote{https://www.reddit.com/}.  

\begin{table*}[!h]
\centering
  \caption{Top ten crypto-currencies that utlise Nakamoto consensus algorithm.}
  \begin{tabular}{p{30mm}|p{25mm}|p{35mm}|p{25mm}|p{15mm}}
    \hline
    \rowcolor[gray]{.6}
       \centering\textbf{Currency } & 
     \centering\textbf{Genesis date (dd.mm.yyyy)}& 
     \centering\textbf{Block reward}& 
     \centering\textbf{Total supply (Million)}& 
     \centering\textbf{Block Time}
    \tabularnewline [2ex]
    \hline
 \hline
        Bitcoin/Bitcoin Cash \cite{bitcoin2018} \cite{bitcoincash2018} &  03.01.2009&   12.5 &  21&  10m \\\hline
%
%
%
    \rowcolor[gray]{.9}Syscoin \cite{syscoin2014}& 16.08.2014& 80.04659537 & 888
    & 1m  \\\hline
    Peercoin \cite{peercoin2012}& 19.08.2012& 55.17265345& 2000 &   10m  \\\hline
    \rowcolor[gray]{.9} Counterparty \cite{counterparty2014}&  01.02.2014&  All currency in circulation& 2.6m & - \\\hline
    Emercoin \cite{emercoin2013}&  11.12.2013 & Smooth emission& 41&  10m \\\hline
   \rowcolor[gray]{.9} Namecoin \cite{namecoin2011}& 19.04.2011&  12.50000000 & 21&  10m  \\\hline
    Steem Dollars \cite{steemdollars2014}& 04.06.2016&  Smooth emission& Unlimited &  3s \\\hline
   \rowcolor[gray]{.9} Crown \cite{crown2014}& 08.09.2014&  1.8& 42&  1m \\\hline
    XP  & 24.08.2016& & 2220 &  NA \\\hline
%
%
  \rowcolor[gray]{.9}  Omni (Mastercoin) \cite{omni2013}& 31.07.2013&  16.71249999 Omni 
    & 0.6&  20s \\\hline
%
%
  \end{tabular}
%
%
  \label{tab:powconsensus}
\end{table*}

\subsubsection{Memory-bound PoW}
To counteract the major criticism of compute-bound PoWs allowing the utilisation of ASIC-based rigs for the mining purpose (see Section \ref{subsubsec:computePoW}), memory-bound PoWs have been proposed. A memory-bound PoW requires the algorithm to access the main memory several times and thus ultimately binds the performance of the algorithm within the limit of access latency and/or bandwidth as well as the size of memory. This restricts ASIC rigs based on a memory-bound PoW to have the manifold performance advantage over their CPU/GPU based counterparts. In addition, the profit margin of developing ASIC with memory and then building mining rigs with them is not viable as of now for these classes of PoWs. Because of these, memory-bound PoWs are advocated as a superior replacement for compute-bound PoWs in de-monopolising mining concentrations around some central mining nodes.

There is a large variety of consensus algorithms belonging to this class, unlike the consensus algorithms of compute-bound PoW which are largely based on Hashcash.  These algorithms can be further categorised as follows: Cryptonight; Scrypt and its variants; Equihash; Ethhash/Dagger; Neoscript; and Timetravel. We now describe each of these different types of memory-bound PoW consensus algorithm.

\vspace{2mm}
\noindent\textbf{\textsc{1) Cryptonight.}} Cryptonight is a class of PoW consensus algorithms that, in principle, is a memory-hard hash function \cite{cryptonight2013}. It utilises the Keccak hashing function~\cite{bertoni2013keccak} internally and relies on a 2MB scratchpad residing on the memory of a computer. The scratchpad is extensively used to perform numerous read/write operations at pseudo-random addresses within that scratchpad. In the final step, the desired hash is generated by hashing the entire scratchpad.

Its reliance on a large scratchpad on the memory of a system makes it resistant towards FPGA and ASIC mining as the economic incentive to create FPGA, and ASIC mining hardware might be too low for the time being. As such, Cryptonight introduces the notion of so called \textit{Egalitarian proof of work \cite{cryptonight2013}} or proof of equality, which enables anyone to join in the mining process using any modern CPU and GPU.

One prominent property of the coins belonging to this class is that all of them support stronger sender-receiver privacy by facilitating anonymous transactions.
%
%

Current currencies utilising Cryptonight according to Coingeko as of July 24, 2019 is presented in Table \ref{tab:cryptonight}. Like Table \ref{tab:powconsensus}, Table \ref{tab:cryptonight} also presents their Block and reward properties as presented in Figure \ref{fig:conBlockTax}.   

\begin{table*}[!h]
\centering
  \caption{Top ten crypto-currencies that utilise Cryptonight, with Bytecoin being the first to use this algorithm.}
  \begin{tabular}{p{25mm}|p{25mm}|p{45mm}|p{35mm}|p{20mm}}
    \hline
    \rowcolor[gray]{.6}
      \centering\textbf{Currency } & 
     \centering\textbf{Genesis date (dd.mm.yyyy)}& 
     \centering\textbf{Block reward}& 
     \centering\textbf{Total supply (Million)}& 
     \centering\textbf{Block Time} \tabularnewline [2ex]
          \hline
          \hline
       Monero& 18.04.2014 &  4.86930501 & Starting at $M = 2^{64} - 1$  infinite supply& 2.0m \\\hline
       \rowcolor[gray]{.9}Bytecoin&  04.07.2012&  666.76  & 184.46 billion & 2.0m  \\\hline
       Aeon& 06.06.2014 & 5.48
  & Starting at $M = 2^{64} - 1$, infinite supply.& 4.0m \\\hline
        \rowcolor[gray]{.9}Boolberry& 17.05.2014&  4.85 & 18.5 Million &  2.0 \\\hline
         Karbowanec & 30.05.2016. & 8.83 
         & Starting with 10 Million, infinite supple &  4.0m \\\hline
        \rowcolor[gray]{.9}Fantomcoin& 06.05.2014&  smooth emission, 50\% coins will be emitted in 6 years and block reward decreases with a similar  Starting at $M = 2^{64} - 1$ & infinite supply &  1.0m \\\hline
        Dashcoin fork of Bytecoin &  05.07.2014 & 1.55& &  2.0m \\\hline
%
%
%
%
\rowcolor[gray]{.9}QuazarCoin& 08.05.2014 &  smooth emission &&   2.0 \\\hline
BipCoin &  20.08.2016& smooth emission &&2.0 \\\hline
\rowcolor[gray]{.9} Cannabis Industry Coin&16.10.2016& 70.00000000 & 21 M& 2.0 \\\hline
\end{tabular}
\label{tab:cryptonight}
\end{table*}
\vspace{1mm}
\noindent\textbf{\textsc{2) Scrypt and its variants.}} Scrypt is a password based key driving function (KDF) that is currently used in many crypto-currencies \cite{scrypt2016}. A KDF is primarily used to generate one or more secret values from another secret key and is widely used in password hashing. Previous key deriving functions such as DES-based UNIX Crypt-function, FreeBSD MD5 crypt, Public-Key Cryptography Standards\#5 (PKCS\#5), and PBKDF2 do not impose any specific hardware requirements. This enables any attacker launch attacks against those functions using specific FPGA or ASIC enabled hardware, the so-called \textit{custom hardware attacks} \cite{customHWWiki2019}. Scrypt has been designed to counteract this threat.

Toward this aim, one of the core characteristics of Scrypt is its reliance on the vast memory of a system, making it difficult to perform using FPGA and ASIC enabled custom hardware. In the underneath, Scrypt utilises Salsa20/8 Core~\cite{bernstein2008salsa20} as its internal hash function. A simplified version of Scrypt is used in the corresponding crypto-currencies, which is much faster and easier to implement, and can be performed using any modern CPU and GPU. Hence, anyone can join in the mining process for crypto-currencies using this function. However, the ever-increasing price of crypto-currencies has incentivised miners to produce custom ASIC hardware for some crypto-currencies utilising Scrypt in recent times. An example of such hardware that can be used to mine different Scrypt crypto-currencies is Antminer L3+ \cite{scryptASIC2019}. 
%
%
%

To tackle this issue of exploiting ASIC for mining, several Scrypt variants have been proposed: Scrypt-N/Scrypt Jane/Scrypt Chacha and Scrypt-OG, each providing particular advantages over others. Scrypt-N and Scrypt Chacha rely on SHA256 and ChaCha \cite{bernstein2008chacha} as their internal hash functions, respectively, whereas Scrypt Jane utilises a combination of different hash functions. All of them support progressive and tunable memory requirements, which can be adjusted after a certain period. This is to ensure that custom ASIC hardware is rendered obsolete once the memory requirement is changed. Finally, Scrypt-OG (Optimised for GPU) is optimised to be eight times less memory intensive than Scrypt \cite{scryptvsx112017}.
%
%
%


Table \ref{tab:cryptoscript} shows the top 10 currencies, which either use Scrypt or one of its variants, as per their market capitalisation according to CoinGecko as of July 24, 2019. 
\begin{table*}[!h]
\centering
   \caption{Top ten crypto-currencies using Scrypt.}
   \begin{tabular}{p{25mm}|p{25mm}|p{30mm}|p{25mm}|p{15mm}}
    \hline
    \rowcolor[gray]{.6}
     \centering\textbf{Currency } & 
     \centering\textbf{Genesis date (dd.mm.yyyy)}& 
     \centering\textbf{Block reward}& 
     \centering\textbf{Total supply (Million)}& 
     \centering\textbf{Block Time} \tabularnewline [2ex]
          \hline
          \hline
	    Litecoin \cite{litecoin2011}&13.10.2011& 25.00&84 million & 2.5m \\\hline
    \rowcolor[gray]{.9}Verge \cite{verge2016}&15.02.2016& 730.00&16.5 billion&0.5m \\\hline
    Bitmark \cite{bitmark2014}&13.07.2014&  (no longer monitored after 2016)&27.58 million&2.0m \\\hline
    \rowcolor[gray]{.9}Dogecoin \cite{dogecoin2013}&06.12.2013& 	10000.00&Total supply&NA\\\hline
    GameCredits \cite{gamecredit2017}&01.06.2015&fixed (12.5 coins)&84 million&1.5m \\\hline
    \rowcolor[gray]{.9}Einsteinium \cite{einsteinnium2014}&01.03.2014 &2&2.9 billion&1.0m \\\hline
    Voxels \cite{voxels2015}&03.11.2015& smooth emission &2.1 billion&2.5m \\\hline
    \rowcolor[gray]{.9}Viacoin \cite{viacoin2014}&18.07.2014& 0.63&23 million&0.5m \\\hline
    Hempcoin \cite{hempcoin2014}&9.03.2014& smooth emission&2.5 billion& 1.0m \\\hline
\end{tabular}
\label{tab:cryptoscript}
\end{table*}


\vspace{1mm}
\noindent\textbf{\textsc{3) Equihash}}
Equihash is one of the recent PoW algorithms that has been well received in the blockchain community \cite{biryukov2017equihash}. It is a memory-bound PoW that requires to find a solution for the Generalised Birthday problem using Wagner's algorithm \cite{wagner2002generalized}. Equihash has been designed to decentralise the mining procedure itself, similar to other memory-bound approaches. However, so far, very small portions of such algorithms have succeeded. One of the crucial reasons for this is that their underlying time-memory complexity trade-off is largely constant. This means that reducing memory requirement in these algorithms have little effect on their corresponding time complexity.

Wagner's solution has a steep time-memory complexity trade-off, reducing memory increases time complexity substantially. This premise has been exploited by Equihash to ensure that mining is exclusively proportional to the amount of memory a miner has. Thus, it is more suitable for a general purpose computer, rather than any ASIC-enabled hardware which can only have relatively small memory in order to make their production profitable for the mining process. Due to this reason, it has been claimed that Equihash can support ASIC resistance, at least for the foreseeable future. In addition, the verification is extremely lightweight and even can be carried out in resource-constrained mobile devices. Table \ref{tab:equihash} shows the eight currencies which utilise Equihash according to CoinGecko as of July 24, 2019. 
%
%
%
%
%

\begin{table*}[!h]
\centering
  \caption{Crypto-currencies utilising Equihash algorithm.}
   \begin{tabular}{p{25mm}|p{25mm}|p{20mm}|p{25mm}|p{15mm}}
    \hline
    \rowcolor[gray]{.6}
     \centering\textbf{Currency} & 
    \centering\textbf{Genesis date (dd.mm.yyyy)}& 
    \centering\textbf{Block reward} & 
    \centering\textbf{Total supply (Million)} & 
    \centering\textbf{Block Time}\tabularnewline [2ex]
          \hline
          \hline
       Zcash \cite{zcash2016}& 28.10.2016&  10& 21 million&   2.5m  \\\hline
\rowcolor[gray]{.9}Bitcoin Gold \cite{bitcoingold2017}& 24.10.2017&12.5& 21 million&  10m \\\hline
Komodo \cite{komodo2016}& 15.10.2016& 3& 200 million&  1m \\\hline
\rowcolor[gray]{.9}Zclassic \cite{zclassic2016}& 6.11.2016&  12.5& 21 million&   2.5m \\\hline

ZenCash \cite{zendcash2017}& 30.05.2017& 7.5& 21 million & 2.5m \\\hline
\rowcolor[gray]{.9}Hush \cite{hush2017}& Genesis date&  11.25& 21 million & 2.5m \\\hline
BitcoinZ \cite{bitcoinz2017}& 10.09.2017&  12500.00& 21 billion&  2.5m \\\hline
\rowcolor[gray]{.9}VoteCoin \cite{votecoin2017}& 31.08.2017& 125& 2.2 billion&  2.5m \\\hline
  \end{tabular}
  \label{tab:equihash}
\end{table*}

\noindent\textbf{\textsc{4) Ethash (Dagger-Hashimoto)/Dagger.}} 
Ethash is a memory-bound PoW algorithm introduced for Ethereum with the goal to be ASIC-resistant for a long period of time \cite{ethash2017}. It was previously known as Dagger-Hashimoto algorithm \cite{daggerHashimoto2017} because of its utilisation of two different algorithms: Dagger \cite{dagger2017} and Hashimoto \cite{dagger2017}.

Dagger is one of the earliest proposed memory-bound PoW algorithm which utilises Directed Acyclic Graph (DAG) for memory-hard puzzle solving with trivial verification that requires less memory to be used in resource constrained devices. However, the Dagger algorithm is proven to be vulnerable towards a shared memory hardware acceleration attack, as discussed in \cite{daggerVul2014}. That is why it has been dropped as a PoW candidate for Ethereum. Hashimoto algorithm, on the other hand, relies on the delay incurred for reading data from memory as the limiting factor and thus, is known as an I-O bound algorithm. 

Ethash combines these two algorithms to be ASIC-resistant and functions as follows. Ethash depends on a large pseudo-random dataset, which is recomputed during each epoch. Each epoch is determined by the time it takes to generate 30,000 blocks in approximately five days. This dataset is essentially a directed acyclic graph and hence, is called DAG. During the DAG generation process, a seed is generated at first, which relies on the length of the chain. The seed is then used to compute a 16 MB pseudo-random cache. Then, each item of the DAG is generated by utilising a certain number of items from the pseudo-random cache. This entire process enables the DAG to grow linearly with the growth of the chain. Then, the latest block header and the current candidate nonce are hashed using Keccak (SHA-3) hash function, and the resultant hash is mixed and hashed several times with data from the DAG. The final hashed digest is compared to the difficulty target and accepted or discarded accordingly.

Every mix operation in Ethash requires to have a read in a pseudo-random fashion from the DAG, which is randomly accessed from the memory. This serves two purposes:
\begin{itemize}
\item The read operation is limited by the speed of the memory access bandwidth, which is thought to be theoretically optimal, and thus, more optimisation is less likely.
\item Even though the mixing circuitry can be built within an ASIC, the bottleneck would still be the memory access delay. 
\end{itemize}
That is why Ethash is thought to be suitable for use on commodity computing capacity with good powerful GPUs. To achieve the same level of performance, an ASIC would require to accommodate as large memory as a general purpose computer providing a financial disincentive.

There are currently two currencies utilising Ethereum according to coingecko as of July 24, 2019 \cite{ethashCoin2017}. Even though Dagger algorithm is proven not to be ASIC resistant, it is being used in 6 currencies \cite{daggerCoin2017}. All of these are presented in Table \ref{tab:ethhash}. 
%
%
\begin{table*}[!h]
\centering
  \caption{Crypto-currencies utilising Ethash algorithm. The block rewards are in the corresponding currencies.}

   \begin{tabular}{p{20mm}|p{25mm}|p{20mm}|p{25mm}|p{15mm}}
    \hline
    \rowcolor[gray]{.6}
\centering\textbf{Currency } & 
\centering\textbf{Genesis date (dd.mm.yyyy)}& 
\centering\textbf{Block reward}& 
\centering\textbf{Total supply (Million)}& 
\centering\textbf{Block Time}\tabularnewline [2ex]
          \hline
	   \hline

Ethereum \cite{ethereum2018}&  30.07.2015& 2& infinite supply&10-20s\\\hline

\rowcolor[gray]{.9}Ethereum Classic \cite{ethereumclassic2017} & 30.07.2015& & 3.88&10-20s\\\hline
    Ubiq \cite{ubiqsmart2017}&   28.01.2017 & 6& NA&88s\\\hline
\rowcolor[gray]{.9}    Shift \cite{shiftnrg2017}&  01.08.2015& 1 & infinite supply&27s\\\hline
    Expanse \cite{expanse2017}& 13.09.2015& 4 &  31.4 Million &1.0m \\\hline
\rowcolor[gray]{.9}    DubaiCoin-DBIX \cite{dubaicoin2017}&   27.03.2017& 6 & Total supply&1.5m \\\hline
    SOILcoin \cite{soilcoin2017}&  03.10.2015& 3.0 & Total supply&52s \\\hline
\rowcolor[gray]{.9}    Krypton \cite{krypton2017}&  15.02.2016& 0.25 & 2.67 Million &1m 44s \\\hline
%
%
\end{tabular}
\label{tab:ethhash}
\end{table*}

\vspace{2mm}
\noindent\textbf{\textsc{5) NeoScrypt.}} NeoScrypt, an extension of Scrypt, is a key derivation function that aims to increase the security and performance on CPUs and GPUs while being strong ASIC resistant \cite{NeoScrypt2014}. Internally it utilises a combination of Salsa 20/20 \cite{bernstein2008salsa20} and ChaCha 20/20 \cite{bernstein2008chacha} along with Blake2s \cite{aumasson2013blake2}. Its constructions impose larger memory segment size, and hence, larger temporal buffer requirements. This makes it 1.25 times more memory intensive than Scrypt. The motivation is that this higher requirement of memory will act as a deterrent towards building ASICs for NeoScrypt.
%
%

Currently, there are 10 currencies utilising NeoScrypt according to Coingecko as of 18 July, 2019 \cite{NeoScryptCoins2017} which are presented in Table \ref{tab:neoscrypt}. 

\begin{table*}[!h]
\centering
 \caption{Crypto-currencies utilising NeoScrypt and Timetravel 10 algorithms. The block rewards are in the corresponding currencies.}
  \begin{tabular}{p{25mm}|p{25mm}|p{25mm}|p{25mm}|p{25mm}|p{15mm}}
    \hline
    \rowcolor[gray]{.6}
    \centering\textbf{Currency } & 
    \centering\textbf{Algorithm } & 
    \centering\textbf{Genesis date (dd.mm.yyyy)}& 
    \centering\textbf{Block reward}& 
    \centering\textbf{Total supply (Million)}& 
    \centering\textbf{Block Time}\tabularnewline [2ex]
          \hline
	   \hline
  Red Pulse \cite{redpulse2017}& &  17.10.2017&  NA& 1.36 Billion&  NA\\\hline
 \rowcolor[gray]{.9} Feathercoin \cite{feathercoin2017}& NeoScrypt&  16.04.2013 & 40& 336 Million &  1.0m \\\hline
     GoByte \cite{gobyte2017}&NeoScrypt&  17.11.2017& 3.71& 31.8 Million &2.5m\\\hline
  \rowcolor[gray]{.9}   UFO Coin \cite{ufocoin2017}& NeoScrypt& 03.01.2014& 625& 4 Billion &  1.5m \\\hline
   Innova \cite{innova2017}& NeoScrypt&  19.10.2017& 2.64&1.29 Million & 2m\\\hline
  \rowcolor[gray]{.9}   Vivo \cite{viacoin2014}& NeoScrypt&  20.08.2017& 4.5& 1.1 Million &  2m18s\\\hline
     Desire \cite{desire2017}& NeoScrypt& Genesis date& 10.45& 1.17 Million & 2.5m\\\hline
  \rowcolor[gray]{.9}   Orbitcoin \cite{orbitcoin2017} & NeoScrypt& 28.07.2013& 0.5& 3.77 Million & 6.0m \\\hline
     Phoenixcoin \cite{phoenixcoin2017}& NeoScrypt& 08.05.2013& 12.5&98 Million &  1.5m  \\\hline
 \rowcolor[gray]{.9}    Bitcore \cite{bitcore2017}&  Timetravel 10 &  April 24, 2017 &   3.13 &  21 Million  &    2.5m\\\hline
\end{tabular}
 
  \label{tab:neoscrypt}
\end{table*}

\vspace{2mm}


\subsubsection{Chained PoW}
A chained PoW utilises several hashing functions chained together in a series of consecutive steps. Its main motivation is to ensure ASIC resistance, which is achieved by the underlying mechanisms by which the corresponding hashing functions are chained together. In addition to this, the PoW algorithms belonging to this series aim to address one particular weakness of any compute-bound and memory-bound PoW algorithm: their reliance on a single hashing function. With the advent of quantum computing, the security of a respective hashing algorithm might be adversely affected, which undermines the security of the corresponding blockchain system. If this happens, the old algorithm needs to be discarded, and a new quantum resistant hashing algorithm needs to be incorporated to the respective blockchain using a mechanism called hard-fork. A hard-fork is a mechanism by which a major update is enforced in a blockchain system. This is quite a disruptive procedure that has negative effect on any blockchain system. In such scenarios, a chained PoW algorithm would continue to function until all its hashing functions are broken.


There are several chained PoW algorithms that are currently available.

\vspace{2mm}\noindent\textbf{\textsc{1) X11/X13/X15.}} X11 is a widely-used hashing algorithm in many crypto-currencies.  In X11, eleven hashing algorithms are consecutively carried our one after another. The hashing algorithms are \emph{blake, bmw, groestl, jh, keccak, skein, luffa, cubehash, shavite, simd, and echo}. 

One advantage of X11 is that it is highly energy efficient: GPUs computing X11 algorithm requires approximately $30\%$ less wattage and remains $30-50\%$ cooler in comparison to Scrypt \cite{miningComp2017}. Even though the algorithm has been designed in such a way that it can only be used with CPUs and GPUs, the economic incentives have allowed the creation of ASIC to be used during the mining process.

It has different variants where the number of chained hashing functions differs. For example, X13 utilises 13 hashing functions, and X15 utilises 15 hashing functions.

Table \ref{tab:x11} presents the top 10 crypto-currencies utilising these three algorithms, as per their market capitalisation as of July 24, 2019 according to CoinGecko. 

\begin{table*}[!h]
\centering
\caption{Crypto-currencies utilising X11/X13 algorithms. The block rewards are in the corresponding currencies.}

  \begin{tabular}{p{25mm}|p{10mm}|p{25mm}|p{20mm}|p{25mm}|p{15mm}}
    \hline
    \rowcolor[gray]{.6}
\centering\textbf{Currency } & 
\centering\textbf{Algorithm } & 
\centering\textbf{Genesis date (dd.mm.yyyy)}& 
\centering\textbf{Block reward}& 
\centering\textbf{Total supply (Million)}& 
\centering\textbf{Block Time} \tabularnewline [2ex]
          \hline
	   \hline
    Dash \cite{dashcoin2017}& X11& January 19, 2014& 1.55& 22 Million &2.5m\\\hline
\rowcolor[gray]{.9}      Stratis \cite{stratis2017}& X13& August 09, 2016&NA&NA&  NA \\\hline
     Cloakcoin \cite{cloakcoin2017}& X13& Genesis date& 496.00&  4.5 Million &  1.0m\\\hline
\rowcolor[gray]{.9}       Stealthcoin \cite{stealthcoin2017}& X13&July 04, 2014& NA& 20.7 Million &  1.0m\\\hline
    DeepOnion \cite{deeponion2017}& X13& July 13, 2017& 4& 18.9 Million & 4m\\\hline
\rowcolor[gray]{.9}     HTMLcoin \cite{htmlcoin2017}&X15& September 12, 2014& NA& 90 Billion& 1.0m\\\hline
     Regalcoin \cite{regalcoin2017}& X11& September 28, 2017& NA& 7.2 Million & NA\\\hline
\rowcolor[gray]{.9}       Memetic \cite{memetic2017}& X11& March 05, 2016 & NA& NA&  NA\\\hline
  ExclusiveCoin \cite{exclusiveCoin2017} &X11&June 12, 2016& NA& NA& NA \\\hline
\rowcolor[gray]{.9}  Creditbit \cite{creditbit2017}&X11&November 02, 2015&   NA& 100 Million & 1.0m\\\hline
\end{tabular}
  \label{tab:x11}
\end{table*}

\vspace{2mm}
\noindent\textbf{\textsc{2) Quark.}} Quark PoW algorithm relies on six different hashing functions: BLAKE \cite{aumasson2013blake2}, Blue Midnight Wish \cite{gligoroski2009cryptographic}, Gr{\o}stl \cite{gauravaram2009grostl, vertCoin2017}, JH \cite{wu2011hash}, Keccak and Skein \cite{QuarkCoin2017}. These functions are implemented in mixed series with nine steps \cite{QuarkCoinWiki2017}. Within these nine steps, three functions are randomly applied in three steps depending on the value of a bit. The main motivations of mixing these six functions in nine steps are as follows:

\begin{itemize}
	\item To alleviate the risk of a compromised system in light of its underlying single hashing algorithm being broken.
    \item To impose restrictions so that Quark can only be mined using CPUs while making it difficult to mine using GPUs and ASICs, because of the usage of ASIC-resistant mechanisms (e.g. Keccak).
\end{itemize}

However, it did not take long before ASIC mining hardware for Quark appeared in the market, so that this could be mined using a GPU and ASIC \cite{QuarkASICMining2016}. However, the profitability and performance of such hardware are not obvious.

The currencies utilising Quark according to CoinGecko as per July 24, 2019 \cite{quarkCoinGecko2017} are presented in Table \ref{tab:quark}. 

\begin{table*}[!h]
\centering
  \caption{Crypto-currencies utilising Quark algorithm. The block rewards are in corresponding currencies.}

  \begin{tabular}{p{30mm}|p{25mm}|p{20mm}|p{25mm}|p{15mm}}
    \hline
    \rowcolor[gray]{.6}
\centering\textbf{Currency } & 
\centering\textbf{Genesis date (dd.mm.yyyy)}& 
\centering\textbf{Block reward}& 
\centering\textbf{Total supply (Million)}& 
\centering\textbf{Block Time}\tabularnewline [2ex]
          \hline
	   \hline
     Quark \cite{QuarkCoin2017}& July 21, 2013& 1& 247 Million& 0.5s\\\hline
     \rowcolor[gray]{.9}PIVX \cite{pivx2017}& NA& 5& NA &  1.0m\\\hline
     MonetaryUnit \cite{monetaryunit2017}& July 26, 2014& 18& 1 Quadrillion &  0.67m\\\hline
      \rowcolor[gray]{.9}ALQO \cite{alqo2017}&October 30, 2017& 3&NA&  1m\\\hline
     Bitcloud \cite{bitcloud2017}& August 15, 2017& 22.5& 200 Million&6.5m\\\hline    
      \rowcolor[gray]{.9}Zurcoin \cite{zurcoin2017}& NA & 12.5&NA& 0.75m\\\hline
     AmsterdamCoin \cite{amsterdamcoin2017}& November 01, 2015& 10& 84 Million & 1.0m\\\hline
     \rowcolor[gray]{.9} Animecoin \cite{animecoin2017}& NA & NA&NA& NA\\\hline
\end{tabular}
  \label{tab:quark}
\end{table*}

\vspace{2mm}
\noindent\textbf{\textsc{3) Lyra2RE.}}
Lyra2RE is a class of chained PoW which utilises five hash functions: BLAKE, Keccak, Lyra2,[13] Skein, and Gr{\o}stl. It has been developed by the developers of Vertcoin, a currency based on Lyra2RE. It was designed to be CPU friendly, however, it was discovered in 2015 that the majority of the hashing power utilised for mining VertCoin in its network was facilitated by a botnet stealing CPU cycles from a large number of infected computers. This motivated the Vertcoin developers to release Lyra2REv2, which utilises six hash functions, BLAKE, Keccak, CubeHash, Lyra2, Skein, and Blue Midnight Wish with GPU only PoW. Currently, there are only three currencies utilising Lyra2REv2 according to CoinGecko as of 31 December 2017 which are presented in Table \ref{tab:lyra2re}. 

\begin{table*}[!h]
\centering
  \caption{Crypto-currencies utilising Lyra2RE algorithm. The block rewards are in corresponding currencies.}

  \begin{tabular}{p{20mm}|p{25mm}|p{30mm}|p{25mm}|p{15mm}}
   \hline
   \rowcolor[gray]{.6}
\centering\textbf{Currency } & 
     \centering\textbf{Genesis date (dd.mm.yyyy)}& 
     \centering\textbf{Block reward}& 
     \centering\textbf{Total supply (Million)}& 
     \centering\textbf{Block Time} \tabularnewline [2ex]
          \hline
	   \hline
     Vertcoin \cite{vertcoin2017}& January 10, 2014& 25& 84 Million & 2.5m\\\hline
    \rowcolor[gray]{.9} Monacoin \cite{monacoin2017}& January 01, 2014& 25& 105 Million &  1.5m\\\hline
     Crypto \cite{crypto2017}& April 30, 2015 &NA&65.8 Million &  0.5m \\\hline
\end{tabular}
  \label{tab:lyra2re}
\end{table*}

\vspace{2mm}
\noindent\textbf{\textsc{4) Magnificent 7.}}
\textit{Magnificent 7} (M7) is a class of chained PoW which utilises seven hash functions to generate the candidate hash during the mining process of Cryptonite coin (not to be confused with the Cryptonight PoW algorithm) \cite{m7Wiki2017}. The utilised hash functions are SHA-256, SHA-512, Keccak, RIPEMD, HAVAL, Tiger and Whirlpool. Internally, the header of the candidate block sequentially hashed by the corresponding functions and then multiplied to generate the final hash, which is then compared against the difficulty threshold. Even though it a not memory-bound PoW, it has been claimed that the multiplication operation enables it to run on a general purpose CPU easily, however, makes it difficult to run on GPUs and ASICs \cite{m7Wiki2017}. Even so, there are is at least one GPU miner available for M7 \cite{cudaMiner2017}. Its performance, though, is not known. The corresponding information for Cryptonite is presented in Table \ref{tab:m7}.

\begin{table*}[h]
\centering
  \caption{Information regarding Cryptonite utilising M7 algorithm. }
  \begin{tabular}{p{20mm}|p{20mm}|p{20mm}|p{20mm}|p{20mm}}
    \hline
        \rowcolor[gray]{.6}
  \centering\textbf{Currency } & 
     \centering\textbf{Genesis date (dd.mm.yyyy)}& 
    \centering\textbf{Block reward}& 
   \centering\textbf{Total supply (Million)}& 
    \centering\textbf{Block Time} \tabularnewline [2ex]
          \hline
          \hline
        Cryptonite & July 28, 2014 & Dynamic & $1.84$ Billion & 1 Minute\\\hline
\end{tabular}
  \label{tab:m7}
\end{table*}

\subsubsection{PoW Limitations}
\label{sec:incentivised:subsec:pow:subsubsec:limitation}

PoW (Nakamoto) consensus algorithm has been widely accoladed for its breakthrough in the distributed consensus paradigm, starting with Bitcoin. It had laid down the foundation for the subsequent advancement, which resulted in different PoW algorithms and crypto-currencies as discussed in the earlier sections. Even so, there are some significant limitations.  Next, we briefly discuss these limitations:

\begin{itemize}
	\item \textbf{Energy consumption:} Each PoW algorithm needs to consume electricity to compute the hash. As the difficulty of the network starts to increase, so does the energy consumption. The amount of consumed energy is quite significant when calculated over the whole network consisting of ASIC/GPU mining rigs all around the world. Digiconomist \footnote{https://digiconomist.net/} website tracks the electricity consumption of Bitcoin and Ethereum. According do it, the energy consumption of Bitcoin and Ethereum are around 40 TWh (Tera-Watt Hour) and 10 TWh, respectively. Their energy consumption graphs for the last one year are presented in Figure \ref{fig:bitcoinEnergy} \cite{bitcoinEnergy2018} and Figure \ref{fig:etherEnergy} \cite{etherEnergy2018}. 
    \begin{figure}[t]
      \centering
      \includegraphics[scale=.5]{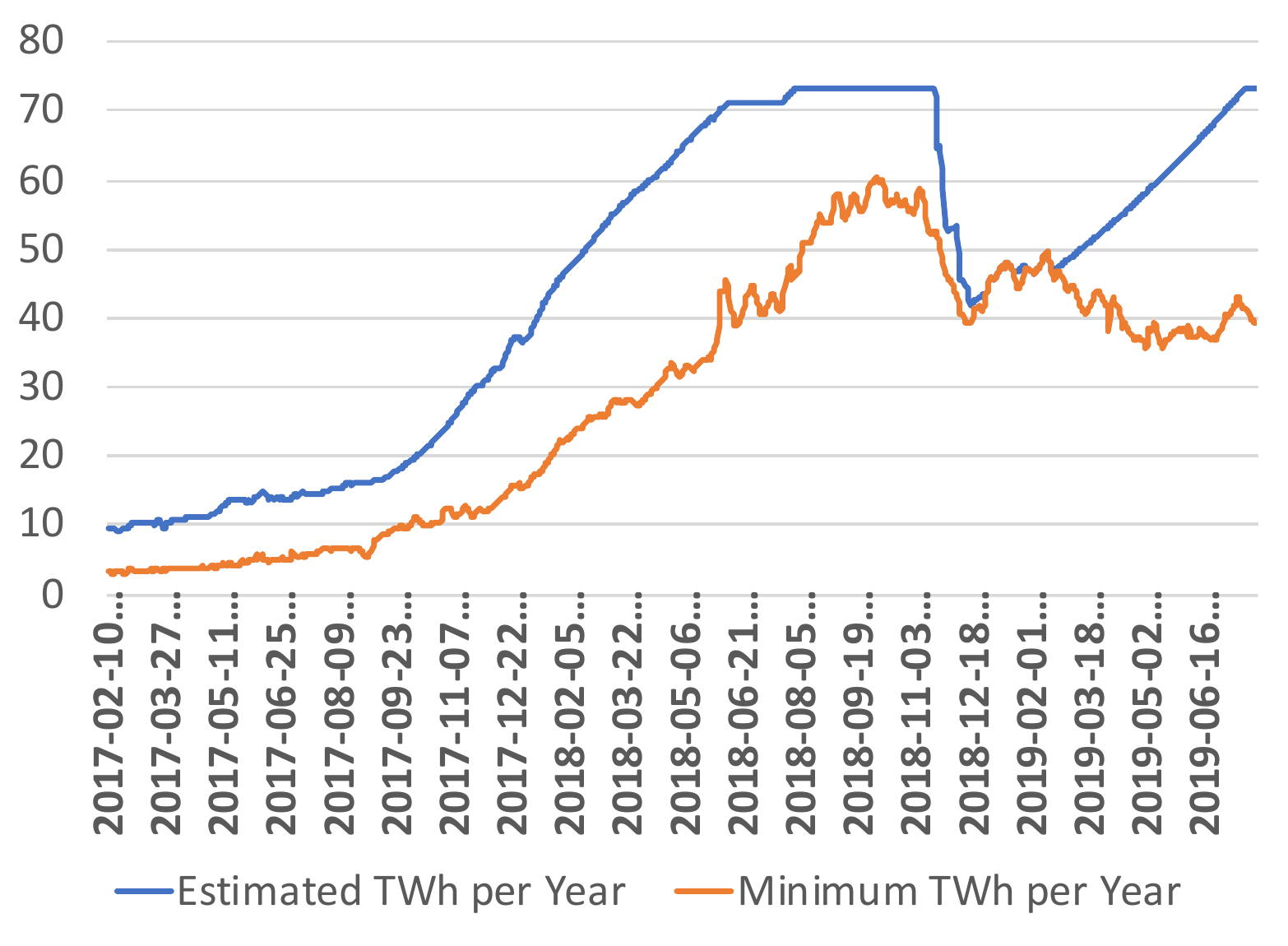}
      \caption{Bitcoin energy consumption over the last years.}
      \label{fig:bitcoinEnergy}
     \end{figure}
     
     \begin{figure}[!h]
      \centering
      \includegraphics[scale=0.75]{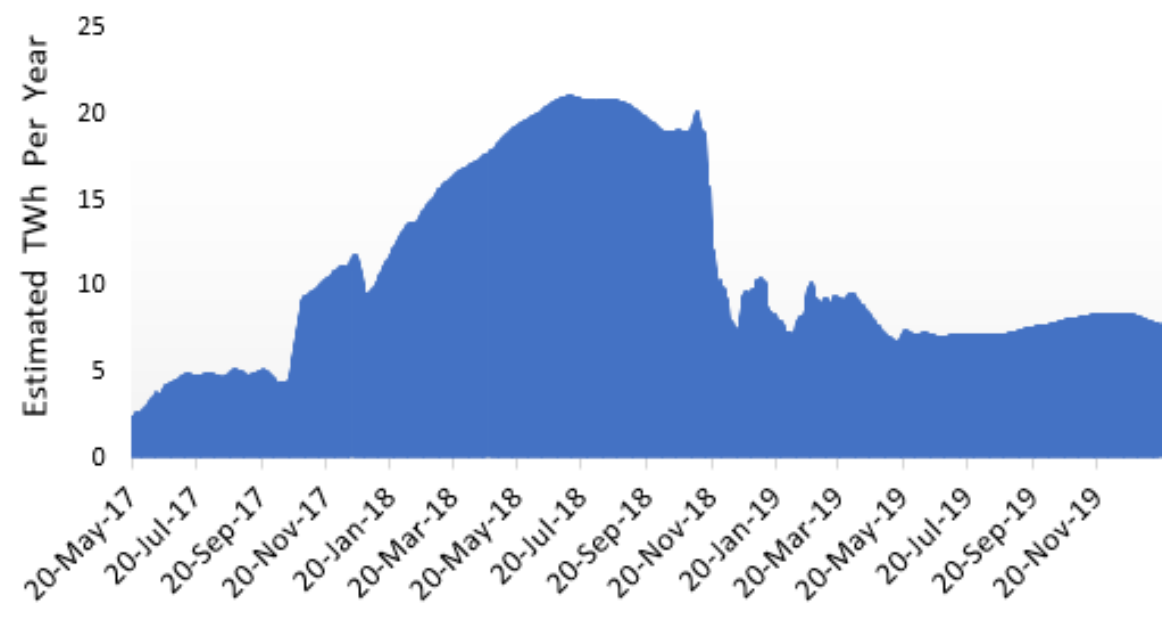}      
      \caption{Ethereum energy consumption over the last year.}
      \label{fig:etherEnergy}
     \end{figure}
     
     To put this into perspective, we present Figure \ref{fig:bitcoinEnergyCountry}, whose data has been collected from \cite{bitcoinEnergy2018}. This figure illustrates Bitcoin's energy consumption relative to the electricity consumption of different countries. For example, the electricity consumed by Bitcoin in a year could power up $6,770,506$ American households and is much more than what Czech Republic consumes in a year \cite{bitcoinEnergy2018}. The utilisation of this huge amount of electricity has raised the question of sustainability of PoW-based crypto-currencies.
     
    \begin{figure}[!h]
      \centering
      \includegraphics[scale=0.54]{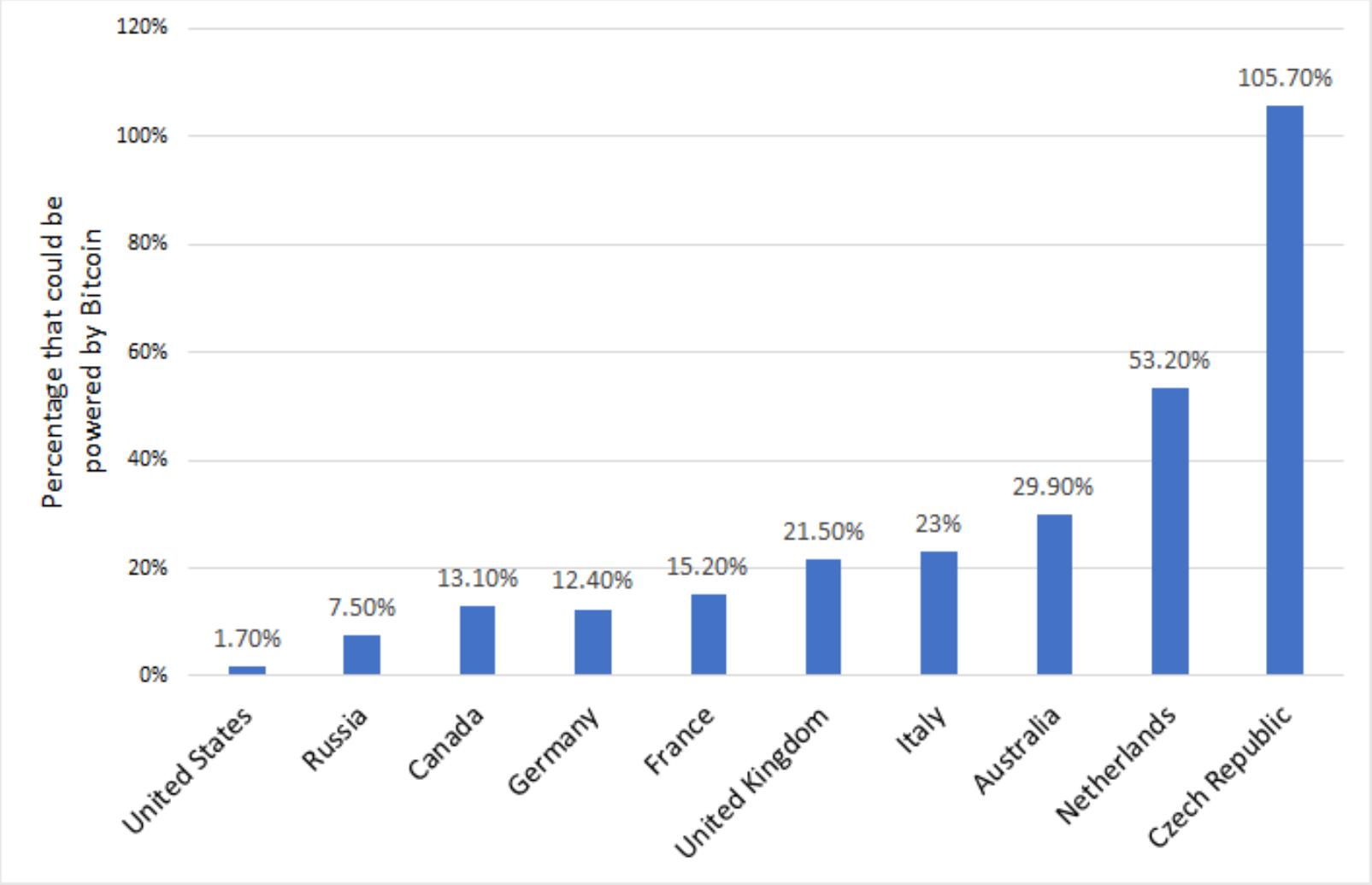}      
      \caption{Bitcoin energy consumption relative to different countries.}
      \label{fig:bitcoinEnergyCountry}
     \end{figure}
     
    \item \textbf{Mining centralisation:} With the ever-increasing difficulty rate, miners within a PoW-based crypto-currency network need to upgrade the capability of their ASIC/GPU mining rigs to increase their chance of creating a new block. Even so, it becomes increasingly difficult for a single miner to join in the mining process without substantial investment in the mining rigs. The consequence is that the \textit{economies of scale} phenomenon strongly impacts the PoW algorithms. The economies of scale in economic theory is the advantage a producer can gain by increasing its output  \cite{economyScaleWiki2017}. This happens because the producer can spread the cost of per-unit production over a larger number of goods, which increases the profit margin. This analogy also applies to PoW mining as explained next. A mining pool can be created where the mining resources of different miners are aggregated to increase the chance of creating a new block. Once a mining pool receives a reward for creating the next block, the reward is then proportionally divided among the participating miners. Unfortunately, this has led to the centralisation problems where block creations are limited among a handful of miners. For example, Figure \ref{fig:bitcoinPool} illustrates the distribution of network hashrate among different miners in Bitcoin \cite{bitcoinHashrate2018}. As evident from the figure, only five mining pools control the $75\%$ of hashrate of the whole network. There is a fear that they could collude with each other to launch the $51\%$ attack to destabilise the whole bitcoin network.
    
    \begin{figure}[!h]
      \centering
      \includegraphics[width=1.0\linewidth]{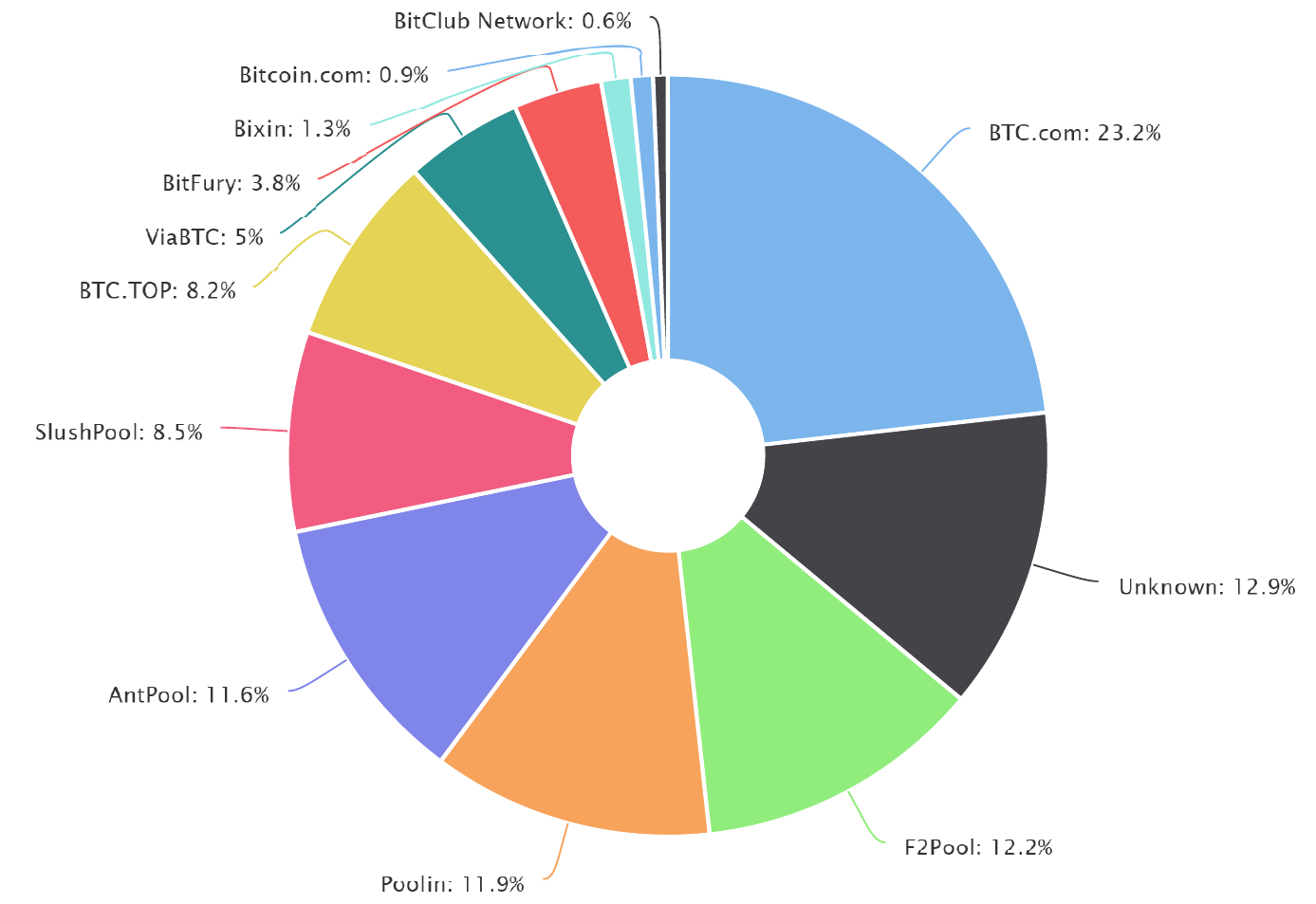}      
      \caption{Bitcoin hashrate distribution of mining pools.}
      \label{fig:bitcoinPool}
     \end{figure}
 
    \item \textbf{Tragedy of commons:} Many PoW algorithms suffer an economic problem called the \textit{Tragedy of the commons}. In economic theory, the tragedy of the commons occurs when each entity rushes to maximise its profit from a depleting resource without considering the well-being of all that share the same resource \cite{commongTragedyWiki2017}. This situation occurs in a crypto-currency if it is deflationary in nature with limited supply, e.g. Bitcoin. It has been argued when the reward of creating a new block in Bitcoin will reach nearly zero; the miners will have to solely rely on the transaction fees to cover their expenses. This might create an unhealthy competition among the miners to include as many transactions as possible, just to maximise one's profit. The consequence of this is that transaction fees will keep decreasing, which might lead to a situation that miners cannot make enough profit to continue the mining process. Eventually, more and more miners will leave the mining process,  which might lead toward $51\%$ attacks or other scenarios that de-stabilise the Bitcoin network.   
    
    \item \textbf{Absence of penalty:} All PoW algorithms (both compute and memory bound) are altruistic in nature in the sense that they reward behaving miners, however, do not penalise a misbehaving  miner. One example is that a miner can collude with a group of miners (a phenomenon known as \textit{selfish mining}) to increase its profitability in an illegitimate way  \cite{Eyal2014bitcoin}. In addition, a miner can engage in Denial-of-Service attack by just not forwarding any transaction or block within the network. Furthermore, such malicious miners can join forces to engage in the \textit{spawn-camping} attack, in which they launch DoS attacks simultaneously over and over again to render the network useless for the corresponding crypto-currency \cite{jonchoi2012}. A penalty mechanism would disincentivize any miner to engage in any type of malicious misbehave.  
%
%
%
%
%
%
\end{itemize}

\subsubsection{Analysis}

In this section, we summarise the properties of different PoW algorithms in Table \ref{tab:powStruct}, Table \ref{tab:powSec} and Table \ref{tab:powPerform} utilising the taxonomies presented in Section \ref{sec:conTaxProp}. In these tables, a `\checkmark' symbol is utilised to indicate if a particular property is supported by the corresponding algorithm. For other properties, explanatory texts have been used for any particular property.

As presented in Table \ref{tab:powStruct}, different types of PoW algorithms share exactly similar characteristics. In these algorithms, they are mainly two types of nodes: clients and miners. Miners are responsible for creating a block using a randomised lottery mechanism. Conversely, clients are the nodes  that are responsible for validating each block as well as utilised to transact bitcoin between different users. Committees in these algorithms represent the set of miners, exhibiting the property of a single open committee structure where anyone can join as a miner. The respective committee is formed implicitly in a dynamic fashion, indicating any miner can join or leave whenever they wish.

\begin{table*}[h]
\centering
\caption{Structural properties of PoW consensus algorithms.}
\label{tab:powStruct}
\begin{tabular}{p{25mm}|c|c|c|p{30mm}}
\hline
\rowcolor[gray]{.75}
 \cellcolor[gray]{.75}& 
 \multicolumn{3}{c|}{\centering\textbf{Single committee}} & \cellcolor[gray]{.75} \\ \cline{3-5}

\multirow{-2}{*}{\cellcolor[gray]{.75}\centering\textbf{Node type}} &
 \cellcolor[gray]{.75}\centering\textbf{Type} &
\cellcolor[gray]{.75}\centering\textbf{ Formation} & 
\cellcolor[gray]{.75}\centering\textbf{ Configuration} & 
 {\multirow{-2}{*}{\cellcolor[gray]{.75}\centering\textbf{Mechanism}} }
 \\ \hline
\hline

Clients \& Miners & Open & Implicit & Dynamic & Lottery, Randomised \\ \hline \hline
\end{tabular}
\end{table*}

\begin{table*}[h]
\centering
\caption{Security properties of PoW consensus algorithms.}
\label{tab:powSec}
\begin{tabular}{c|c|p{15mm}|c|c|c}
\hline
\rowcolor[gray]{.75}
 & 
 & 
 & 
 \multicolumn{3}{c|}{\textbf{Attack Vectors}} \\ \cline{4-6}

 {\multirow{-2}{*}{\cellcolor[gray]{.75}\textbf{Authentication}}}& 
 {\multirow{-2}{*}{\cellcolor[gray]{.75}\textbf{Non-repudiation}}} & 
 {\multirow{-2}{*}{\cellcolor[gray]{.75}\textbf{\begin{tabular}[c]{@{}c@{}}Censorship \\ resistance\end{tabular}}}} & 
 \cellcolor[gray]{.75}\textbf{Adversary tolerance} & 
 \cellcolor[gray]{.75}\textbf{Sybil protection} &
 \cellcolor[gray]{.75}\textbf{DoS Resistance} \\ \hline
\hline
x & \checkmark & High & $2f + 1$ & \checkmark & \checkmark \\ \hline \hline
\end{tabular}
\end{table*}

\begin{table*}[!h]
\centering
  \caption{Performance properties of PoW consensus algorithms.}

   \begin{tabular}{p{20mm}|p{20mm}|p{15mm}|p{25mm}|p{30mm}}
    \hline
   \rowcolor[gray]{.75} 
\centering\textbf{Fault tolerance}& 
\centering\textbf{Throughput}& 
\centering\textbf{Scalability}& 
\centering\textbf{Latency}&
\centering\textbf{Energy consumption} \tabularnewline [2ex]
          \hline
          \hline
	    \centering $2f + 1$ & \centering Low & \centering Low & \centering Medium-High & \centering High \tabularnewline [0ex] \hline \hline
\end{tabular}
\label{tab:powPerform}
\end{table*}
As per Table \ref{tab:powSec}, none of the algorithms requires any node to be authenticated to participate in the algorithm. All of them have strong support for non-repudiation in the form of digital signature as part of every single transaction. These algorithms have a high level of censorship resistance, which means that it will be difficult for any regulatory agency to impose any censorship on these algorithms. As for the attack vector, each PoW algorithm requires every miner node to invest substantially for mining hardware in order to participate in these consensus algorithms. This feature, thus, acts as a deterrent against any Sybil or DoS attack in any PoW algorithm. The adversary tolerance is based on the assumption that PoW suffers from $51\%$ attacks, and thus, adversary nodes need to have less than $50\%$ of the total hashing power of the network.

According to Table \ref{tab:powPerform}, these algorithms have low throughput, and unfortunately, do not scale properly. Furthermore, most of the algorithms require a considerable time to reach finality, and their energy consumption is considerably high, as explained in Section \ref{sec:incentivised:subsec:pow:subsubsec:limitation}. The fault tolerance in these algorithms is $2f+1$ like any BFT algorithm, implying they can achieve consensus as long as more than $50\%$ of nodes function correctly.
\subsection{Proof of Stake}
\label{sec:incentivised:subsec:pos}
%
%
To counteract the limitations of any PoW algorithm, another type of consensus algorithm, called Proof of Stake (PoS) has been proposed. The earliest proposal of a PoS algorithm can be found on the \emph{bitcointalk} forum in 2011 \cite{posFirst2011}. Soon after, several projects started experimenting with the idea. Peercoin~\cite{peercoin2012}, released in 2012, was the first currency to utilise the PoS consensus algorithm. 

The core idea of PoS evolves around the concept that the nodes who would like to participate in the block creation process must prove that they own a certain number of coins at first. Besides, they must lock a certain amount of its currencies, called \textit{stake}, into an escrow account in order to participate in the block creation process. The stake acts as a guarantee that it will behave as per the protocol rules. The node escrows its stake in this manner is known as the stakeholder, leader, forger, or minter in PoS terminology. The minter can lose the stake, in case it misbehaves.

In essence, when a stakeholder escrows its stake, it implicitly becomes a member of an exclusive group. Only a member of this exclusive group can participate in the block creation process. In case the stakeholder gets the chance to create a new block,  the stakeholder will be rewarded in one of the two different ways. Either it can collect the transaction fees within the block, or it is provided a certain amount of currencies that act as a type of interest against their stake. 

It has been argued that this incentive, coupled with any punitive mechanism, can provide a similar level of security of any PoW algorithm. Moreover, it can offer several other advantages. Next, we explore a few benefits of a PoS mechanism \cite{jonchoi2012}.

\begin{itemize}
	\item \textbf{Energy Efficiency:} A PoS algorithm does not require any node to solve a resource-intensive hard cryptographic puzzle. Consequently,  such an algorithm is extremely energy efficient compared to their PoW counterpart. Therefore, a crypto-currency leveraging any PoS algorithm is likely to be more sustainable in the long run.
    \item \textbf{Mitigation of Centralization:} A PoS algorithm is less impacted by the economies of scale phenomenon. Since it does not require to build up a mining rig to solve any resource-intensive cryptographic puzzle, there is no way to maximise gain by increasing any output. Therefore, it is less susceptible to the centralisation problem created by the mining pool.
    \item \textbf{Explicit Economic Security:} A carefully designed penalty scheme in a PoS algorithm can deter any misbehaving attack, including  spawn-camping. Anyone engaging in such attacks will lose their stake and might be banned from any block creation process in the future, depending on the protocol. This eventually can strengthen the security of the system.
\end{itemize}

\textbf{Initial supply:} One of the major barriers in a PoS algorithm is how to generate the initial coins and fairly distribute them among the stakeholders so that they can be used as stakes. We term this barrier as the \textit{bootstrap} problem. There are two approaches to address the bootstrap problem:

\begin{itemize}
	\item Pre-mining: A set of coins are pre-mined, which are then sold before the launch of the system in an IPO (Initial Public Offering) or ICO (Initial Coin Offering). 
    
    \item PoW-PoS transition: The system starts with a PoW system to fairly distribute the coins among the stakeholders. Then, it slowly transitions towards the PoS system.
\end{itemize}

\textbf{Reward process:} Another important aspect is the rewarding process to incentivise the stakeholder to take part in the minting process. Unlike any PoW, where a miner is rewarded with new coins for creating a valid block, there is no reward for creating a valid block. Instead, to incentivise a minter, two types of reward mechanisms are available within a PoS algorithm:

\begin{itemize}
	\item Transaction Fee: The minter can collect fees from the transactions included within the minted block.
    \item Interest rate: A lower interest rate is configured, which allows the currency to inflate over time. This interest is paid to the minter as a reward for creating a valid block.
\end{itemize}

\textbf{Selection process:} A crucial factor in any PoS algorithm is how to select the stakeholder who can mint the next block. In a PoW algorithm, a miner is selected based on who can find the resource-intensive desired hash. Since PoS does not rely on hind such a hash as the mechanism to find the next block, there must be a mechanism to select the next stakeholder. 

Currently, there three different approaches to Proof of Stake: Chained, BFT, and Delegated. 
%
%
%

\vspace{2mm}
\noindent\textbf{\textsc{Chained PoS.}} The general idea of a chained PoS is to deploy a combination of PoW and PoS algorithms chained together to achieve any consensus. Because of this, there can be two types of blocks, PoW and PoS blocks, within the same blockchain system. To accomplish this, the corresponding algorithm relies on different approaches to  select/assign a particular miner for creating a PoW block or select a set of  validators for creating a PoS block in different epochs or after a certain number of blocks created. In general, a chain based PoS can employ any of the following three different approaches to select the miner/stakeholder:


\begin{itemize}
	\item \textit{Randomised PoW Mining:} A miner who can solve the corresponding cryptographic PoW puzzle is selected in a random fashion.
	\item \textit{Randomised Stakeholder Selection:} A randomised PoS utilises a probabilistic formula that takes into account the staked currencies and other parameters to select the next stakeholder. The other parameters ensure that a stakeholder is not selected only based on the number of their staked coins and act as a pseudo-random seed for the probabilistic formula.
    \item \textit{Coin-age based selection.} A coin-age is defined as the holding period of a coin by its owner. For example, if an owner receives a coin from a sender and holds it for five days then the coin-age of the coin can be defined as five coin-days. Formally,
			\[
				\mathit{coin-age} = \mathit{coin} * \mathit{holding period}
			\]
     	   Algorithms belonging to this class select the stakeholder using staked coins of the stakeholders and their corresponding coin-age.	       
\end{itemize}

In general, a chained PoS algorithm favours towards availability over consistency when network partition  occurs, as per the CAP theorem.

\vspace{2mm}
\noindent\textbf{\textsc{BFT PoS.}} BFT PoS is a multi-round PoS algorithm. In the first step, a set of validators are pseudo-randomly selected to propose a block. However, the consensus regarding committing this block to the chain depends on the $>2/3$ quorum of super-majority among the validators on several rounds. It inherits the properties of any BFT consensus, and as such, it tolerates up to $1/3$ of byzantine behaviour among the nodes.

In general, a BFT PoS algorithm favours towards consistency over availability when network partition occurs, within the setting of CAP theorem. 

\vspace{2mm}
\noindent\textbf{\textsc{Delegated Proof of Stake.}} Delegated Proof of Stake (or DPoS in short) is a form of consensus algorithm in which reputation scores or other mechanisms are used to select the set of validators \cite{dposWhat2017}. Even though it has the name Proof of Stake associated with it, it is quite different from other PoS algorithms. 

In DPoS, users of the network vote to select a group of delegates (or witnesses) who are responsible for creating blocks. Users utilise reputations scores or other mechanisms to choose their delegates. Delegates are the only entities who can propose new blocks. For each round, a leader is selected from the set of delegates who can propose a block. How such a leader is chosen depends on the respective system. The leader gets rewards for creating a new block, and is penalised and de-listed from the set of validators if it misbehaves.

The delegates themselves compete with each other to get included in the validator list. In such, each validator might offer different levels of incentives for the voters who vote for it. For example, if a delegate is selected to propose a block, it might distribute a certain fraction of its reward among the users who have selected it. Since the number of validators is small, the consensus finality can be fast.

Next, we explore several crypto-currencies or mechanisms that use the above mentioned PoS approaches.

\subsubsection{Chained PoS}
\label{sec:incentivised:subsec:pos:subsubsec:chainedPoS}
Next, we present two examples of a chained PoS algorithm to illustrate how this approach has been applied in practice.

\vspace{2mm}
\noindent\textbf{\textsc{1) PeerCoin (PPCoin).}} Peercoin is the first crypto-currency to formalise the notion of PoS by utilising a hybrid PoW-PoS protocol \cite{PeerCoinPaper2012}. The Peercoin protocol is based on the assumption that \textit{coin-age} can be leveraged to create a PoS algorithm which is as secure as any PoW algorithm while minimising the disadvantages associated with a PoW algorithm. 

Peercoin protocol recognises two different kinds of blocks: PoW blocks and PoS blocks, within the same blockchain. These blocks are created by two separate entities: miners and minters. Miners are responsible for creating PoW blocks, similar to Bitcoin whereas minters are responsible for creating PoS blocks. Irrespective of the last block type, the next block either can be a PoW block or a PoS block, and these entities compete with each other to create the next block \cite{peercoinDiscussion12014}. Miners compete with other miners to find a valid PoW block that matches the PoW difficulty target, similar to Bitcoin. Similarly, minters compete among themselves  to find a valid PoS block that matches the PoS difficulty target (similar to a PoW algorithm but requires much less computation). As soon as any PoW or PoS block is found, it is broadcast to the network, and other nodes validate it.

Within a PoS block, a minter utilise their holding coins as a stake, and the minter is rewarded approximately 1\% per annum based on the coin-age of the staked coins. The reward is paid out for each block in a newly created special transaction called the \textit{coinstake} transaction. Each coinstake transaction consists of the number of staked inputs and a \textit{kernel}, containing the hash that meets the PoS difficulty. The hash itself is calculated over a small space and hence not computationally intensive at all. It utilises the number of staked inputs and a probabilistic variable,  whereas the difficulty condition is calculated utilising the coin-age of the staked inputs as well as a difficulty parameter. This parameter is adjusted dynamically to ensure that one block is created in 10 minutes. In other words, the valid kernel depends on the coin-age of the staked inputs, and the higher the coin-age, the higher is the probability to match the difficulty.


The coinstake transaction is paid to the minter, which contains the coins staked along with the reward.  Once a PoS block is added to the chain, the coin-age of the staked coins is reset to zero. This indicates that all the stacked coins are consumed. This ensures that the same coins cannot be used over and over again to create a PoS block within a short period of time. The main chain in Peercoin is selected based on the highest total coin-age consumed in all blocks. That means, if a PoW block and PoS block are received simultaneously as the next block by a node, the algorithm dictates the PoS block to be selected over the PoW block.

The block reward for a PoW block in Peercoin decreases and will cease to be significant after a certain period of time. It is currently used for the coin generation and distribution purpose and will be completely phased out in the future \cite{peercoinDiscussion22014}. It has no role whatsoever on securing the network, which is largely based on the PoS algorithm. Once the PoW algorithm is phased out, it is suggested that the energy consumption of Peercoin will be significantly low while providing similar security as any PoW algorithm.

Peercoin is highly regarded for formalising the first alternative mechanism to PoW, however, it suffers from all the attack vectors of PoS, as presented in Section \ref{subsubsec:posLimitations}. Two other coins Black and Nxt removes age from the equation in order to avoid the exploitation of the system by the dishonest entities having a significant amount of coins. 

\vspace{2mm}
\noindent\textbf{\textsc{2) CASPER FFG.}} Casper the Friendly Finality Gadget (CFFG) is a PoW-PoS hybrid consensus algorithm proposed to replace the Ethereum's PoW consensus algorithm \cite{casperFFGPaper2014}. In fact, CFFG provides an intermediate PoS overlay on top of its current PoW algorithm so that Ethereum is transformed to a pure PoS protocol called Casper the Friendly Ghost (CTFG) described below (Section \ref{sec:incentivised:subsec:pos:subsubsec:bft}).  

The PoS layer requires the participation of validators. Any node can become a validator by depositing some Ethereum's native crypto-currency called \textit{Ether} to a designated smart-contract, which acts as a security bond. The network itself will mostly consist of PoW miners who will mine blocks according to its current PoW algorithm. However, the finalisation/check-pointing of blocks will be carried out by PoS validators. The check-pointing/finalisation is the process to ensure that the chain becomes irreversible up to a certain block and thus, short and low range attacks (particular types of PoS only attacks presented in Section \ref{subsubsec:posLimitations}) as well as the $51\%$ attack cannot be launched beyond the check-pointing block.

The check-pointing occurs  every 50 blocks, and this interval of 50 blocks is called an \textit{epoch} \cite{caspervten2017}. The finalisation process requires two rounds of voting in two successive epochs. The process is as follows. In an epoch, the validators vote on a certain checkpoint $c$ (a block). A super-majority (denoted as $+2/3$) occurs when more than $2/3$ of the validators vote for the checkpoint $c$. In such a case, the checkpoint is regarded as \textit{justified}. If in the next epoch, ($+2/3$) of the validators vote on the next checkpoint $c'$ (a block which is a child of the block belonging to $C$), $c'$ is considered justified whereas $c$ is considered finalised. A checkpoint created in this manner for each epoch is assumed to create a checkpoint tree where $c'$ is a direct child of $c$. The process can be summarised in the following way:
	$+2/3$ Vote $c$ $\rightarrow$ Justify $c$ $\rightarrow$ $+2/3$ Vote $c'$ $\rightarrow$ Finalize $c$ and Justify $c'$

Once a checkpoint is finalised, the validators are paid. The payment is interest-based and is proportional to the number of ethers deposited. If it occurs that there are two checkpoints, it signifies that a fork has occurred. This can only happen when a validator or a set of validators has deviated from the protocol. In such cases, a penalty mechanism is imposed in which the deposit of the violating validator(s) is destroyed.

In essence, CGGF is a combination of Chained and BFT consensus mechanisms with strong support for availability over consistency. Its properties ensure that block finalisation occurs quickly, and the protocol is mostly secure against all PoS attacks except the cartel formation attack (a particular type for PoS only presented in Section \ref{subsubsec:posLimitations}). However, it is to be noted that this consensus mechanism has not been implemented yet. Therefore, it is yet to be seen how it performs in reality.

\subsubsection{BFT PoS}
\label{sec:incentivised:subsec:pos:subsubsec:bft}
In this section we describe three notable BFT PoS algorithms that have had significant uptake in practice: Tendermint, CTFG and Ouroboros. 

\vspace{2mm}
\noindent\textbf{\textsc{1) Tendermint.}} Tendermint is the first to showcase how the BFT consensus can be achieved within the PoS setting of blockchain systems \cite{tenderIntro2017,tendermintPaper2014,tenderWiki2018}. It consists of two major components: a consensus engine known as Tendermint Core and its underlying application interface, called the \textit{Application BlockChain Interface} (ABCI). The Tendermint core is responsible for deploying the consensus algorithm,  whereas  the ABCI can be utilised to deploy any blockchain application using any programming language.

The consensus algorithm relies on a set of validators. It is a round-based algorithm where a proposer is chosen from a set of validators. In each round , the proposer proposes a new block for the blockchain at the latest height. The proposer itself is selected using a deterministic round-robin algorithm, which ultimately relies on the voting power of the validators. The voting power, on the other hand, is proportional to the security deposit of the validators.

\begin{figure*}[t]
      \centering
      \includegraphics[scale=.35]{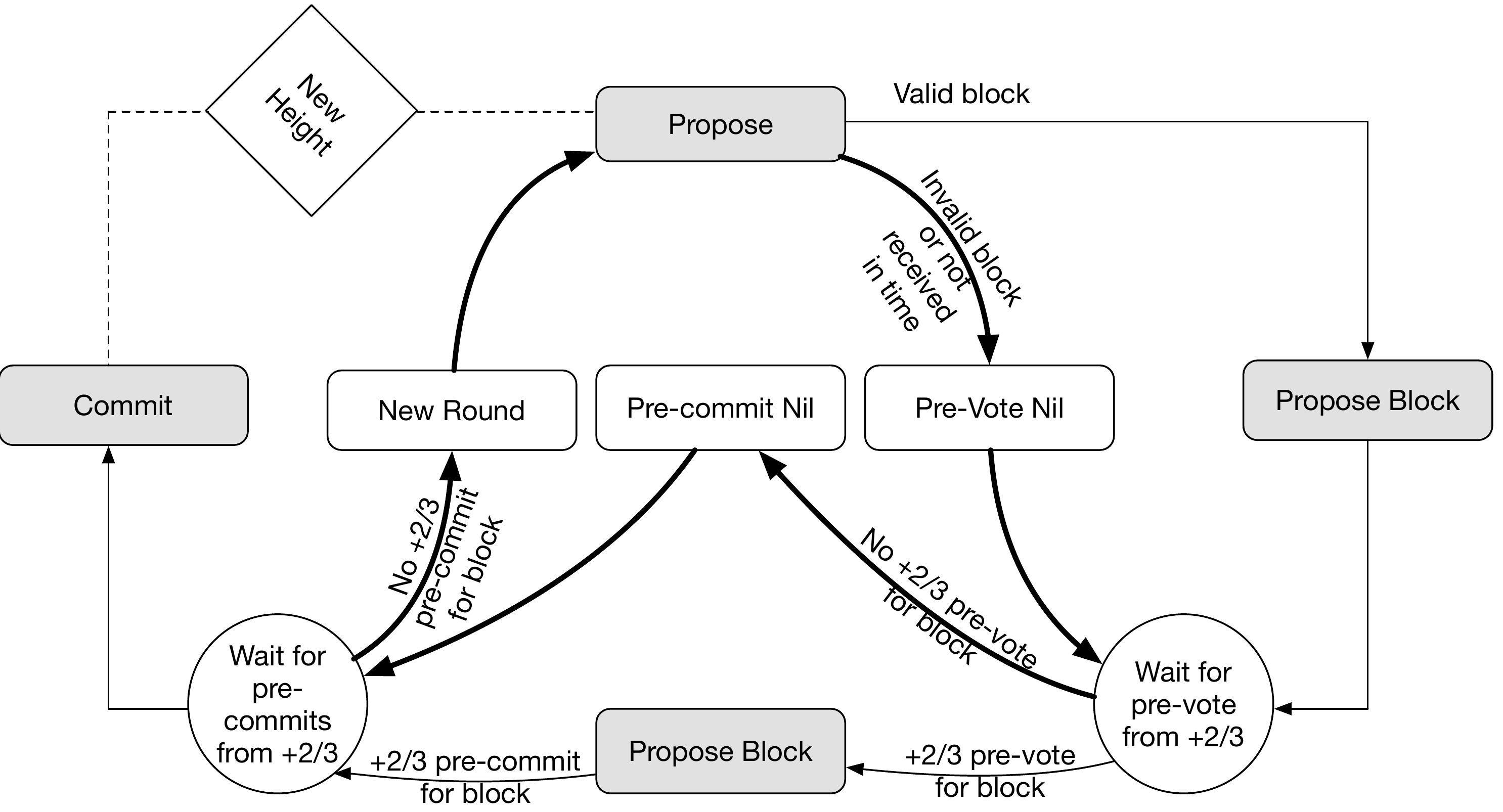}      
      \caption{Tendermint consensus steps.}
%
%
      \label{fig:tendermint}
     \end{figure*}

The consensus algorithm consists of three steps (propose, pre-vote, and pre-commit) in each round bound by a timer equally divided among the three steps, thus making it a weakly synchronous protocol. These steps signify the transition of states in each validator. Figure \ref{fig:tendermint} illustrates the state transition diagram for each validator. At the beginning of each round, a new proposer is chosen to propose a new block. The proposed block needs to go through a two-stage voting mechanism before it is committed to the blockchain.

When a validator receives the proposed block, it validates the block at first, and if okay, it pre-votes for the proposed block. If the block is not received within the \textit{propose} timer or the block is invalid, the validator submits a special vote called \textit{Prevote nil}. Then, the validator waits for the \textit{pre-vote} interval to receive pre-votes from the super-majority (denoted as $+2/3$) of the validators. A $+2/3$ pre-votes signifies that the super-majority validators have voted for the proposed block, implying their confidence on the proposed block and is denoted as a \textit{Polka} in Tendermint terminology. At this stage, the validator pre-commits the block. If the validator does not receive enough pre-votes for the proposed block, it submits another special vote called \textit{Precommit nil}. Then, the validator waits for the \textit{pre-commit} time-period to receive $+2/3$ pre-commits from the super-majority of the validators. Once received, it commits the block to the blockchain. If $+2/3$ pre-commits not received within the \textit{pre-commit} time-period, the next round is initiated where a new proposer is selected, and the steps are repeated.

To ensure the safety guarantee of the algorithm, Tendermint is also coupled with locking rules. Once a validator pre-commits a block after a polka is achieved, it must lock itself onto that block. Then, it must obey the following two rules:

\begin{itemize}
	\item it must pre-vote for the same block in the next round for the same blockchain height,
    \item the unlocking is possible only when a newer block receives a polka in a later round for the same blockchain height.
\end{itemize}

With these rules, Tendermint guarantees that the consensus is secure when less than one-third validators exhibit byzantine behaviour, meaning conflicting blocks will never be committed at the same blockchain height. In other words, Tendermint guarantees that no fork will occur under this assumption. Since Tendermint favours safety over availability, it has one particular weakness. It requires $100\%$ uptime of its $+2/3$ (super-majority) validators. If more than one-third ($+1/3$) are validators are offline or partitioned, the system will stop functioning \cite{tenderIntro2017}. In such cases, out-of-protocol steps are required to tackle this situation.

Unlike PoW or other PoS algorithms that come with defined reward mechanisms and crypto-currency applications, the latest version of Tendermint more likely acts as the consensus plugin, which can be retro-fit to other blockchain systems. For example, Tendermint has been integrated with a private instantiation of Ethereum in a Hyperledger project called Burrows \cite{hyperledgerBurrow2018}. That is why there is no reward/punishment mechanism defined in Tendermint. However, it can be easily introduced in the application layer via the ABCI. For example, a reward mechanism can be introduced for the proposer and the validator to motivate them to engage in the consensus process. A node can become a validator by bonding a certain amount of security deposit. The deposit is destroyed, in case the corresponding validator misbehaves, and thus acts as a deterrent for the validator to launch any attack in the network. Together with the consensus algorithm and a carefully designed reward and punishment mechanism, all PoS attacks can be effectively handled. 

\vspace{2mm}
\noindent\textbf{\textsc{2) Casper the Friendly Ghost (CTFG).}} CTFG is a pure BFT PoS algorithm that aims to transform Ethereum to a PoS-only blockchain system in the future \cite{casperCTFGPaper2018}. As described above, CFFG is geared towards a gentle transition from a PoW to a PoS model for Ethereum, where CTFG will take control of the consensus mechanism ultimately. 

CTFG is based upon a rigorous formal model called Correction by Construction (CBC) that utilises the GHOST (Greedy Heaviest-Observed Subtree) primitive as its consensus rule during fork \cite{ghost2015}. The idea is that the CTFG protocol will be partially specified at the initial stage along with a set of desired properties. Then, the rest of the protocol is dynamically derived in such a way that it satisfies the desired properties - hence the name correction by construction. This is in contrast to the traditional approach for designing a protocol where a protocol is fully defined at first, and then it is tested to check if it satisfies the desired properties \cite{jonchoi2012}.

To achieve this, CTFG introduces a safety oracle, acting as an ideal adversary, which raises exceptions when a fault occurs and also approximates the probability of any future failure. Based on this, the oracle can dynamically fine-tune the protocol  as required to evolve it towards its completion.

Similar to CFFG, CTFG also requires a set of bonded validators that will bond ethers as a security deposit in a smart-contract. However, unlike any other PoS mechanisms, the validators will bet on the block, which has the highest probability to be included in the main chain according to their own perspective. If that particular block is included in the main chain, the validators receive rewards for voting in favour of the block. Otherwise, the validators receive certain penalties. 

Like any PoW algorithm, CTFG favours availability over consistency. This means that blocks are not finalised instantly, like Tendermint. Instead, as the chain grows and more blocks are added, a previous block is considered implicitly final. A major advantage of CTFG over Tendermint is that it can accommodate dynamic validators. This is because the finality condition in Tendermint requires that its block interval is short,  which in turn demands a relatively lower number of pre-determined validators. Since CTFG does not rely on any instant finality, it can theoretically accommodate a higher number of dynamic validators.

CTFG is currently is the most comprehensive proposal which addresses all PoS attack vectors. However, it is to be noted that this is just a proposal at the current stage. Therefore, its performance in real settings is yet to be analysed.


\vspace{2mm}
\noindent\textbf{\textsc{3) Ouroboros.}} Ouroboros is a provably secure PoS algorithm \cite{ouroboros2017, ouroborosIntro2018} utilised in the Cardano platform \cite{cardano2017}. Cardano is regarded as third-generation blockchain system supporting smart-contract and decentralised application without relying on any PoW consensus algorithm. 

In Ouroboros, only a stakeholder can participate in the block minting process. A stakeholder is any node that  holds the underlying crypto-currency of the Cardano platform called \textbf{Ada}. Ouroboros is based on the concept of \textit{epoch}, which is essentially a predefined time period. Each epoch consists of several slots. A stakeholder is elected for each slot to create a single block, meaning a block is created in each slot. The selected stakeholder is called a slot leader and is elected by a set of \textit{electors}. An elector is a specific type of stakeholder which has a certain amount Ada in its disposal.

In each epoch, the electors select the set of stakeholders for the next epoch using an algorithm called \textit{Follow the Satoshi} (FTS). The FTS algorithm relies on a random seed to introduce a certain amount of randomness in the election process. A share of the random seed is individually generated by all electors who participate in a multiparty computation protocol. Once the protocol is executed, all electors posses the random seed, constructed with all of their shares. The FTS algorithm utilises the random seed to select a coin for a particular slot. The owner of the coin is then elected as the slot leader. Intuitively, the more coins a stakeholder possesses, the higher is its probability of being selected as the slot leader.

Ouroboros is expected to provide a transaction fee based reward to incentivise stakeholders to participate in the minting process. However, the details are in the process of being finalised. It has been mathematically proven to be secure against almost all PoS attack vectors except the cartel formation \cite{ouroboros2017}. Nevertheless, how it will perform once deployed is yet to be seen. 





\subsubsection{DPoS}

There are several mechanisms deployed by different crypto-currencies under the general category of DPoS. Next, we present a few prominent approaches of some well-known DPoS based crypto-currencies. Our analysis of these crypto-currencies are summarised in Table \ref{tab:dpos}.

\vspace{2mm}
\noindent\textbf{\textsc{1) EOS.}} EOS is the first and the most widely known DPoS crypto-currency and smart-contract platform as of now \cite{eos2019}. 
With the promise of greater scalability and higher transactions per second than Ethereum, it raised 4 billion USD in the highest ever ICO event to date \cite{eosICONews2019}. Initial EOS currency was created on the Ethereum platform, and later migrated to their own blockchain network. The DPoS consensus algorithm of EOS utilises $21$ validators, also known as \textit{Block Producers} (BPs). These $21$ validators are selected with votes from EOS token (currency) holders. The number of times a particular BP is selected to produce a block is proportional to the total votes received from the token holders.

Evey DPoS currency must create an initial supply before the network is operational. This supply is used to select $21$ BPs (with voting) as well as to reward the BPs for creating blocks, and thus, securing the network. EOS had an initial supply of $1$ Billion EOS tokens with an annual inflation of $5\%$. Among the inflated currencies, $1\%$ is used to reward the block producers,  whereas the rest of the $4\%$ are kept for future R\&D for EOS \cite{eosExpAnother2019}. Currently, an EOS block is created in $0.5$s. Blocks in EOS are produced in rounds where each round consists of 21 blocks \cite{eosExp2019}. At the beginning of each round, 21 BPs are selected. Next, each of them gets a chance to create a block in pseduo-random fashion within that particular round. Once a BP produces a block, other BPs must validate the block and reach into a consensus. A block is confirmed only when ($+2/3$) majority of the BPs reach the consensus regarding the validity of the block. Once this happens, the block and the associated transactions are regarded as confirmed or final, so no fork can happen. 

\vspace{2mm}
\noindent\textbf{\textsc{2) Tron.}} Tron is another popular DPoS based crypto-currency \cite{tron2019}. With an initial supply of $99$ Billion Tron tokens (represented with \textbf{TRX}), it is another smart-contract supported blockchain platform, very similar to Ethereum and EOS in functionality. Its consensus mechanism utilises $27$ validators, known as Super Representatives (SRs) \cite{tronWP2019}. The SRs are selected in every six hours with votes by TRX holders who must freeze a certain amount of TRX to vote for an SR. The deposits amount can be frozen back after three days once the voting is cast \cite{tronGuide2019}. A block in Tron is created in every $3$s for the corresponding SR receives a reward of $32$ TRX. Another important feature of Tron is that there is no in-built inflation mechanism in the protocol, which implies that the total supply will remain constant throughout its lifespan.

\vspace{2mm}
\noindent\textbf{\textsc{3) Tezos.}} Tezos is, like EOS and Tron, a smart-contract platform which utilises a variant of DPoS consensus algorithm \cite{tezos2019}. With a block reward of $16$ XTZ (Tezos currency) and block creation time of $60$s, Tezos does not require any pre-defined number of stakeholders (or \textit{Bakers} as defined in Tezos) \cite{tezosWP2019}. This differs Tezos from other DPoS currencies. Instead, the consensus mechanism utilises a dynamic range of stakeholders where anyone holding a substantial amount of XTZ can be a stakeholder. This limits general users to participate in the consensus mechanism. To rectify this problem, Tezos provides a mechanism by which anyone can delegate their XTZ to someone so that it can accumulate the required number of XTZ to be a baker. In return, the baker would return a certain proportion of their received block reward to the delegating party. Tezos started with an initial supply of $765$ Million XTZ tokens. It relies on an annual inflation of $5.51\%$ and the inflated currencies are used to reward the bakers.

\vspace{2mm}
\noindent\textbf{\textsc{4) Lisk.}} Lisk is a unique DPoS blockchain platform which, enables the development of DApps using JavaScript \cite{lisk2019}. Another unique feature of Lisk is its ability to accommodate and then to operate with multiple blockchains, known as \textit{sidechains} along with a central blockchain called \textit{mainchain}. Each sidechain can be deployed and maintained by a particular application provider, which needs to be synced with the mainchain as per the Lisk's protocol rule. In this way, different applications can leverage different sidechains simultaneously without burdening off the mainchain. Even though the responsibility of maintaining a sidechain relies on the particular application provider, the mainchain must be maintained with the Lisk DPoS consensus protocol, which utilises $101$ delegates \cite{liskWP2019}. Only these delegates can produce a block. These delegates are selected using votes from Lisk currency (denoted with $LSK$) owners, where each holder has $101$ votes. The weight of each vote is proportional to the amount of LSK owned by the respective owner. The selection of delegates happens before a round, where each round consists of $101$ block generation cycle. Thus, in a round, each delegate is randomly selected to create a block. It has a block creation time of $10$ seconds and block reward of $5$ LSK. Started with an initial supply of $100$ million LSK, Lisk has a current supply of $132$ million with an annual inflation of $5.65\%$.

\vspace{2mm}
\noindent\textbf{\textsc{5) Ark.}} Ark is yet another DPoS based blockchain platform \cite{ark2019}. It utilises $51$ delegates to create $51$ blocks in each round \cite{arkWP2019}. With a block creation time of $8$s, each round lasts for $408$s. Each delegate receives $2$ ARK (the native currency of the ARK platform) for creating a block. It had an initial supply of $125$ million. With an annual inflation of $5.55$, the  supply was around $142$ million (as of June 2019). Like other DPoS blockchains, the delegates in Ark are also selected with votes by Ark currency owner, where the weight of each is proportional to the amount of ARK owned by the voter.
\begin{table*}[!h]
\centering
\caption{Comparison of DPoS Currencies with `-' signifying not applicable.}
  \begin{tabular}{l|p{20mm}|p{20mm}|p{15mm}|p{25mm}|p{25mm}|l|p{15mm}}
    \hline
      \rowcolor[gray]{.6} 
    \centering\textbf{Currency} & \centering\textbf{Genesis date (dd.mm.yyyy)} & \centering\textbf{Initial supply} & \centering\textbf{Inflation} & \centering\textbf{Current supply (23.05.2019)} &  \centering\textbf{Block reward} & \centering\textbf{Block Time} & \centering\textbf{Validator nos} \tabularnewline [2ex]\hline
	   \hline
     EOS & $01.07.2017$ & $1$ Billion & $5\%$ & $1.04$ Billion & $1\%$ of inflated currency divided among 21    validators & $0.5$s & $21$\\\hline
    \rowcolor[gray]{.9} Tron & $28.08.2017$ & $99$ Billion & - & $99$ Billion & $32$ TRX & $3$s & $27$\\\hline
    Tezos & $30.06.2018$ & $765$ Million & $5.51\%$ & $795$ Million & $16$ XTZ  & $60$s & Not pre-defined\\\hline
    \rowcolor[gray]{.9}  Lisk & $24.05.2016$ & $100$ Million & $5.67\%$ & $132$ Million & $5$ LSK & $10$s & $101$ \\\hline
     Ark & $21.03.2017$ & $125$ Million & $5.55\%$ & $142$ Million & $2$ ARK & $8$s & $51$\\\hline
\end{tabular}
  \label{tab:dpos}
\end{table*}
\subsubsection{Limitations of PoS}
\label{subsubsec:posLimitations}
Even though the variants of different PoS algorithms offer several significant advantages, there are still a few disadvantages in these classes of algorithms. We explore these disadvantages below.

\begin{itemize}
    \item \textbf{Collusion:} If the number of validators is not large enough, it might be easier to launch a $51\%$ attack on the corresponding consensus algorithm by colluding with other validators.
    \item \textbf{Wealth effect:} The sole reliance on coin-wealth in a consensus algorithm or for the selection of validators creates an environment where people with a large portion of coins can exert greater influence.
\end{itemize}

In addition to these disadvantages, there have been a few other attack vectors identified for the PoS algorithms:
\begin{itemize}
	\item \textbf{Nothing-at-stake (NAS) attack \cite{bitfury2015}:} During a blockchain fork, an attacker might attempt to add its newly created block in all forked branches to increase their probability to add their block as the valid block. Such scenario is unlikely to occur in any PoW algorithm. This is because a miner would need to share their resources in order to mine at different branches. This would eventually decrease their chance of finding a new block because of the resources shared in multiple branches. Since it does not cost anything for a minter in a PoS algorithm to add blocks in multiple parallel branches, the attacker is motivated to do so. Applying a penalty for such misbehaviour could effectively tackle this problem. 
    \item \textbf{Bribing (short-range, SR) attack \cite{bitfury2015,bentov2016cryptocurrencies}:} In this attack, an attacker tries to double spend by creating a fork. An example of this attack would be as follows. The attacker pays to a seller to buy a good. The seller waits for a certain number of blocks (e.g., six blocks) before the good is delivered to the attacker. Once delivered, the attacker forks the main chain at the block (e.g., six blocks back, which is relatively short and hence the name) in which the payment was made. Then, the attacker bribes other minters to mint on top of the forked branch. As long as the bribed amount is lower than the price of the delivered good, it is always profitable for the attacker. The colluding minter has nothing to lose if it is coupled with the nothing-at-stake attack on their part but can gain from the bribery. Again, it can be tackled by introducing a penalty mechanism for all misbehaving parties.
    \item \textbf{Long-range (LR) attack \cite{bitfury2015}:} In this attack, the attacker attempts to build an alternative blockchain starting from the earliest blocks if the attacker can collude with the majority of the stakeholders. The motivation might be similar to double spending or related issues providing advantages to the attacker as well as the colluded stakeholders. As explained above, the colluded stakeholder has nothing to lose if it can be coupled with the nothing-at-stake attacks. Check-pointing is one of the methods by which it can be tackled. The check-pointing codifies a certain length of the blockchain to make it immutable up to that point, and thereby undermining the attack. This is because the attacker cannot fork the blockchain before that check-point.
    \item \textbf{Coin-age accumulation (CAC) attack \cite{bitfury2015,bentov2016cryptocurrencies}:}  The PoS algorithms that rely on the uncapped coin-age parameter are susceptible to this attack. In this attack, the attacker waits for their coins to accumulate enough coin-age to exploit the algorithm for launching double spends by initiating a fork. This attack can be tackled by introducing a cap on the coin-age  which minimises the attack vector. 
    
    \item \textbf{Pre-computing (PreCom) attack \cite{bitfury2015,posWiki2018}:} A pre-computing attack, also known as Stake-grinding attack, would allow an attacker to increase the probability of generating subsequent blocks based on the information of the current block. If there is not enough randomness included in the PoS algorithm, the attacker can attempt to pre-compute subsequent blocks by fine-tuning information of the current block. For a particular set of information (e.g., a set of transactions), if the attacker finds that the probability of minting a few subsequent blocks is less than desired, the attacker can update the set of transactions to increase their probability of determining the next few blocks. It can be effectively tackled by introducing a secure source of randomness in the algorithm.
    \item \textbf{Cartel formation (CAF) attack \cite{caspervten2017}:} In economic theory, an oligopoly market is dominated by a small set of entities having greater influence or wealth than other entity. They can collude with one another by forming a cartel to control price or reduce competition within the market. It has been argued that "\textit{Blockchain architecture is mechanism design for oligopolistic markets.}" \cite{casperHistory2017} which affects both PoW and PoS algorithms. Such a cartel can launch $51\%$ attacks on the PoS algorithm or exploit the stakes to monopolise the PoS algorithm. 
\end{itemize}

\subsubsection{Analysis}

In  this  section,  we  summarise  the  properties  of  different PoS algorithms utilising the taxonomies and PoS attack vectors in Table \ref{tab:posStruct}, Table \ref{tab:posSec}, Table \ref{tab:poSSecOther}  and Table \ref{tab:poSPerform}. Like before, a `\checkmark' symbol has been utilised to indicate if  the corresponding algorithm supports a particular property, and the `X' symbol signifies that the particular property is not supported. The `-' symbol implies that the property is not applicable, whereas the symbol `?' indicates that no information has been found for that particular feature. For other properties, explanatory texts have been used as well. 

From Table \ref{tab:posStruct}, only chained algorithms are based on multiple committee utilising a flat topology with a dynamic configuration. These algorithms also use a probabilistic lottery to select a minter. Conversely, other PoS algorithms, except Tendermint, are based on the single committee having an open type and explicit formation with a dynamic configuration and mostly rely on voting mechanisms. Tendermint uses a closed committee with a static configuration. 

As per Table \ref{tab:posSec}, none of the algorithms, except Tendermint requires any node to be authenticated to participate in the algorithm. All of them have strong support for non-repudiation in the form of digital signature as part of every single transaction. These algorithms have a high level of censorship resistance, as do all PoW algorithms. As for the attack vector, each PoS algorithm requires every miner node to invest substantially to participate in this algorithm. This feature, thus, acts as a deterrent against any Sybil or DoS attack in any PoS algorithm. The adversary tolerance for Chained systems can be calculated using this formula: $min(2f+1, 3f+1) = 3f +1$. This is because a chained algorithm utilises both PoW and PoS algorithms and thus needs to consider the adversary tolerance for both of them. We consider the minimum of these two ($3f+1$). The supported adversary tolerance for other algorithms is $3f+1$ except BFT Ouroboros whose adversary tolerance is $2f+1$.

According to Table \ref{tab:poSPerform}, all BFT, and DPoS algorithms have considerably high throughput, low latency, and high scalability. Their energy consumption is negligible. However, the chained algorithms have a comparatively lower throughput, lower scalability, and higher latency with respect to their BFT and DPoS counterparts. The fault tolerance of chained and BFT algorithms is $2f+1$ like any BFT algorithm, implying they can achieve consensus as long as more than $50\%$ of nodes function properly. However, DPoS algorithm requires a $3f+1$ fault tolerance. 

Table \ref{tab:poSSecOther} outlines a comparison of additional attack vectors with symbols representing the usual semantics. CTFG, Tentermint, and Ouroboros have mitigation mechanisms against these attack vectors. However, Casper FFG, and any DPoS algorithms cannot successfully defend against the cartel formation attack. Peercoin, on the other hand, has mechanism against this cartel formation attack, unfortunately, suffers from all other attack vectors.

Finally, a comparison of the selected DPoS crypto-currencies is presented in Table \ref{tab:dpos}.

\begin{table*}[h]
\centering
\caption{Comparing structural properties of PoS Consensus Algorithms.}
\label{tab:posStruct}
\begin{tabular}{P{20mm}|P{20mm}|c|c|c|c|c|P{25mm}}
\hline
\rowcolor[gray]{.75} 
 & 

&
\multicolumn{3}{c|}{\textbf{Single committee}} & 
\multicolumn{2}{c|}{\textbf{Multiple committee}} & 

\\ \cline{3-7} 
\multirow{-2}{*}{\cellcolor[gray]{.75}\textbf{\begin{tabular}[c]{@{}c@{}}Consensus\\ /System\end{tabular}}}&
\multirow{-2}{*}{\cellcolor[gray]{.75}\textbf{Node type}}  & 
\cellcolor[gray]{.75}\textbf{ Type} &
\cellcolor[gray]{.75}\textbf{ Formation} &
\cellcolor[gray]{.75}\textbf{ Configuration} &
\cellcolor[gray]{.75} \textbf{Topology} & 
\cellcolor[gray]{.75}\textbf{ Configuration} & 
{\multirow{-2}{*}{\cellcolor[gray]{.75}\textbf{Mechanism}}}
 \\ \hline
 \hline
Chained (PeerCoin) & Clients, Miners \& Minters & - & - & - & Flat & Dynamic & Probabilistic lottery \\ \hline
\rowcolor[gray]{.9} Chained (CFFG) & Clients, Miners \& Validators & - & - & - & Flat & Dynamic & Probabilistic lottery  \\\hline
BFT (Tendermint) & Clients \& Validators & Open (Close) & Explicit & Dynamic (Static) & - & - & Voting \\ \hline
\rowcolor[gray]{.9} BFT (CTFG) & Clients \& Validators & Open & Explicit & Dynamic & - & - & ? \\ \hline
BTFG (Ouroboros) & Clients, Electors \& Stakeholders & Open & Explicit & Dynamic & & & Voting \\ \hline
\rowcolor[gray]{.9} DPoS & Clients \& Validators & Open & Explicit & Dynamic & - & - & Voting \\ \hline
\end{tabular}
\end{table*}

\newcommand{\specialcell}[2][c]{%
  \begin{tabular}[#1]{@{}c@{}}#2\end{tabular}}
  
\begin{table*}[h]
\centering
\caption{Comparing security properties of PoS Consensus Algorithms.}
\label{tab:posSec}
\begin{tabular}{P{20mm}|P{20mm}|c|P{15mm}|c|c|c}
\hline
\rowcolor[gray]{.75} 
 & 
&
 & 
 & 
 \multicolumn{3}{c|}{\textbf{Attack Vectors}} \\ \cline{5-7}

{\multirow{-2}{*} {\cellcolor[gray]{.75}\textbf{\begin{tabular}[c]{@{}l@{}}Consensus \\ /System\end{tabular}}}} & 
{\multirow{-2}{*}{\cellcolor[gray]{.75}\textbf{Authentication}}}  & 
{\multirow{-2}{*}{\cellcolor[gray]{.75}\textbf{Non-repudiation}}} & 
{\multirow{-2}{*}{\cellcolor[gray]{.75}\textbf{\specialcell{Censorship\\resistance}}}}& 
\cellcolor[gray]{.75}\textbf{Adversary tolerance} &
\cellcolor[gray]{.75}\textbf{Sybil protection} & 
\cellcolor[gray]{.75}\textbf{DoS Resistance}

\\ \hline
\hline

Chained (PeerCoin) & X & \checkmark & High & $3f + 1$ & \checkmark & \checkmark \\ \hline
\rowcolor[gray]{.9} Chained (CFFG) & X & \checkmark & High & $3f + 1$ & \checkmark & \checkmark \\ \hline
BFT (Tendermint) & \checkmark (In close type), X (In open type) & \checkmark & High & $3f + 1$ & \checkmark & \checkmark \\ \hline
\rowcolor[gray]{.9} BFT (CTFG) & X & \checkmark & High & $3f + 1$ & \checkmark & \checkmark \\ \hline
BFT (Ouroboros) & X  & \checkmark & High & $2f + 1$ & \checkmark & \checkmark \\ \hline
\rowcolor[gray]{.9} DPoS  & X & \checkmark & High & $3f + 1$ & \checkmark & \checkmark \\ \hline
\end{tabular}
\end{table*}

\begin{table*}[!h]
\centering
  \caption{Comparison of additional attack vectors protection among PoS Consensus Algorithms}

   \begin{tabular}{P{28mm}|P{25mm}|P{15mm}|P{15mm}|P{15mm}|P{20mm}|P{25mm}}
    \hline
       \rowcolor[gray]{.6} 
    \centering\textbf{Consensus\textbackslash System} & 
    \centering\textbf{Nothing-at-Stake}& 
    \centering\textbf{Bribing}& 
    \centering\textbf{Long-range}& 
    \centering\textbf{Coin-age }&
    \centering\textbf{Pre-computing}&
    \centering\textbf{Cartel formation} \tabularnewline [2ex]
          \hline
\hline
	    Chained (PeerCoin) & X & X & X & X & X & \checkmark \\\hline
	   \rowcolor[gray]{.9}  Chained (Casper FFG) & \checkmark & \checkmark & \checkmark & \checkmark & \checkmark & X \\\hline
        BFT (Tendermint) & \checkmark & \checkmark & \checkmark & \checkmark & \checkmark & \checkmark \\\hline
        \rowcolor[gray]{.9} BFT (CTFG) & \checkmark & \checkmark & \checkmark & \checkmark & \checkmark & \checkmark \\\hline
        BFT (Ouroboros) & \checkmark & \checkmark & \checkmark & \checkmark & \checkmark & \checkmark \\\hline
      \rowcolor[gray]{.9}   DPoS & \checkmark & \checkmark & \checkmark & \checkmark & \checkmark & X \\\hline
\end{tabular}
\label{tab:poSSecOther}
\end{table*}

\begin{table*}[!h]
\centering
  \caption{Comparing performance properties of PoS Consensus Algorithms.}

   \begin{tabular}{P{25mm}|P{15mm}|P{20mm}|P{15mm}|P{15mm}|P{30mm}}
    \hline
       \rowcolor[gray]{.6} 
      \centering\textbf{Consensus\textbackslash System} & 
     \centering\textbf{Fault tolerance}& 
     \centering\textbf{Throughput}& 
     \centering\textbf{Scalability}& 
     \centering\textbf{Latency}&
    \centering\textbf{Energy consumption} \tabularnewline [2ex]
          \hline
 \hline
	    Chained (PeerCoin, CFFG) & $2f + 1$ & Medium & Medium & Medium & Medium \\\hline
       \rowcolor[gray]{.9}  BFT (Tendermint, CTFG, Ouroboros) & $2f + 1$ & High & High & Low & Low \\\hline
        DPoS & $3f + 1$ & High & High & Low & Low \\\hline
\end{tabular}
\label{tab:poSPerform}
\end{table*}

\section{Incentivised Consensus: Beyond PoW and PoS}
\label{sec:hybrid}
Some consensus algorithms take a different approach in which they do not solely rely on any PoW or PoS mechanism. Instead, they use an approach in which a PoW/PoS mechanism is combined with another approach. We consider such algorithms as hybrid algorithms which are presented in Section \ref{sec:hybrid:subsec:hybrid}. Other approaches adopt a more drastic approach in which they do not leverage any type PoW/PoS algorithm whatsoever. Such algorithms are tagged as \textit{N-POS/POW} (to symbolise Non-PoS/PoW) algorithms and discussed in Section \ref{sec:hybrid:subsec:npos}.

\subsection{Hybrid Consensus}
\label{sec:hybrid:subsec:hybrid}
In this section, we outline a new breed of consensuses algorithms that combine either a PoW or PoS algorithm or both with another novel algorithm or mechanism, thus creating a hybrid mechanism. 

\vspace{2mm}
\noindent\textbf{\textsc{1) Proof of Research (PoR).}} Proof of research is a hybrid approach that combines proof-of-stake with the proof-of-BOINC \cite{Gridcoin2019}. BOINC stands for Berkeley Open Infrastructure for Network Computing \cite{boinc2019}. It is a grid computing platform widely used by scientific researchers in different domains by allowing them to exploit the idle computing resources of personal computers around the world. With the proof-of-BOINC, a researcher has to prove his contribution for the BOINC research work. 

The PoR mechanism is leveraged by Gridcoin \cite{Gridcoin2019, gridcoinWP2019}, a crypto-currency that can be earned by anyone by sharing their computing resources with the BOINC project. The mechanism by which PoS and Proof-of-BOINC are tied together for the PoR is explained next \cite{gridcoinWP2019}. The PoS mechanism is similar to the traditional PoS algorithm. Anyone can become a minter, known as \textit{Investor} in Gridcoin terminology, by owning a certain amount of Gridcoin and participating in the minting process. In addition to this, other users, known as \textit{Researchers} in Gridcoin terminology, can also participate in the minting process. Interestingly, an investor can also be a researcher and thus, can increase their amount of grid coin earned.

For this, a researcher installs the BOINC software and registers a project from the \textbf{BOINC} whitelist with his email address. The researcher is assigned a unique cross project identifier (CPID) and starts downloading the work share. Once the computation is completed, the researcher returns the result with a credit recommendation for the completed workload.  The recommendation is compared with that of another researcher, and the minimum credit is rewarded. This workload credit data is stored in the header of each block and the researcher is rewarded with the corresponding amount of Gridcoin. To summarise, the consensus mechanism is mostly dominated by the PoS mechanism with Proof-of-BOINC acts as a reward mechanism for sharing unused computing resources available to the researchers. Hence, its security is similar to that of the traditional PoS algorithm.

%
%

%
%
%

\vspace{2mm}
\noindent\textbf{\textsc{2) Slimcoin's Proof-of-Burn (PoB).}} The Proof-of-Burn is a consensus algorithm proposed by Ian Stewart as an alternative to PoW \cite{pob2019}. In PoW, miners need to invest in building a mining rig in order to participate in the mining process. In PoB, miners need to burn their coins in order to participate in the mining process. Burning coins mean that sending coins to an address without the private key and thus never usable. Thus, burning coins is an analogous idea to the investment for building a mining rig. The amount of burning has a positive correlation with the possibility of being selected for mining the next block. This is similar to the PoW system, where the miners increasingly invest in modern equipment to maintain the hash power, as the incentive decays with the complexity. 

Slimcoin is a crypto-currency which utilises the idea of PoB in combination with PoW and PoS \cite{slimcoin2019, slimcoinWP2019}, thus creating a hybrid consensus mechanism. Algorithmically, their idea is similar to the chained PoS algorithm of Peercoin as presented in in Section \ref{sec:incentivised:subsec:pos:subsubsec:chainedPoS} with additional PoB mechanism sandwiched in between PoW and PoS algorithms. The PoW is used to generate the initial coin supply using the mechanism of Bitcoin. When the system has sufficient amount of money supply, it plans to switch to a hybrid of PoW and PoS mechanism similar to Peercoin where PoB will be used to select the miner. As this happens, the minters will need to burn their accumulated coins in order to be eligible to participate in the PoS minting process. Since PoB algorithm is mostly used for minter selections, it has hardly any effect on the security of the system. Hence, its security and other properties are mostly similar to that of Peercoin.

\vspace{2mm}
\noindent\textbf{\textsc{3) Proof of Stake-Velocity (PoSV).}} One of the  major limitations of coin-age based PoS is that there is no incentive (or lack of penalty thereof) for the minters to be online to participate in the staking process. This is because that the coin-age increases linearly over time, without the need for the stakeholders to be online and participate in the staking process. They can, therefore, choose to participate for a short period and then collect the reward and may go offline again. The lack of participants may facilitate attacks at a certain time.

To counteract this problem, a crypto-currency called Reddcoin proposed a novel hybrid algorithm called Proof of Stake-Velocity (PoSV) \cite{reddcoin2019, reddcoinWiki2019}. The central to the PoSV is the idea of a mechanism called the \textit{velocity of stakes} coupled with any traditional PoS algorithm. Conceptually, the velocity of stake mirrors the notion of the velocity of money, a terminology from Economics implying the frequency of money flow within the society \cite{moneyVelocity2019}. Indeed, the velocity of stakes evolves around the idea of increasing the flow of stakes during the PoS consensus mechanism \cite{reddcoinWP2019}. This (the flow of stakes) can be achieved if the minters are encouraged to actively participate in the consensus mechanism by staking their crypto-currency, instead of holding their coins offline.  This process in a way will also increase the overall security of the system and counteract the lack of participant issue in PoS.

To facilitate this PoSV introduces a non-linear coin-ageing function in which the coin-age of a particular coin is gained much faster in the first few days and weeks than the gain in later weeks. For example, it has been estimated that minters who stake their coins every two weeks or less, can earn up to $20\%$ more than people who do not participate in the staking process \cite{reddcoinWP2019}. Such incentives encourage the minters to increase the velocity of stakes in the whole network. Note that PoSV is similar to any PoS mechanism along with its properties and hence, not explored in detail here.

\subsection{N-POS/POW}
\label{sec:hybrid:subsec:npos}
The consensuses algorithms presented in this category do not rely any way on either PoW or PoS algorithms. Instead, they rely on completely novel mechanisms. Therefore, we call them N-POS/PoW algorithms for the convenience of group naming.

\noindent\textbf{\textsc{1) Proof-of-Cooperation (PoC).}} The Proof-of-Cooperation is a consensus algorithm introduced by the FairCoin crypto-currency \cite{faircoin2019, faircoinWP2019}. This consensus algorithm relies on several special nodes known as Certified Validating Nodes (CVNs). CVNs function similar to the way validators act in a DPoS consensus algorithm as utilised by EOS or Tron crypto-currencies, as they are nodes which can create blocks in Faircoin using the PoC consensus algorithm. However, unlike any DPoS validators, each CVN node is authenticated by their corresponding Faircoin identifier as well as trusted following a set of community-based rules and technical requirements \cite{faircoinWP2019}. The community rules state that a candidate node willing to be a CVN must participate in Faircoin community activities by performing some tasks. Examples of these tasks are running a local node or contributing to any technical or management issue related to Faircoin which must be confirmed by at least two active members of the community. Besides, the candidate node must follow a set of technical requirements such as 24/7 network availability and a special cryptographic hardware used for signature generation. 

With the involvement of CVNs selected in the previously discussed manner, the core mechanism for PoC consensus algorithm is briefly discussed next. Blocks in Faircoin are created in a round-robin fashion in every
three minutes of epoch by one of the CVNs. To create a new block, a CVN needs to be selected using a deterministic voting mechanism individually carried out by every single CVN in the network.  The steps of this mechanism are:
\begin{itemize}
    \item Each CVN finds the CVN,  which has created a block furthest in the chain by traversing backwards through the chain. 
    \item Next, it is checked if the found CVN has been active recently in the network by looking for its signature in the last few blocks. If so, this CVN will be selected as the next CVN.
    \item Then, each node creates a data set consisting of the hash of the last block, the ID of the selected CVN for the next block, and its own CVN ID, which is then signed by the specified cryptographic hardware. The created dataset, along with the signature, is then propagated through the network.
    \item The selected CVN receives this dataset along with their signature from multiple CVNs and verifies each signature. As soon as the selected CVN finds that more than $50\%$ CVNs have selected it to be the next block creator, it can be certain that its turn is next at the end of the current epoch, i.e., three minutes.
    \item The selected CVN adds all pending transactions into a new block, along with all the received signatures, and propagates the block in the network.
    \item Upon receiving the block, other CVNs verify the block by checking the if the CVN who created the block is actually the one selected as the block creator as well as validating all signatures in it and its transactions.  If the verification is successful, the block is added to the blockchain and the same mechanism continues.
\end{itemize}

\vspace{2mm}
\noindent\textbf{\textsc{2) Proof of Importance (PoI).}} 
PoS gives an unfair advantage to coin hoarders. The more coins they keep in their accounts, the more they earn. This means the rich get richer and everyone has an incentive to save coins instead of spending them. To solve these issues NEM has introduced a novel consensus mechanism called ``Proof of Importance (PoI)" \cite{NEMWP2019}. It functions similarly to PoS: nodes need to 'vest' an amount of currency to be eligible for creating blocks and are selected for creating a block roughly in proportion to some score. In Proof-of-stake, this 'score' is one's total vested amount, but in PoI, this score includes more variables. All the nodes that have more than $10000$ XEM (the corresponding crypto-currency of XEM) are theoretically given equal positive importance and  with $9$B XEM coins there can be maximum $900K$ such nodes. However, the actual number of such nodes and their importance vary with time and their amount of transaction in XEM.

The calculations borrow from the math of network clustering and page ranking. At a high level, the primary inputs are:

\begin{itemize}
    \item Net transfers: how much has been ‘spent’ in the past 30 days, with more recent transactions weighted more heavily.
    \item Vested amount of currency for purposes of creating blocks.
    \item Cluster nodes: accounts that are part of interlinked clusters of activity are weighted slightly more heavily than outliers or hubs (which link clusters but not part of them).
\end{itemize}{}
  
In NEM, the importance of an account depends only on the net transfers of XEMs from  that account. To be considered for the importance estimation at a certain block height, $h$ , a node must have transferred at least $100$ XEMs during the last $30$ days or $43,200$ blocks. The ``importance score" addresses two primary criticisms of proof-of-stake.

One risk is that people hoard many coins as possible and reap the rewards from block creation. This concentrates wealth while discouraging transactions. The importance score means that hoarding will result in a lower score while spreading XEM around will increase it. Being a merchant pays better than having a hoard. 

\subsection{Analysis}

In this section, we summarise the properties of different Hybrid and N-Pow/PoS algorithms utilising the taxonomies in Table \ref{tab:nopospowStruc}, Table \ref{tab:nopospowSec}, Table \ref{tab:nopoSSecOther} and Table \ref{tab:nopospowPerform}. Like before, `-' signifies that the corresponding property is not applicable for the respective consensus algorithm, `?' indicates that the information the property has not been found, a `\checkmark'\, is used to indicate an algorithm satisfies a particular property and `X' is used to imply the reverse (not satisfied).

Table \ref{tab:nopospowStruc} presents the comparison of structural properties for the corresponding consensus algorithms. Among them, PoR and PoB depend on a multiple committee formation with a flat topology and dynamic configuration. Conversely, PoSV and PoI use an open single committee with a dynamic configuration, and probabilistic lottery as their underlying mechanism. PoC has an implicit, open,  and dynamic single committee, which relies on voting mechanism. 

All these algorithms have an adversary tolerance of $3f+1$ with the support of non-repudiation, Sybil protection, DoS resistance, and high censorship resistance as reported in Table \ref{tab:nopospowSec}. Entities in PoB, PoSV, and PoI do no require to be authenticated while PoC entities must be authenticated, and researchers in PoR need to be authenticated. However, other entities in PoW can remain non-authenticated, as indicated with the `X' symbol in the table. All of them except PoC and PoI have $3f+1$ adversary tolerance because of their usage of PoS algorithms. We have not found any regarding adversary tolerance for PoC and PoI.

Table \ref{tab:nopoSSecOther} presents the comparison of some additional attack vectors for the Hybrid algorithms. As evident from the table, since these algorithms utilise PoS as one of their consensus algorithms, they suffer from the similar limitations of any PoS algorithm. For example, none of them has any guard against most of these additional attack vectors. The only exception is PoB which is because of its use of Peercoin like functionality, can resist the cartel formation attack.

The comparison of the performance properties for these algorithms is presented in Table \ref{tab:nopospowPerform}. All of them have $2f+1$ fault tolerance except PoC and PoI as we have not found any information fault tolerance for PoC and PoI. In terms of Scalability, Latency and Energy, every algorithm except PoB exhibits similar characteristics: they have high throughput, consume low energy, and have low latency, meaning they reach finality quickly. Because of its reliance on PoW, PoB has low scalability, low latency, and also consume meidum energy. In terms of throughput, PoR, PoSV and PoI have high throughput, whereas PoC has a low throughput and PoB has a medium throughput.

Finally,   a   comparison   of   the   selected   Hybrid and N-PoW/PoS crypto-currencies is presented in Table \ref{tab:dash}. 

\begin{table*}[h]
\centering
\caption{Comparing structural properties of Hybrid and N-POS/POW Consensus Algorithms.}
\label{tab:nopospowStruc}
\begin{tabular}{P{20mm}|P{20mm}|c|c|c|c|c|P{25mm}}
\hline
\rowcolor[gray]{.75}
 & & \multicolumn{3}{c|}{\textbf{Single committee}} & \multicolumn{2}{c|}{\textbf{Multiple committee}} &  \\ \cline{3-7}
\multirow{-2}{*}{\cellcolor[gray]{.75}\textbf{\begin{tabular}[c]{@{}c@{}}Consensus\\ /System\end{tabular}}} & \multirow{-2}{*}{\cellcolor[gray]{.75}\textbf{Node type}}  & \cellcolor[gray]{.75}Type & \cellcolor[gray]{.75}\textbf{Formation} & \cellcolor[gray]{.75}\textbf{ Configuration} & \cellcolor[gray]{.75} \textbf{Topology} & \cellcolor[gray]{.75} \textbf{Configuration} & {\multirow{-2}{*}{\cellcolor[gray]{.75}\textbf{Mechanism}}}\\ \hline
\hline

PoR & Clients (Researchers) \& Minters & - & - & - & Flat & Dynamic & Probabilistic lottery 
\\ \hline

\rowcolor[gray]{.9}PoB & Clients, Miners \& Minters & - & - & - & Flat & Dynamic & Probabilistic lottery  \\ \hline
PoSV & Clients \& Minters & Open & Implicit & Dynamic & & & Probabilistic lottery \\ \hline
\rowcolor[gray]{.9}PoC & Clients \& CVNs & Open & Explicit & Dynamic & - & - & Voting \\ \hline
PoI & Clients \& transaction partners & Open & Implicit & Dynamic & - & - & Probabilistic lottery \\ \hline
\end{tabular}
\end{table*}

\begin{table*}[h]
\centering
\caption{Comparing security properties of Hybrid and N-POS/POW Consensus Algorithms.}
\label{tab:nopospowSec}
\begin{tabular}{P{20mm}|P{20mm}|c|P{15mm}|c|c|c}
\hline
\rowcolor[gray]{.75}
 & 
 &
 & 
 &
\multicolumn{3}{c|}{Attack Vectors} 
\\ \cline{5-7}

{\multirow{-2}{*}{\cellcolor[gray]{.75}\textbf{Consensus}}} & {\multirow{-2}{*}{\cellcolor[gray]{.75}\textbf{Authentication}}} &
{\multirow{-2}{*}{\cellcolor[gray]{.75}\textbf{Non-repudiation}}} & 
{\multirow{2}{*}{\cellcolor[gray]{.75} \textbf{\specialcell{Censorship\\resistance}}}} & 
\cellcolor[gray]{.75}\textbf{Adversary tolerance} &
\cellcolor[gray]{.75}\textbf{Sybil protection} & 
\cellcolor[gray]{.75}\textbf{DoS Resistance} \\ \hline
\hline
PoR & X/\checkmark & \checkmark & High & $3f + 1$ & \checkmark & \checkmark \\ \hline
\rowcolor[gray]{.9}PoB & X & \checkmark & High & $3f + 1$ & \checkmark & \checkmark \\ \hline
PoSV & X  & \checkmark & High & $3f + 1$ & \checkmark & \checkmark \\ \hline
\rowcolor[gray]{.9}PoC  & \checkmark & \checkmark & High & ? & \checkmark & \checkmark \\ \hline
PoI  & X & \checkmark & High & ? & \checkmark & \checkmark \\ \hline
\end{tabular}
\end{table*}

\begin{table*}[!h]
\centering
  \caption{Comparison of additional attack vectors protection for Hybrid and N-POS/POW Consensus Algorithms}

   \begin{tabular}{P{20mm}|P{25mm}|P{15mm}|P{15mm}|P{15mm}|P{20mm}|P{25mm}}
    \hline
    \rowcolor[gray]{.6} 
      \centering \textbf{\specialcell{Consensus\\/System}} & 
      \centering \textbf{Nothing-at-Stake} & 
      \centering \textbf{Bribing} & 
      \centering \textbf{Long-range} & 
      \centering \textbf{Coin-age} & 
      \centering \textbf{Pre-computing} & 
      \centering \textbf{Cartel formation} \tabularnewline [2ex] 
      \hline
      \hline
	    PoR & X & X & X & X & X & X \\\hline
	  \rowcolor[gray]{.9}   PoB & X & X & X & X & X & \checkmark \\\hline
        PoSV & X & X & X & X & X & X \\\hline
\end{tabular}
\label{tab:nopoSSecOther}
\end{table*}

\begin{table*}[!h]
\centering
  \caption{Comparing performance properties of Consensus Algorithms of Hybrid and N-POS/POW.}

   \begin{tabular}{p{20mm}|p{15mm}|p{30mm}|p{15mm}|p{15mm}|p{20mm}}
    \hline
      \rowcolor[gray]{.6} 
      \centering\textbf{Consensus} & 
      \centering\textbf{Fault tolerance} & 
      \centering\textbf{Throughput} & 
      \centering\textbf{Scalability} & 
      \centering\textbf{Latency} & 
      \centering\textbf{Energy} \tabularnewline [2ex]
      \hline
       \hline
	     PoR & $2f + 1$ & High & Medium & Low & Low \\\hline
          \rowcolor[gray]{.9}PoB & $2f + 1$ & Medium & Low & Medium & Medium \\\hline
        PoSV & $2f + 1$ & High & Medium & Low & Low \\\hline
         \rowcolor[gray]{.9} PoC & ? & LoW (10.6 TPS \cite{faircoinTPS2019}) & Medium & Low & LoW \\\hline
        PoI & ? & High & Medium & Low & Low \\\hline
\end{tabular}
\label{tab:nopospowPerform}
\end{table*}

\begin{table*}[!h]
\caption{Hybrid \& Non-PoW/PoS currencies}
\centering
  \begin{tabular}{p{20mm}|p{20mm}|p{30mm}|p{20mm}|p{20mm}|p{20mm}}
   \hline
  \rowcolor[gray]{.6} 
\centering\textbf{Currency } & 
\centering\textbf{Genesis date (dd.mm.yyyy)}& 
\centering\textbf{Block reward}& 
\centering\textbf{Total supply}& 
\centering\textbf{Consensus}& 
\centering \textbf{Block Time} \tabularnewline [2ex]
          \hline
	   \hline

	 Gridcoin & 24 Mar 2016&  Minting & 42 Million& PoR, PoS & 1 minute\\\hline
      \rowcolor[gray]{.9}  Slimcoin & May 2014& 50-250 coins & 133 Million & PoB, PoW, PoS &1.5 minutes\\\hline
     Reddcoin & January 20, 2014& Block reward& 2.8 Billion & PoSV & 1 minute\\\hline
      \rowcolor[gray]{.9}  Faircoin & 6th of March, 2014. & Block reward& 5.3 Million & PoC &Depends on Time-weight Parameter\\\hline
     Burst & 11 August 2014 & Reduces at a fixed rate of 5 percent each month& 204 Million & PoC &4 minutes\\\hline
     \rowcolor[gray]{.9}   NEM & March 31st, 2015 & transaction fees only + node rewards& 899 Million & PoI &1 minute\\\hline
     
\end{tabular}
\label{tab:dash}
\end{table*}
\section{Non-incentivised Consensus}
\label{sec:nonIncentivConsensus}
In this section, we present non-incentivised consensus algorithms that are used in private blockchain systems well-suited for non-crypto-currency applications. These algorithms are mostly based on classical consensus algorithms with special features added for their adoption for the corresponding blockchain systems.

One of the major initiatives within the private blockchain sphere is the Hyperledger project, which is an industry-wide effort \cite{hyperledger2018}. Founded by the Linux Foundation, it is a consortium of some of the major tech vendors of the world. It provides an umbrella to facilitate the development of different types of open source projects utilising private blockchains with a specific focus to address issues involving business and governmental use-cases. Currently, there are six major projects within Hyperledger: Hyperledger Fabric \cite{hyperledgerFabric2018}, Hyperledger Sawtooth \cite{hyperledgerSawtooth2018}, Hyperledger Burrow \cite{hyperledgerBurrow2018}, Hyperledger Iroha \cite{hyperledgerIroha2018} and Hyperledger Indy \cite{hyperledgerIndy2018}. Each of them is analysed below with a brief introduction.

\subsection{Hyperledger Fabric}
Hyperledger Fabric is the first major private blockchain system that  originated from the Hyperledger ecosystem \cite{hyperledgerFabric2018}. It has been designed with strong privacy in mind to ensure that different businesses organisations, including governmental entities, can take advantage of a blockchain system in different use-cases. A crucial capability of Fabric is that it can maintain multiple ledgers within its ecosystem.  This is a useful feature, which separates Fabric from other blockchain systems consisting of only one ledger in each of their domains.

A key strength of Fabric is its modular design and pluggable features. For example, Fabric is not dependant on a particular format of ledger data, which is useful in several use-cases. In addition, the consensus mechanism is fully pluggable. Therefore, different types of consensus algorithms can be used in different situations. 

As part of its consensus process, Fabric utilises a special entity called Orderer, which is responsible for creating a new block and extending the ledger by adding the block in the appropriate order. In addition, there are other entities known as endorsers. Each endorser is responsible for validating and endorsing a transaction where it checks if an entity is allowed to perform a certain action in a ledger encoded within the transaction. Other participating entities are general users who create transactions. All the  entities, including the Orderer(s) and the endorsers,  are registered and authenticated via a Fabric specific special entity called \textit{Membership Service Provider} (MSP). The MSP is responsible for managing the identities of all participants in the ledger. Using this identity layer, it is possible to create security policies that dictate which entities can perform what actions within a specific ledger. A simple flow of a consensus process in Fabric is illustrated in Figure \ref{fig:hpledgercs}.
\begin{figure}
\begin{mdframed}[backgroundcolor=black!7,rightline=false,leftline=false]
\begin{enumerate}[leftmargin=2mm,label=\Alph*]
	\item All required entities are registered in the MSP.
	\item A channel with a ledger is initiated. In addition, a policy is created containing the endorsement criteria as well as other security and privacy criteria.
	\item A chaincode (smart-contract written either in Java or Go) is deployed in the ledger.
    \item When an entity wishes to invoke certain functions in the chaincode to read data from the ledger or to write data into the ledger, it submits a transaction proposal to all the required endorsers as dictated in the policy.
    \item Each endorser validates the proposal, executes the chaincode and returns a proposal response consisting of other ledger data.
    \item The proposal, its response,  and other ledger data are encoded as a transaction and sent to the Orderer.
    \item The Orderer creates a block using the transaction and returns the block to the endorsers.
    \item Each endorser validates the block and, if validated, extends the ledger by attaching the new block. This essentially updates the state of the ledger.
\end{enumerate}
\end{mdframed}
\caption{A simple flow of a consensus process in Fabric.}
\label{fig:hpledgercs}
\end{figure}

The number of Orderer can be increased to distribute the ordering service. Currently, it supports SOLO and Kafka. A SOLO ordering service consists of just one single orderer and hence, cannot provide any type of fault tolerance. That is why it is not recommended to utilise the SOLO model in the deployed system and has only been provided for initial testing. On the other hand, the Kafka Orderer utilises a Kafka cluster for deploying distributed Orderers. Kafka is a distributed streaming platform with a pub-sub architecture \cite{Kafka2018} and is coupled with Zookeeper, a distributed coordination service \cite{zookeeper2018}. At this point, the Kafka Orderer is the only recommended setting for achieving consensus in Fabric. An SBFT (Simplified Byzantine Fault Tolerance) based consensus algorithm is currently being developed and is to be released soon.

\subsection{Hyperledger Sawtooth}
Hyperledger Sawtooth, initially developed by Intel, is a software framework for creating distributed ledgers suitable for a variety of use cases \cite{hyperledgerSawtooth2018}. Sawtooth utilises a novel consensus algorithm called Proof-of-Elapsed-Time (PoET),  which depends on Intel SGX (Software Guard Extension). Intel SGX is a new type of Trusted Execution Environment (TEE) integrated into the new generation of Intel processors. SGX enables the execution of code within a secure enclave inside the processor, whose validity can be verified using a remote attestation process supported by the SGX.

PoET, similar to the Nakamoto consensus algorithm in Bitcoin, relies on the concept of electing a leader in each round to propose a block to be added in the ledger. The difference is that the Nakamoto algorithm and its variants select a leader by a lottery mechanism, which utilises computing power to generate a proof, as described previously. However, PoET solely relies on the Intel SGX capability to elect a leader. During each round, every validator node in the network, requests for a wait time from a trusted function in the SGX enclave. The validator that is assigned the shortest waiting time is elected as the leader for that round. The winning validator then can propose a block, consisting of a series of transactions from the defined transaction family. Other validators can utilise a trusted function supported by SGX to assess whether a trusted function has assigned the shortest time to the winning validator, and the winning validator has waited the specified amount of time. Furthermore, other validators verify the validity of the block before it is included in the ledger. The inclusion of the PoET as a consensus algorithm enables Sawtooth to achieve massive scalability as it does not need to solve a hard, computationally intensive cryptographic puzzle. In addition, it allows Sawtooth to be used not only for a permissioned ledger, but also for a public ledger.
\subsection{Hyperledger Burrow}
Hyperledger Burrow is a private (permissioned) deployment of the Ethereum platform \cite{hyperledgerBurrow2018}. It has been created and then deposited to the Hyperledger code-base by Monax Industries Limited \cite{Monax2018}. The core component in Burrow is a permissioned version of the EVM (Ethereum Virtual Machine) to ensure that only authorised entities can execute code. Two additional components have been added: Byzantine fault-tolerant Tendermint protocol \cite{tendermintPaper2014, burrowGithub2018} and the RPC gateway.

The Tendermint consensus falls under the category of a Byzantine Fault Tolerance (BFT) algorithm, which can be used to achieve consensus even under the Byzantine behaviour of a certain number of nodes as presented in Section \ref{sec:incentivised:subsec:pos:subsubsec:bft}. 

Burrow depends on several validators, which are known (authorised) entities with the duty to validate each block utilising the Tendermint consensus algorithm. This algorithm allows consensus to be achieved in Burrow with 1/3 nodes exhibiting Byzantine behaviour, either acting maliciously or having been down due to network or system failure.

Since Burrow utilises the EVM, a wide-range of smart-contracts and DApps (Decentralised Applications) could be deployed. Using the Tendermint algorithm with a set of known validators allows Burrow to scale at a much faster rate than Ethereum while preserving the privacy of transactions by allowing only known entities to participate in the network.

\subsection{Hyperledger Iroha}
Hyperledger Iroha is a private blockchain system initially developed by Soramitsu, Hitachi, NTT Data, and Colu and is currently hosted by Linux foundation under the Hyperledger Project \cite{hyperledgerIroha2018,IrohaCodeBase2018}. Iroha aims to create a simple blockchain infrastructure which can be incorporated into any system which requires a blockchain architecture underneath to function. The major emphasis while designing Iroha is on a simpler construction with a strong focus on mobile-friendly application development using a novel consensus mechanism called YAC (Yet Another Consensus) \cite{IrohaCodeBase2018,IrohaTutorial2018}. One fundamental different of Iroha from other Hyperledger project is its fine-grained permission control mechanism which allows defining  permissions for all relevant commands, queries, and even joining in the network.

The core architecture consists of several components \cite{IrohaTutorial2018,irohaDocu2018}. A brief description of its major components is presented below: 
\begin{itemize}
	\item \textbf{Troii} represents the entry point of any application to the Iroha network. It utilises gRPC (gRPC Remote Procedure Calls \cite{grpc2018}), an open source RPC framework, to interact with different peers and entities within the blockchain network.
 	\item \textbf{Model} represents how different entities are represented within the system and defines the mechanism to interact with them. 	
	\item \textbf{Network} provides the network functionalities required to maintain the P2P network and to propagate transactions in the network.
    \item \textbf{Consensus} facilitates the functionalities related to achieving consensus in the network using the YAC consensus protocol, a practical byzantine fault-tolerant algorithm (discussed below).	
	\item \textbf{Simulator} provides a mechanism to simulate the effects of transactions on the chain by creating a temporary snapshot of the chain state.
    \item \textbf{Validator} allows the validation of transactions by verifying its formats and signature along with the verification of business rules and policies involved in the transactions. There are two types of validations in Iroha:
    	\begin{itemize}
    		\item \textit{Stateless} validation checks for transaction formats and signature.
            \item \textit{Stateful} validation checks the business rules and policies, e.g.,  if a certain action is allowed by an entity.
    	\end{itemize}
    \item \textbf{Synchroniser} is a part of the consensus component and is responsible for synchronising the chain to a new or disconnected node.
	\item \textbf{Ametsuchi} is the storage component of Iroha and is used to store the blocks and the chain state known as World State View (WSV).
\end{itemize}

These components are used by three core entities within the architecture \cite{IrohaTutorial2018}:
\begin{itemize}
\item Clients are applications that they can query data from the allowed Iroha chain as well as can perform certain actions, called commands, by which the state of the chain is updated. For each of these, clients need to interact with the peer.
\item Peers are nodes that have the following two functionalities:
	\begin{itemize}
	\item To maintain a copy of the ledger. Applications can thus interact with a peer to query a chain or to submit transactions to update the chain.
    \item To participate in the consensus process by maintaining its address, identity and trust as a single entity in the network. 
	\end{itemize}
\item Ordering service node(s): Like Fabric, ordering service nodes are responsible for ordering transactions and creating a proposal of a block.
\end{itemize}

With these components and entities, a flow of transactions in \textit{Iroha} is briefly presented in Figure \ref{fig:irohatrans} \cite{IrohaTutorial2018}.

\begin{figure}[h]
\begin{mdframed}[backgroundcolor=black!7,rightline=false,leftline=false]
\begin{enumerate}[leftmargin=4mm,label=\Alph*]
	\item A client prepares and sends a transaction to a peer using Troii.
    \item The peer performs stateless validation to the transaction and forwards the transaction to the ordering service using an ordering gate.
    \item The ordering service combines and orders transactions from different peers in a transaction proposal which is then broadcast to the peers. 
    \item Each peer performs a stateful validation of the proposal using the simulator and creates a block consisting of only verified transactions. Each peer signs the block, generates a hash of the proposed block and finally, creates a tuple containing the hash and the signature. Such a tuple is called a \textit{vote}. The block and the vote are then internally sent to the consensus gate to initiate the YAC mechanism.    
    \item The YAC mechanism in each peer prepares an ordered list of voting peers utilising the hashes created in the previous step. The first peer in the list is regarded as the \textit{leader} and is responsible for aggregating votes from other voting peers.    
    \item After aggregating all votes from the voting peers, the leader computes the supermajority (usually 2/3rd) of votes for a certain hash (signifying a block).
    \item Once a supermajority for a proposed block is achieved, the leader propagates a commit message for this particular block to all voting peers.
    \item Each voting peer verifies the commit message and adds the block to the blockchain.
\end{enumerate}
\end{mdframed}
\caption{A flow of transactions in \textit{Iroha}.}
\label{fig:irohatrans}
\end{figure}

\subsection{Hyperledger Indy}
Hyperledger Indy is a private blockchain system purposefully built for providing an ecosystem for blockchain-based self-sovereign identity \cite{hyperledgerIndy2018, FerdousSelfSovreign2019}. The concept of Self-Sovereign Identity has been initially promoted by the Sovrin foundation \cite{Sovrin2018}, a non-profit international entity consisting of several private organisations to promote the notion of Self-sovereign Identity. The Indy project is closely associated with the Sovrin foundation focusing on materialising this notion of a self-sovereign identity system as a public identity utility.  

Currently, Indy consists of the following two major components:
\begin{itemize}[leftmargin=3mm]
  \item \textbf{Indy-plenum}:    Plenum is the underlying distributed ledger (blockchain) construct of the Indy platform. Like any distributed ledger, the Plenum ledger is fundamentally an ordered log of transactions. In addition, it consists of several nodes, among which a single or a few chosen ones act as the leader responsible for ordering the transactions. The nodes execute a consensus protocol which utilises a three-phase commit to reach agreement among themselves regarding the order of the transactions.

  \item \textbf{Indy-SDK}: This provides the required software APIs and tools to enable other software to interact with the Plenum ledger. It hides all the intricate internals from the users of the platform so that the platform can be utilised without even knowing the complexities of the ledger and its associated consensus protocol.
\end{itemize}

The consensus protocol utilised in Indy is called RBFT (Redundant Byzantine Fault Tolerance) \cite{RBFT2013}. Like any other byzantine fault tolerance protocol, it relies on $3f + 1$ nodes (a participant in the consensus protocol) in order to handle $f$ byzantine nodes \cite{RBFT2013, IrohaConsensus2018}. For example, it requires four deployed nodes in order to handle a single byzantine node. Each participating node in RBFT deploys two (or more) protocol instances, aptly called Master and Backup protocol instance, each of which is executed in parallel. A separate primary node (also called a leader) is selected from the master and the backup protocol instance. The leader is responsible for ordering the transactions. Its performances, i.e., latency and throughput, are periodically observed by the other instances. If its performance degrades, a different leader is selected from the backup instance.

Indy maintains a number of ledgers for different purposes, unlike many other blockchain systems which employ a solo ledger. For example, separate ledgers are maintained for node maintenance, for identity transactions and so on. Clients (users via their appropriate software interfaces) can interact with these ledgers via different nodes for updating the ledger via transactions and for reading from the ledger via queries. A fine-grained permission mechanism can be used to dictate which client has to write permissions, however, any client can read from the ledger. 

Once a node receives a transaction from a client, it performs some validation and then broadcasts the transaction to other nodes in the network. When the transaction reaches enough nodes, the primary node starts a new consensus round using a three-phase commit mechanism. In the end, all nodes agree to the order proposed by the primary node and add the transaction into the corresponding ledger.
\subsection{Analysis}
In this section, we analyse the non-incentive consensus protocols against the criteria selected before.
Block and reward properties are not considered as they are not relevant for non-incentivised consensus protocols. We use the notation `\checkmark'\, to indicate an algorithm satisfies a particular property and the notation `-'\, to indicate that there is no information regarding that specific property. For other properties, explanatory texts are added.

\vspace{2mm}
\noindent\textbf{\textsc{Structural properties.}} Tn Table \ref{tab:nonIncentiveHypPerStruc}, we present the comparison of structural properties among the non-incentivised consensus algorithms discussed in this section. As evident from the table, different algorithms use different types of nodes, and all algorithms are based on single committee with closed committee type and explicit committee formation. Only YAC relies on a dynamic configuration which utilises the reputations of the nodes from previous interactions; all others have the static configuration.

Voting is the predominantly used underlying mechanism, which is utilised by Tendermint Burrow, YAC, and RBFT, whereas PoET relies on a lottery mechanism. Fabric currently utilises the ordering services by the orderer. In the future, it might utilise SBFT, which leverages the voting mechanism.

\begin{table*}[h]
\centering
\caption{Comparing structural properties of Consensus Algorithms of Hyperledger Systems.}
\label{tab:nonIncentiveHypPerStruc}
\begin{tabular}{c|c|p{35mm}|c|c|c}
\hline
\rowcolor[gray]{.75}
\cellcolor[gray]{0.75} & \cellcolor[gray]{0.75}  & \cellcolor[gray]{0.75}  & \multicolumn{3}{c|}{ \textbf{Single committee}} \\ \cline{4-6}

\multirow{-2}{*}{\cellcolor[gray]{0.75} \textbf{\begin{tabular}[c]{@{}c@{}}Consensus\\ /System\end{tabular}}} & {\multirow{-2}{*}{\cellcolor[gray]{0.75} \textbf{Mechanism}}} & \multirow{-2}{*}{\cellcolor[gray]{0.75} \textbf{Node type}} & \cellcolor[gray]{0.75} \textbf{Type} & \cellcolor[gray]{0.75} \textbf{Formation} & \cellcolor[gray]{0.75}\textbf{Configuration} \\ \hline
\hline

Fabric & Ordering/Voting (SBFT) & Client (Regular peer), Endorsing peer \& Orderer & Close & Explicit & Static \\ \hline
\rowcolor[gray]{.9}PoET & Lottery & Client, Transaction processor \& Validator & Close & Explicit & Static \\ \hline
Tendermint Burrow & Voting & As Tendermint (3) & Close  & Explicit & Static \\ \hline
\rowcolor[gray]{.9}YAC & Voting & Client, Peer \& Ordering Service node & Close & Explicit & Dynamic \\ \hline
RBFT & Voting & Client \& Node & Close  & Explicit & Static \\ \hline
\end{tabular}
\end{table*}

\vspace{2mm}
\noindent\textbf{\textsc{Security properties.}} The comparison of security properties among the non-incentivised consensus algorithms is presented in Table \ref{tab:nonIncentiveHypPerSec}. All algorithms support non-repudiation via digital signature and have a significantly low censorship resistance. This is because the identities of all participating nodes are known. In case any node starts misbehaving, because of an attacker taking control of that node, it can be easily identified, and proper actions can be taken. The same logic applies for the Sybil protection and towards DoS resistance. Being mostly based BFT algorithms, all algorithms, except PoET, have $3f + 1$ adversarial tolerance. It has been found that PoET has an adversarial tolerance of $\Theta \left ( \frac{log\text{ } log\, n}{log\ n}\right )$ \cite{PoetAdversary2017}, where $n$ is the number of nodes.
\begin{table*}[h]
\centering
\caption{Comparing security properties of Consensus Algorithms of Hyperledger Systems.}
\label{tab:nonIncentiveHypPerSec}
\begin{tabular}{c|c|c|c|c|c}
\hline
\rowcolor[gray]{.75}
  \cellcolor[gray]{0.75}& \cellcolor[gray]{0.75} & \cellcolor[gray]{0.75} & \multicolumn{3}{c|}{Attack Vectors} \\  \cline{4-6} 

   \multirow{-2}{*}{ \cellcolor[gray]{0.75}\textbf{Consensus}} & \multirow{-2}{*}{ \cellcolor[gray]{0.75}\textbf{Non-repudiation}} & \multirow{-2}{*}{ \cellcolor[gray]{0.75}\textbf{Censorship resistance}} & \cellcolor[gray]{0.75} \textbf{Adversary tolerance} & \cellcolor[gray]{0.75} \textbf{Sybil protection} & \cellcolor[gray]{0.75} \textbf{DoS Resistance} \\ \hline
\hline
Fabric & \checkmark & Low & $3f + 1$ & \checkmark & \checkmark \\ \hline
\rowcolor[gray]{.9}PoET & \checkmark & Low & $\Theta \left ( \frac{log\text{ } log\, n}{log\ n}\right )$ & \checkmark & \checkmark \\ \hline
Tendermint Burrow  & \checkmark & Low & $3f + 1$ & \checkmark & \checkmark \\ \hline
\rowcolor[gray]{.9}YAC  & \checkmark & Low & $3f + 1$ & \checkmark & \checkmark \\ \hline
RBFT  & \checkmark & Low & $3f + 1$ & \checkmark & \checkmark \\ \hline
\end{tabular}
\end{table*}

\vspace{2mm}
\noindent\textbf{\textsc{Performance properties.}} The comparison of performance properties among the non-incentivised consensus algorithms is presented in Table \ref{tab:nonIncentiveHypPerform}. All algorithms can provide a good throughput and do not require to consume any significant amount of energy. PoET, utilising a lottery mechanism, can be scaled with a large number of validators, however, this will increase the latency (finality) of transactions \cite{HyperArchi2017}. All other algorithms employing a voting mechanism cannot be scaled with a large number of validators, providing low latency for the transactions. Fabric, YAC, and RBFT provide a $2f + 1$ fault tolerance, whereas the information regarding the fault tolerance for PoET and Tendermint Burrow is not specified formally. 
\begin{table*}[!h]
\centering
  \caption{Comparing performance properties of Consensus Algorithms of Hyperledger Systems.}

   \begin{tabular}{c|c|c|c|c|c}
    \hline
    \rowcolor[gray]{.6}
    \centering\textbf{Consensus} & 
   \centering\textbf{Fault tolerance}& 
    \centering\textbf{Throughput}& 
    \centering\textbf{Scalability}& 
    \centering\textbf{Latency}& 
    \centering\textbf{Energy} \tabularnewline [2ex]
          \hline
\hline
	    \centering Fabric & \centering  $2f + 1$ & \centering Good & \centering Medium & \centering Low & Low \\\hline
         \rowcolor[gray]{.9}   PoET & - & Good & Good & \centering Medium & Low \\\hline
        Tendermint Burrow & - & Good & Medium & \centering Low & Low \\\hline
        \rowcolor[gray]{.9} YAC & $2f + 1$ & Good &  Medium & \centering Low & Low \\\hline
        RBFT & $2f + 1$ & Good & Medium & \centering Low & Low \\\hline
\end{tabular}
\label{tab:nonIncentiveHypPerform}
\end{table*}
    
    
\section{Discussion}
\label{sec:disccusion}

As per our analysis in different sections, it is clear that PoW consensus algorithms have major limitations, specifically in terms of power consumption and scalability. Many regard PoS, and it is variant DPoS, to be the most suitable alternatives. To understand the applications of these algorithms in public blockchain systems, we have analysed the top 100 crypto-currencies, as reported on CoinMarketCap \footnote{https://coinmarketcap.com/} as of 18 July, 2019.

\begin{figure*}[h]
    \centering
    \includegraphics[width=.8\linewidth]{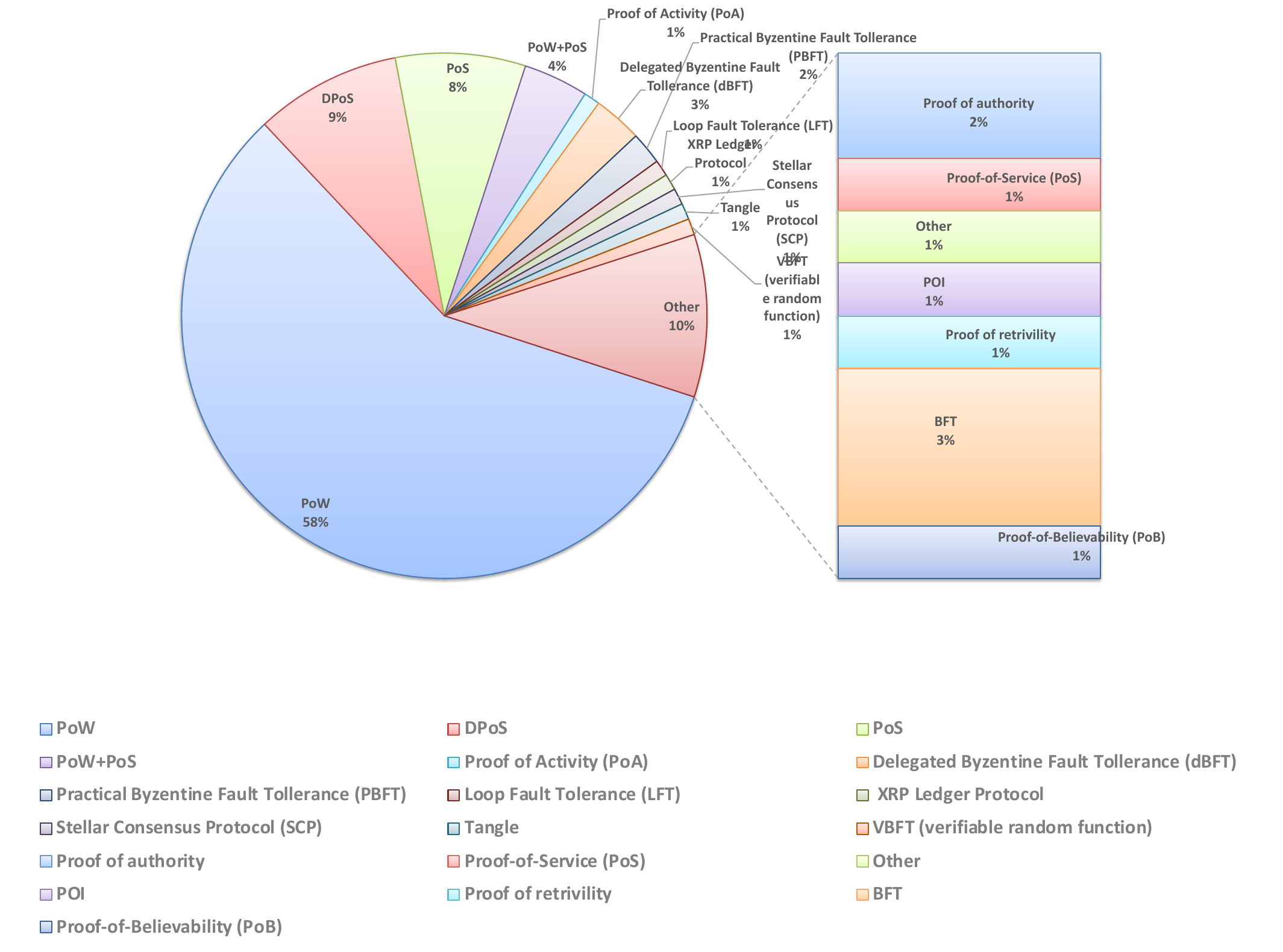}      
    \caption{Consensus algorithms in Top 100 Crypto-currencies}
    \label{fig:conAlgo}
\end{figure*}

In the first analysis, we have calculated the number of consensus algorithms used in these (top 100) crypto-currencies. The distribution of consensus algorithms is presented in Figure \ref{fig:conAlgo} . As per our analysis, PoW is still the most widely used ($57\%$) consensus algorithms to date, whereas DPoS is the second most with $11\%$, and PoS is the third most with $6\%$ used consensus algorithms. All other consensus algorithms represent the remaining $26\%$. This means that, even though many consider that PoS and DPoS are the best alternatives to PoW, their adoption is still far behind PoW.

\begin{figure*}[h]
    \centering
    \includegraphics[width=0.7\linewidth]{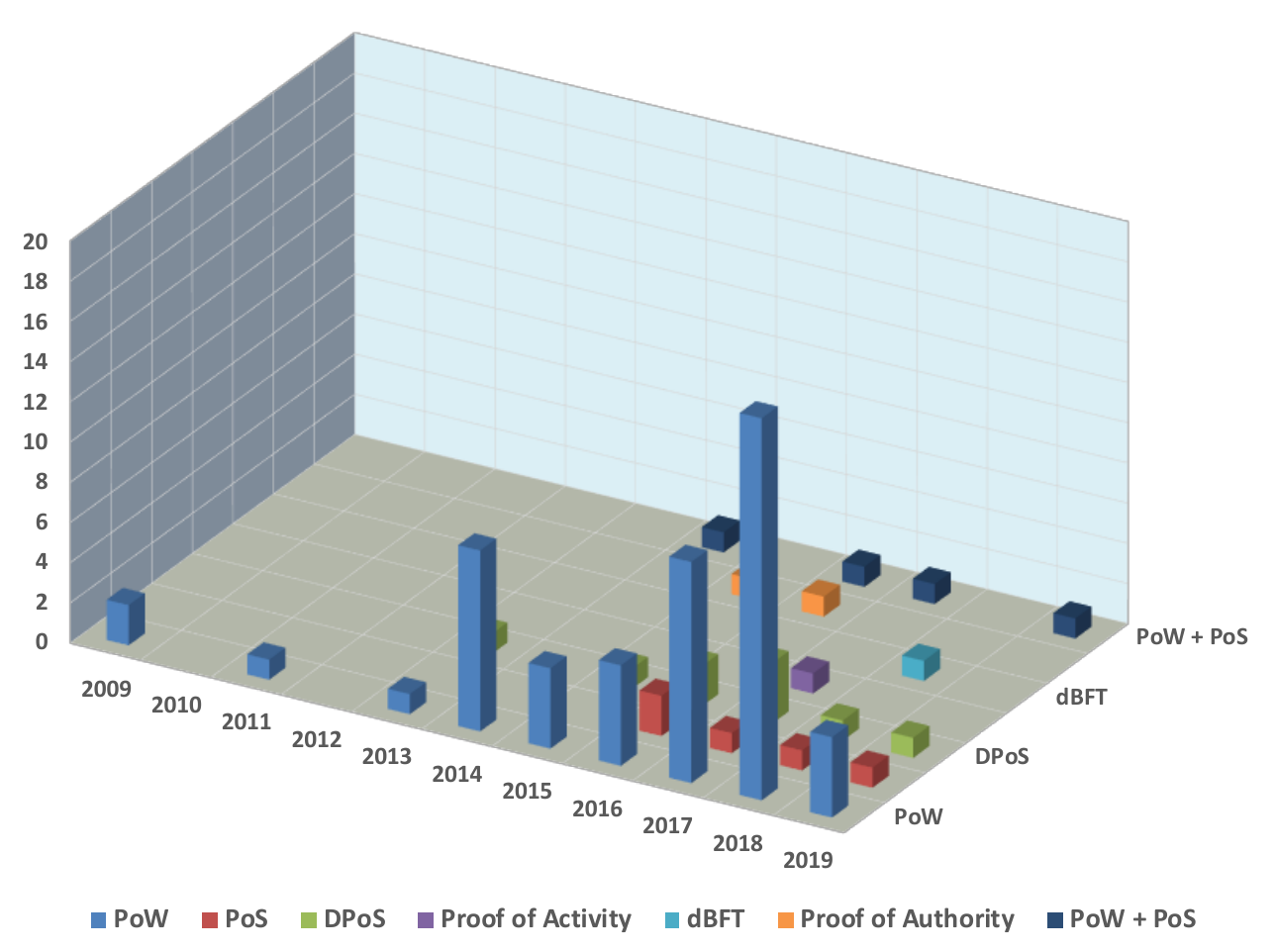}      
    \caption{Year-wise distribution of consensus algorithms in Top 100 Crypto-currencies}
    \label{fig:conAlgoYear2}
\end{figure*}

To investigate it further, we have analysed a year-wise distribution of the genesis dates of different crypto-currencies. It is to understand if there is any inclination towards an alternative consensus algorithm over PoW in recent years. The distribution is illustrated in Figure \ref{fig:conAlgoYear2}, which represents a surprising observation: PoW is still the most widely used algorithms for crypto-currencies which have been created in recent years. For example, the numbers of crypto-currencies created with PoW algorithms in last three years (2017, 2018 \& 2019) are $11$, $19$ and $4$ respectively, in comparison to $4$, $2$ and $2$ for PoS and DPoS combinedly. This implies that PoW is still the most popular consensus algorithm among the crypto-currency community. A deeper investigation reveals another insight though. The top 100 list retrieved from Coinmarketcap also contains crypto-tokens generated on top of any smart-contract platform such as Ethereum, EOS and Tron with majority tokens are built on top of Ethereum. Most of these tokens have emerged after 2016 with Ethereum utilising PoW . This could be the reason why the most recent crypto-currencies have been found to utilise PoW.

\begin{figure*}[h]
    \centering
    \includegraphics[width=0.7\linewidth]{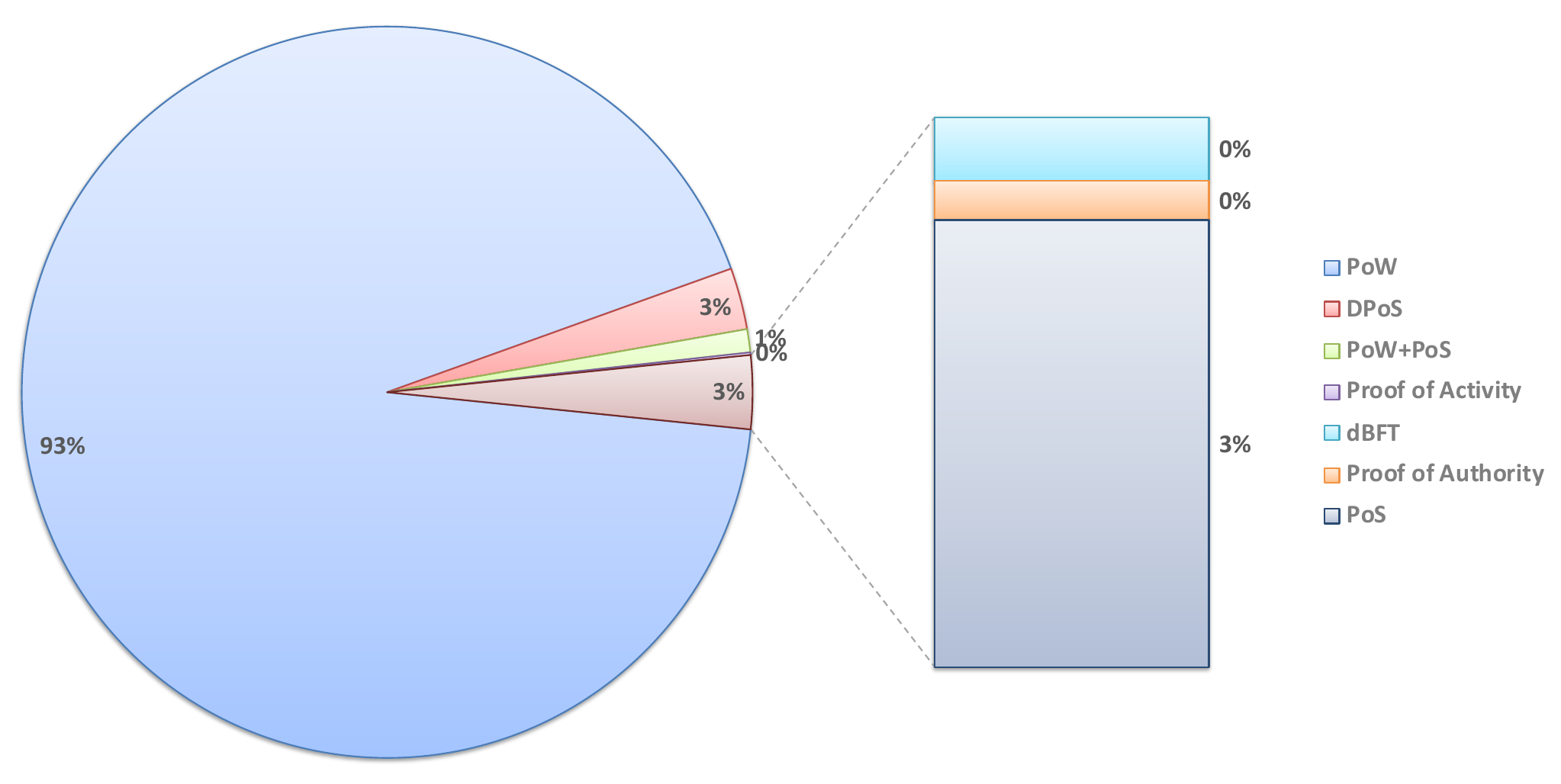}      
    \caption{Percentage of market capitalisation of consensus algorithms in top 100 Crypto-currencies}
    \label{fig:conAlgoMarketCap}
\end{figure*}

Another indication of PoW domination over other algorithms is the market-cap distribution of their corresponding crypto-currencies. The distribution is presented in Table \ref{tab:conAlgoMarketCap} and illustrated in Figure \ref{fig:conAlgoMarketCap}. Not surprisingly, PoW currencies with a market-cap of around $221$ Billion USD have a massive $93\%$ dominance over other currencies. DPoS and PoS currencies are the nearest rivals with a market-cap of around $6$ Billion USD and dominance of only $3\%$ for each group.

\begin{table}[h]
\centering
\caption{Market capitalisation of major consensus algorithms in top 100 Crypto-currencies}
\label{tab:conAlgoMarketCap}
\begin{tabular}{c|c}
  \hline
   \rowcolor[gray]{.6} 
    \centering\textbf{Consensus Algorithms} & 
     \centering\textbf{Market-cap (USD)}\tabularnewline [2ex]
          \hline
 \hline
PoW & $221,238,526,412$ \\ \hline
 \rowcolor[gray]{.9} DPoS & $6,483,606,020$ \\ \hline
PoS & $6,287,224,485$ \\ \hline
 \rowcolor[gray]{.9} PoW+PoS & $2,436,683,929$ \\ \hline
Proof of Authority & $572,188,935$ \\ \hline
 \rowcolor[gray]{.9} Proof of Activity & $274,066,240$ \\ \hline
\end{tabular}
\end{table}

From our investigation, it is clearly evident that PoW algorithm, even with its major limitations, is still the most popular consensus algorithm to be utilised in different crypto-currencies. Currencies which utilise PoW algorithms consume a significant amount of energy as illustrated in Section \ref{sec:incentivised:subsec:pow:subsubsec:limitation}. Besides, they have a reduced throughput (in terms of transaction number) compared to PoS and DPoS currencies. For example, the reported TPS (Transactions Per Second) for Bitcoin and Ethereum are $7$ and $15-25$, respectively \cite{BCVSETH2019}, while DPoS currencies EOS has a reported and estimated TPS of 50 and $4000$ respectively \cite{BCVSETH2019} and Tron has a claimed TPS of $2000$ \cite{TronTPS2019}. Clearly, DPoS currencies have better performance, at least in terms of TPS, over any PoW currency. Therefore, one might ask the underlying reason behind this counter-intuitive trend of PoW being the most popular consensus algorithm. We have identified a few reasons behind this as presented below:

\begin{itemize}
    \item Bitcoin is the most dominant crypto-currency in terms of market-cap. As of 18 July, it has a market cap of around $171$ Billion USD. In addition to this, its different forked variants (Bitcoin Cash \footnote{https://www.bitcoincash.org/} and Bitcoin Satoshi Vision \footnote{https://bitcoinsv.io/}) also have a combined market-cap of $8$ Billion USD. If we exclude Bitcoin and its variants, we have a slightly different distribution of market-cap,  as illustrated in Figure \ref{fig:conAlgoMarketCap2}. Here, the market-cap percentage of PoW algorithm is reduced from $93\%$ to $71\%$ percent, which is still significant in comparison to DPoS and PoS, its nearest rivals.
    \begin{figure*}[h]
        \centering
        \includegraphics[width=0.7\linewidth]{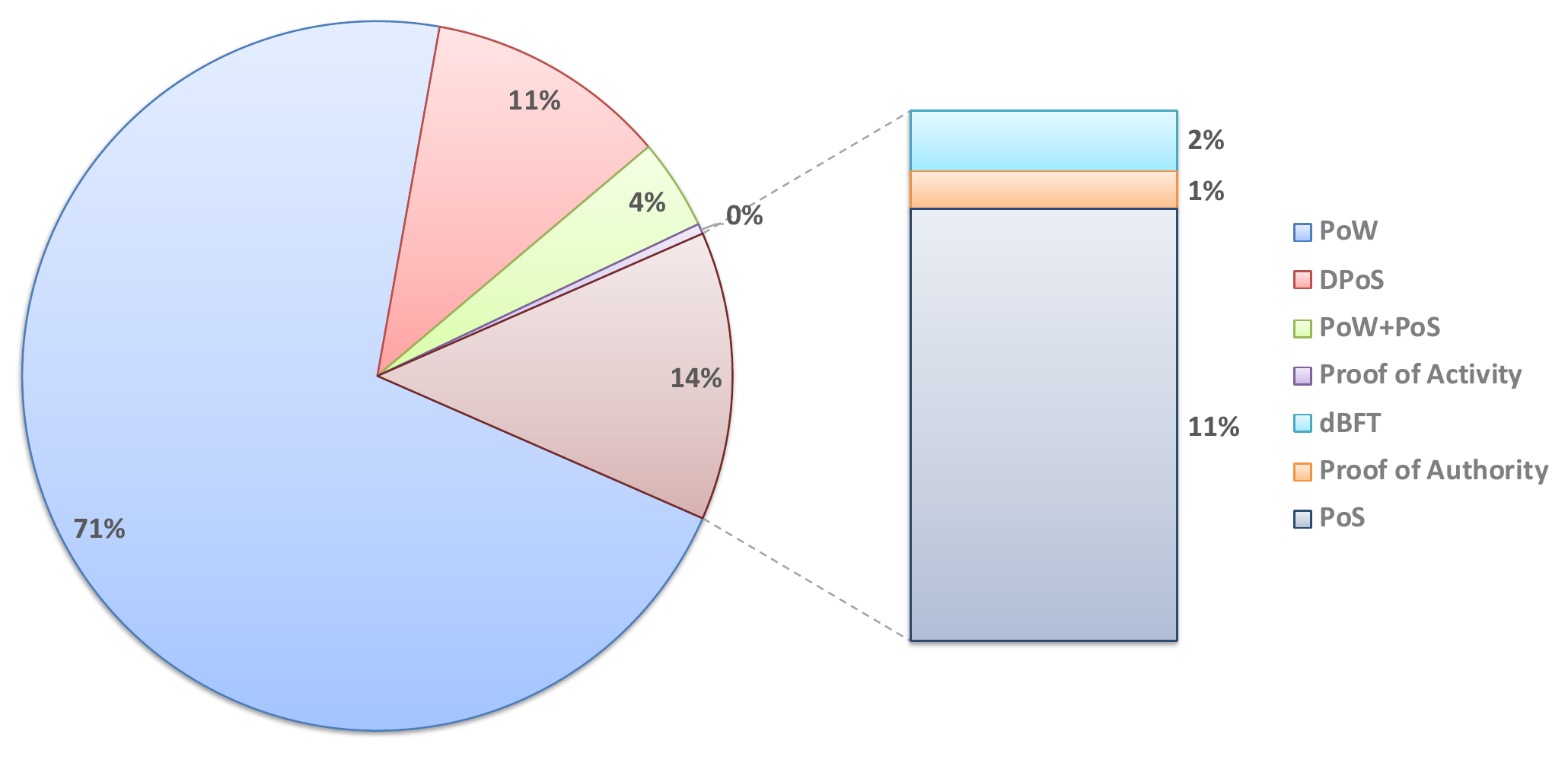}      
        \caption{Percentage of market capitalisation excluding Bitcoin and its variants}
        \label{fig:conAlgoMarketCap2}
    \end{figure*}
    \item PoW has the first-mover advantage because of Bitcoin and Ethereum, both being the pioneer in their respective domain. Bitcoin has been the first successful crypto-currency, while Ethereum is the first blockchain-based smart-contract platform. Other crypto-currencies, being motivated by their success, might have adopted the approach of utilising PoW as their corresponding consensus algorithm.
    \item Another strong argument in favour of PoW is its underlying security. The number of miners is far greater in Bitcoin than the number of validators in PoS and DPoS. This implies a better decentralisation in Bitcoin than PoS or DPoS. For example, EOS has only 21 validators, while Tron has 27 validators. The probability of collusion among these validators is far greater than that of any popular PoW currency. For this reason, many in the blockchain community have been doubtful of the security of any PoS/DPoS currency. However, there is a counter argument against this. Because of the mining centralisation issue ( highlighted in Section \ref{sec:incentivised:subsec:pow:subsubsec:limitation}), many point out that a PoW algorithm might also be prone to centralisation. Therefore, a PoW currency might also suffer from collusion attack. 
\end{itemize}

With the dominance of PoW over other consensus algorithms, one might wonder what lies ahead and might ask if there will be any shift of balance among the consensus algorithms. We believe that we will most definitely experiment with a shifting of balance in the near future. In this regard, the PoS transformation process of Ethereum will be a crucial factor. The proposed Ethereum PoS consensus mechanisms, both CFFG and CTFG, are highly regarded by the academics and industrial enthusiasts for their strong guarantee of security. With their strong focus on economic incentive and game-theoretic based approach, it is believed that their security will be as close as PoW and much better than any current PoS/DPoS algorithm can provide. In particular, the number of validators will be much higher than any number leveraged in the current PoS/DPoS algorithms. However, it is yet to be seen how they will perform once deployed in real-life settings.

\begin{figure*}[t]
	\centering
	\includegraphics [width=1.0\textwidth]{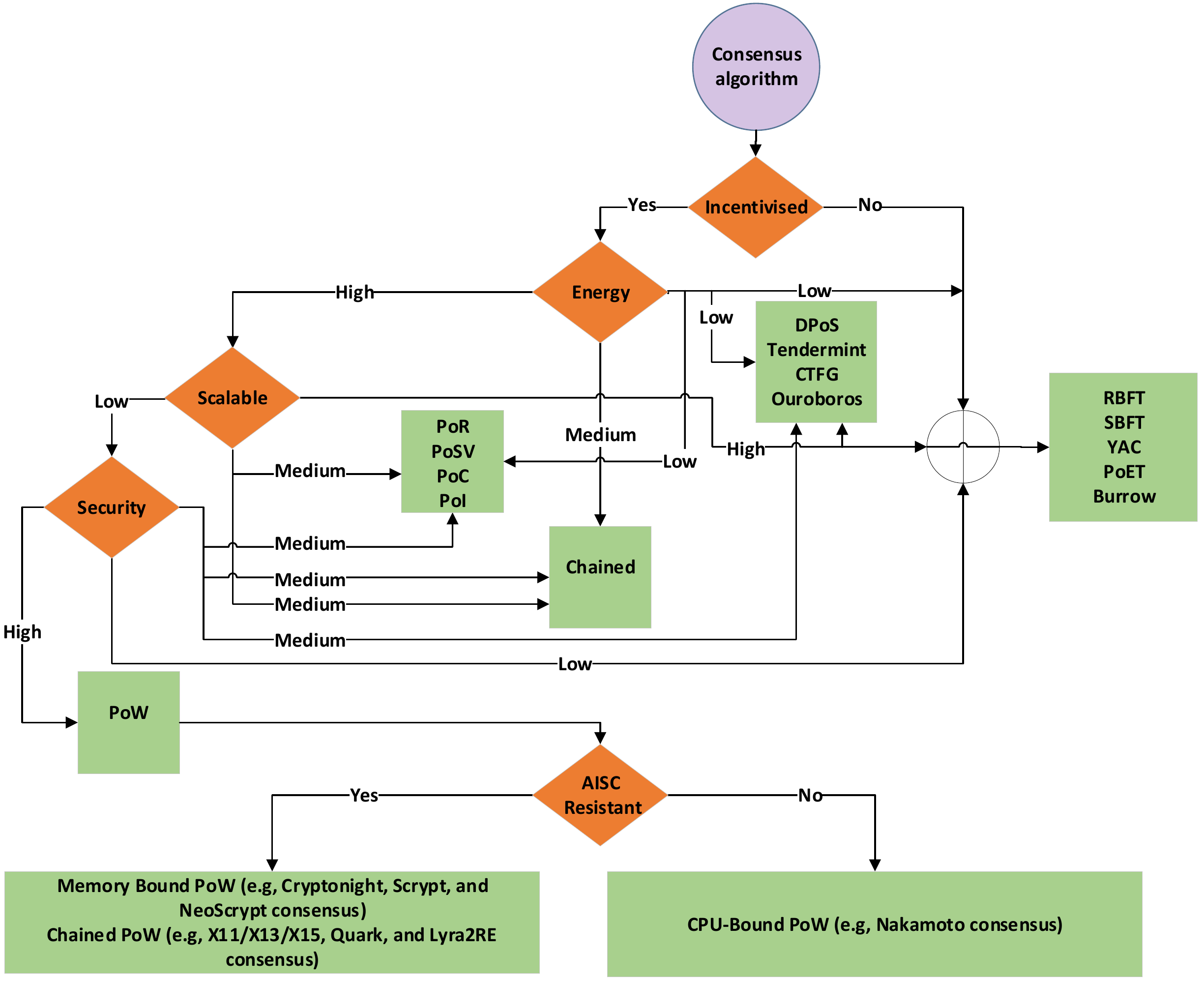}
	\caption{Decision tree to choose appropriate consensus algorithms}
	\label{fig:decisiontree}
\end{figure*}

The existence of numerous algorithms and wide variations in their properties impose a major challenge to comprehend them properly. In particular, it is often difficult to test the suitability of a particular algorithm under certain criteria. A visual tool would be a great help in this regard. Towards this aim, we present a decision tree in Figure \ref{fig:decisiontree}, which can be used to determine the suitable consensus algorithms under certain criteria in different scenarios. For example, such a decision tree diagram can be leveraged to select a particular consensus algorithm while designing/developing a new blockchain system. 

The tree utilises five critical criteria to achieve its goal: incentives, energy consumption, scalability, security (with respect to adversary tolerance),  and ASIC-resistance. If the system needs to incentivise the miner/validating nodes, then proof-of-work(PoW) and proof-of-stake (PoS) consensus are appropriate choices. Because of their underlying incentives mechanisms, the primary applications of these consensus algorithms are public crypto-currencies. On the other hand, a private blockchain network usually does not rely on any crypto-currencies to motivate or incentivise any validators to run the blockchain network. In addition to incentives, energy consumption is another determining factor in  choosing appropriate consensus algorithms. PoW-type algorithms consume high energy, whereas PoS algorithms and their derivatives consume a moderate amount of energy. PoW-types algorithms are very slow as of now and can process only a limited number of transactions. However, compromising a popular PoW-based blockchain network is very difficult, and therefore, they are more secure than their counterparts. PoW-based algorithms can also be classified based on computational complexity. As discussed earlier, ASIC is a specialised hardware, designed and used to solve hash-based computational problems. ASIC  is expensive and hinders common people from participating in the blockchain network. Therefore, memory-based PoW has been designed. Now it is widely used in different crypto-currencies. Non-incentivised consensus algorithms are mostly used in private blockchain systems. They consume a very low amount of energy compared to other types of consensus algorithms and are also very scalable. That means the miners can verify the transactions and create blocks really fast. However, a comparatively low number of validating nodes makes these algorithms more vulnerable to attacks. 

For clarity, we provide a few examples to utilise the decision tree diagram presented in Figure \ref{fig:decisiontree}. If an incentivised algorithm is required for a highly scalable blockchain system that aims to consume low energy DPoS and BFT derivatives such as Tendermint, CTFG, and Ouroboros are the preferred options. However, they will have moderate security as described earlier. On the other hand, if security is of the highest priority, PoW algorithms are more suitable. In this scenario, there are two options: memory-bound or CPU bound. If ASIC resistance is desired, one should opt for memory-bound PoW algorithms. However, in such a case, one has to sacrifice scalability, and such algorithms will consume high energy. 

Note that this is just an example of how such a diagram can be developed using our selected four criteria. Other criteria can be utilised to generate a different diagram which might be suitable for other specific scenarios. Whenever such a diagram is to be developed, the tables (Table \ref{tab:powStruct}, Table \ref{tab:powSec}, Table \ref{tab:powPerform}, Table \ref{tab:posStruct}, Table \ref{tab:posSec}, Table \ref{tab:poSSecOther}, Table \ref{tab:poSPerform}, Table \ref{tab:nopospowStruc}, Table \ref{tab:nopospowSec}, Table \ref{tab:nopoSSecOther}, Table \ref{tab:nopospowPerform}, Table \ref{tab:nonIncentiveHypPerStruc}, Table \ref{tab:nonIncentiveHypPerSec} and Table \ref{tab:nonIncentiveHypPerform}) utilised to compare different consensus algorithms against the defined properties in the taxonomy will be crucial as the these tables will provide the required template by which such a diagram can be created.

\section{Conclusion}
With the popularisation of crypto-currencies, and blockchain in general, there has been a renewed interest in the practical implications of different distributed consensus algorithms. Most of the existing systems struggle to properly satisfy the need for any wide-scale real-life deployment as they have serious limitations. Many of these limitations are due to the underlying consensus algorithm used in a particular system. Therefore, in the quest to create more suitable practical blockchain systems, the principal focus has been on distributed consensus. This has led to the explorations; either existing consensus algorithms have been exploited or novel consensus mechanisms have been introduced. The ultimate consequence of this phenomenon is a wide-range of consensus algorithms currently in existence. To advance the knowledge of this domain, it is essential to synthesise these consensus algorithms under a systematic study, which is the main motivation of this article.

Even though there have been several similar works, this is the first paper to introduce a taxonomy of properties desirable for a consensus algorithm and then utilise that taxonomy to analyse each algorithm in a detailed fashion. In addition, different consensus algorithms have been grouped into two major categories: Incentivised and Non-incentivised consensus algorithms. An incentivised consensus algorithm, exclusively utilised by public blockchain systems and crypto-currencies, relies on incentives for the participants in order to motivate them to behave as intended. On the other hand, in any non-incentivised algorithm, the participants are considered as trusted, and hence, it is assumed that no incentives are required to ensure intended behaviour. As such, these algorithms are mostly used in the private blockchain sphere. We have again grouped incentivised algorithms into three major sub-categories: PoW (Proof of Work), Proof of Stake (PoS) and consensus algorithms beyond PoW and PoS.

A PoW algorithm relies on computational complexities or memory size/performance to solve a cryptographic puzzle. There are three major approaches followed  by PoW consensus algorithms: i) a compute-bound PoW leveraging the capabilities of the processing unit, ii) a memory-bound PoW which is more reliant on the size and performance of the main memory, and iii) a chained PoW utilises a number of hashing algorithms executed consecutively one after another. Blockchain systems utilising such a mechanism has special nodes, called miner nodes, who are responsible for solving this puzzle and creating a new and valid block and extending the chain by appending this block in the existing chain. The probability to solve this puzzle depends on a network parameter, called difficulty, which is adjusted automatically after a certain period of time. As more miners participate in the network, the network parameters are adjusted in such a manner that requires more computational power to mine a new block. As the corresponding systems become more popular, it attracts more miners,  which increases the security of the system. However, the increased computational power results in more energy being consumed. Apart from this, PoW systems generally have a low throughput and do not scale properly. PoS algorithms and their corresponding mechanisms have been analysed in greater detail in Section \ref{sec:incentivised:subsec:pow}.

To alleviate the major issues of PoW, Proof of Stake (PoS) has been proposed. In PoS, the nodes who would like to participate in the block creating process are called minters, and they need to own and lock a certain amount of the corresponding crypto-currency, called stake. Such a stake is used to ensure that the minters will act as required since they will lose their stakes when acting maliciously. PoS has several variants: Chained PoS, BFT PoS and DPoS. The core  idea of a  chained  PoS  is to leverage a combination of PoW and PoS algorithms chained together to achieve consensus. BFT PoS uses a multi-round PoS mechanism in which a validator (minter) is selected, from a set of validators, by the agreement of super-majority quorum among other validators. On the other hand, DPoS selects a minter, from a set of minters, using votes from other clients of the network. PoS algorithms are generally fast and scalable, having high throughput. However, they also need to consider several other attack vectors such as Nothing-at-stake, bribing, long-range attack, cartel formation, and so on. Detailed analysis of different aspects of PoS algorithms has been presented in Section \ref{sec:incentivised:subsec:pos}.

There are also some Hybrid consensus algorithms that  combine the mechanisms of PoW and/pr PoS with another novel algorithm. Proof of Research, Proof of Burn, Proof of Stake-Velocity are examples of such an algorithm. Again, there are mechanisms that are novel and have no reliance on PoW/PoS whatsoever. Proof of Cooperation and Proof of Importance are examples of such novel algorithms. The discussion and analysis of these consensus algorithms have been presented in Section \ref{sec:hybrid}.

Finally, there are also a few non-incentivised consensus algorithms which are exclusively utilised in private blockchain systems. Hyperledger is the leading private blockchain foundation under which different private blockchain systems such as Hyperledger Fabric, Hyperledger Sawtooth, Hyperledger Burrow, Hyperledger Iroha, Hyperledger Indy, and so on. These systems rely on different other consensus mechanisms such as SBFT, PoET, Tendermint Burrow, YAC, and RBFT. Key characteristics of these consensus algorithms are high throughput and low latency with acceptable scalability. Also, the algorithms require that every entity that participates in the network must be properly authenticated. A detailed analysis of these algorithms has been presented in Section \ref{sec:nonIncentivConsensus}.

Our analysis in Section \ref{sec:disccusion} suggests that PoW, with its many disadvantages, still is the most dominant in terms of market capitalisation (indicating its adoption) and crypto-currency in the world. As discussed earlier, DPoS and PoS algorithms, PoW's closest rivals, aim to tackle many of PoW's limitations. However, their adoption is still limited. In addition to this analysis, we have presented an exemplary decision tree-based figure which can be utilised to filter out or select consensus algorithms that fit certain criteria. Such a figure will be a useful tool for any who would like to test the suitability of a certain consensus algorithm under certain criteria.

There is one issue that must be highlighted before we conclude this article. The principal focus of this article has been to explore and synthesise the consensus algorithms available in different blockchain systems. However, there are other distributed ledger systems, which do not rely on any blockchain-type structure. Instead, they utilise other structures to represent their respective ledgers. Examples of two such prominent crypto-currencies are IoTA \footnote{https://www.iota.org/} and NANO \footnote{https://nano.org/en}. Both of their ledgers are based on DAG (Directed Acyclic Graph), a specific type of directed graph with no cycle. However, IoTA uses a novel consensus algorithm called Tangle \cite{TangleWP2019} while NANO utilises a representative based consensus mechanism \cite{NanoWP2019}. These two systems have received significant attention because of their fee-less structure and fast transaction rates. However, we do not consider these systems any further as they are out of scope for this article. We plan to investigate such novel systems in the future in a different exploration.

There is high anticipation among the blockchain enthusiasts that blockchain technology will disrupt many existing application domains. However, to unlock its true potential, a blockchain system must adopt a suitable consensus that can enable it to satisfy its intended properties. This is because a consensus algorithm is the core component of any blockchain system, and it dictates how a system behaves and the performance it can achieve. However, as our analysis in this article suggests, an ideal consensus algorithm is still elusive as almost all algorithms have significant disadvantages in one way or another with respect to their security and performance. Until a consensus algorithm finds the correct balance between these crucial factors, we might not see the wide-scale adoption as many crypto-currency enthusiasts are hoping.

\bibliographystyle{unsrt}








\end{document}